\documentclass[10pt, conference, letterpaper,keeplastbox]{IEEEtran}
\usepackage{cite}
\usepackage{amsmath,amssymb,amsfonts}
\usepackage{algorithmic}
\usepackage{graphicx}
\usepackage{textcomp}
\usepackage{xcolor}
\usepackage{float}
\usepackage{subfigure}
\PassOptionsToPackage{hyphens}{url}\usepackage{hyperref}
  
\usepackage{bm}
\usepackage{multicol}

\usepackage{fancyhdr}
 \usepackage{lastpage}

\makeatletter
\newcommand*\titleheader[1]{\gdef\@titleheader{#1}}
\AtBeginDocument{%
  \let\st@red@title\@title
  \def\@title{%
    \bgroup\normalfont\large\centering\@titleheader\par\egroup
    \vskip1.5em\st@red@title}
}
\makeatother

\usepackage{graphicx}  

\usepackage{caption}
 
\DeclareCaptionLabelFormat{lc}{{#1}~#2}
\makeatletter
\def\endthebibliography{%
  \def\@noitemerr{\@latex@warning{Empty `thebibliography' environment}}%
  \endlist
}
\makeatother

\usepackage{environ}         
\usepackage{etoolbox}        
\usepackage{graphicx}        
\usepackage{caption}
 \usepackage{epsfig,endnotes}

\usepackage{floatpag,enumitem}

\usepackage{relsize}
 
 \usepackage{tikz}

\usepackage{multirow}
\usepackage{setspace}
\usepackage{mathrsfs}
\usepackage{bbm}
\usepackage{pifont}
\usepackage{widetext}

\usepackage{amsfonts}

\usepackage{amsmath, amsthm}

\usepackage{tabularx}
\usepackage{listings}
\usepackage{graphicx}
\usepackage{amssymb,booktabs}
\usepackage{array}
\usepackage{cases}

 \usepackage{supertabular}
\usepackage{mathrsfs}

\usepackage{psfrag}
\usepackage{bbding}

\usepackage{float}

\usepackage{amsbsy}

\makeatletter
\providecommand{\leftsquigarrow}{%
  \mathrel{\mathpalette\reflect@squig\relax}%
}
\newcommand{\reflect@squig}[2]{%
  \reflectbox{$\m@th#1\rightsquigarrow$}%
}
\makeatother

\usepackage{algorithm}
 \usepackage{algorithmic}

\makeatletter
\newcommand{\newalgname}[1]{%
  \renewcommand{\ALG@name}{#1}%
}

\def\centerhack#1{\hbox to 0pt{\hss\footnotesize #1\hss}}
\def\centerhackn#1{\hbox to 0pt{\hss #1\hss}}
\def\dchack#1{\vbox to 0pt{\vss{\hbox to 0pt{\hss#1\hss}}\vss}}

\usepackage{etoolbox}
\AtBeginEnvironment{align}{\setcounter{subeqn}{0}}
\newcounter{subeqn} %

\usetikzlibrary{positioning}

\newcounter{mysub}

\newtheorem*{proposition1.1}{Proposition 1.1}
\newtheorem*{proposition1.2}{Proposition 1.2}
\newtheorem*{proposition1.3}{Proposition 1.3}
\newtheorem*{proposition2.1}{Proposition 2.1}
\newtheorem*{proposition2.2}{Proposition 2.2}
\usepackage{slashbox}

\usepackage{subfigure}

\newcommand{\HH}{\bm H}

\newcommand{\W}{\bm W}

\newcommand{\boldPhi}{\boldsymbol{\Phi}}

\newcommand{\tr}{\text{trace}}
\newcommand{\Find}{\text{Find}}

\newsavebox{\minpowerunicast}
\sbox{\minpowerunicast}{%
\parbox{.2\textwidth}{%
\begin{equation} 
\resizebox{.38\textwidth}{!}{$\begin{aligned}
&\text{(P2):}\nonumber\\
&\min_{\W, \boldPhi}~\sum\limits_{k=1}^{K}\|\bm{w}_k\|^2 \\
&\mathrm{s.t.}~\frac{|\bm{h}^H_{k}(\boldPhi)\bm{w}_k |^2}{\sum\limits_{j\in\{1,\ldots,K\}\setminus\{k\}}|\bm{h}^H_{k}(\boldPhi)\bm{w}_j |^2 +  \sigma^2_k}\geq \gamma_k, \nonumber\\
&~~~~~~~~~~~\forall k\in\{1,\ldots,K\},  \nonumber\\
 &~~~\text{Constraints on $\boldPhi$;} \nonumber\\
 &\text{e.g.,~$\boldPhi:= \text{diag}( e^{j \theta_1}, \ldots, e^{j \theta_N})$, with}\nonumber\\
 &\text{~~~\textbullet~Continuous phase shifts $\theta_n|_{n \in \{1,\ldots,N\}} \in [0,2\pi)$}\nonumber\\&~~~~~\text{(in~[Wu and Zhang, Globecom '18]}\nonumber\\&~~~~~\text{[Wu and Zhang,  arXiv 1810.03961 '18])} 
 \nonumber\\
 &\text{or~\textbullet~Discrete phase shifts $\theta_n|_{n \in \{1,\ldots,N\}}\hspace{-2pt} \in \hspace{-2pt} \textstyle{\Big\{0,\hspace{-1pt} \frac{2\pi}{\tau},\hspace{-1pt}...,\hspace{-1pt}\frac{2\pi (\tau-1)}{\tau} \hspace{-1pt}\Big\}} $}\nonumber\\&~~~\text{for some $\tau$ (in~[Wu and Zhang, arXiv 1906.03165  '19]}. \nonumber
\end{aligned}$}
\end{equation}}}

\newsavebox{\minpowerbroadcast}
\sbox{\minpowerbroadcast}{%
\parbox{.2\textwidth}{%
\begin{equation} 
\resizebox{.18\textwidth}{!}{$\begin{aligned}
&\text{(P3):}\nonumber\\
&\min_{\bm{w}, \boldPhi}~ \|\bm{w}\|^2 \nonumber \\
&\mathrm{s.t.}~\frac{|\bm{h}^H_{i}(\boldPhi)\bm{w}  |^2}{\sigma^2_i}\geq \gamma_i,\nonumber\\
&~~~~~~~~~\forall i \in \{1,\ldots,K\}, \nonumber\\
 &~~~\text{Constraints on $\boldPhi$}. \nonumber
\end{aligned}$}
\end{equation}}}

\newsavebox{\minpowermulticast}
\sbox{\minpowermulticast}{%
\parbox{.2\textwidth}{%
\begin{equation} 
\resizebox{.3\textwidth}{!}{$\begin{aligned}
&\text{(P1):}\nonumber\\
&\min_{\W, \boldPhi}~\sum\limits_{k=1}^{g}\|\bm{w}_k\|^2  \nonumber \\
&\mathrm{s.t.}~\frac{|\bm{h}^H_{i}(\boldPhi)\bm{w}_k |^2}{\sum\limits_{j\in\{1,\ldots,g\}\setminus\{k\}}|\bm{h}^H_{i}(\boldPhi)\bm{w}_j |^2 +  \sigma^2_i}\geq \gamma_i,   \nonumber\\
&~~~~~~~~~~\forall k \in\{1,\ldots,g\},~\forall i \in \mathcal{G}_k, \nonumber\\
 &~~~\text{Constraints on $\boldPhi$}. \nonumber
\end{aligned}$}
\end{equation}}}

\newsavebox{\fairunicast}
\sbox{\fairunicast}{%
\parbox{.2\textwidth}{%
\begin{equation} 
\resizebox{.38\textwidth}{!}{$\begin{aligned}
&\text{(P5):}\nonumber\\
&\max_{\W, \boldPhi} \min_{k\in\{1,\ldots,K\}}   \frac{|\bm{h}^H_{k}(\boldPhi)\bm{w}_k |^2}{\gamma_k\left[\sum\limits_{j\in\{1,\ldots,K\}\setminus\{k\}}|\bm{h}^H_{k}(\boldPhi)\bm{w}_j |^2 +  \sigma^2_k\right]}  \nonumber\\
&\mathrm{s.t.}~\sum\limits_{k=1}^{K}\|\bm{w}_k\|^2 \leq P ,   \nonumber\\
 &~~~\text{Constraints on $\boldPhi$}. \nonumber
\end{aligned}$}
\end{equation}}}

\newsavebox{\fairbroadcast}
\sbox{\fairbroadcast}{%
\parbox{.2\textwidth}{%
\begin{equation} 
\resizebox{.2\textwidth}{!}{$\begin{aligned}
&\text{(P6):}\nonumber\\
&\max_{\bm{w},\boldPhi} \min_{i \in \{1,\ldots,K\}}   \frac{|\bm{h}^H_{i}(\boldPhi)\bm{w}  |^2}{\gamma_i\sigma^2_i}  \nonumber\\
&\mathrm{s.t.}~\|\bm{w}\|^2 \leq P ,    \nonumber\\
 &~~~\text{Constraints on $\boldPhi$}. \nonumber
\end{aligned}$}
\end{equation}}}

\newsavebox{\fairmulticast}
\sbox{\fairmulticast}{%
\parbox{.2\textwidth}{%
\begin{equation} 
\resizebox{.41\textwidth}{!}{$\begin{aligned}
&\text{(P4):}\nonumber\\
&\max_{\W, \boldPhi} \min_{k\in\{1,\ldots,g\}}  \min_{i \in \mathcal{G}_k}  \frac{|\bm{h}^H_{i}(\boldPhi)\bm{w}_k |^2}{\gamma_i\left[\sum\limits_{j\in\{1,\ldots,g\}\setminus\{k\}}|\bm{h}^H_{i}(\boldPhi)\bm{w}_j |^2 +  \sigma^2_i\right]}  \nonumber\\
&\mathrm{s.t.}~\sum\limits_{k=1}^{g}\|\bm{w}_k\|^2 \leq P ,  \nonumber\\
 &~~~\text{Constraints on $\boldPhi$}. \nonumber
\end{aligned}$}
\end{equation}}}
 
\newsavebox{\weightedunicast}
\sbox{\weightedunicast}{%
\parbox{.2\textwidth}{%
\begin{equation} 
\resizebox{.38\textwidth}{!}{$\begin{aligned}
&\text{(P8):}\nonumber\\
&\max_{\W, \boldPhi} \sum_{k=1}^K c_k \log\left(1 +  \frac{|\bm{h}^H_{k}(\boldPhi)\bm{w}_k |^2}{\sum\limits_{j\in\{1,\ldots,K\}\setminus\{k\}}|\bm{h}^H_{k}(\boldPhi)\bm{w}_j |^2 +  \sigma^2_k}\right) \nonumber\\
&\mathrm{s.t.}~\sum\limits_{k=1}^{K}\|\bm{w}_k\|^2 \leq P ,   \nonumber\\
 &~~~\text{Constraints on $\boldPhi$; } \nonumber\\
 &\text{e.g.,~\textbullet~General  $\boldPhi:= \text{diag}(\beta_1 e^{j \theta_1}, \ldots, \beta_N e^{j \theta_N})$ for} \nonumber\\
 &~~~~~\text{where $\beta_n \in [0,1]$, $\theta_n \in [0,2\pi)$ for $n \in \{1,\ldots,N\}$,} 
\nonumber\\
 &\text{~~or~\textbullet~Continuous phase shifts presented in (P2) above,} \nonumber\\
 &\text{~~or~\textbullet~Discrete phase shifts presented in (P2) above,}\nonumber\\&\text{(all three cases in~[Guo~\textit{et~al.}, '19: arXiv 1905.07920])}. \nonumber 
\end{aligned}$}
\end{equation}}}     

\newsavebox{\weightedmulticast}
\sbox{\weightedmulticast}{%
\parbox{.2\textwidth}{%
\begin{equation} 
\resizebox{.41\textwidth}{!}{$\begin{aligned}
&\text{(P7):}\nonumber\\
&\max_{\W, \boldPhi} \sum_{k=1}^g c_k \log\left(1 +   \min_{i \in \mathcal{G}_k}  \frac{|\bm{h}^H_{i}(\boldPhi)\bm{w}_k |^2}{\sum\limits_{j\in\{1,\ldots,g\}\setminus\{k\}}|\bm{h}^H_{i}(\boldPhi)\bm{w}_j |^2 +  \sigma^2_i} \right) \nonumber\\
&\mathrm{s.t.}~\sum\limits_{k=1}^{g}\|\bm{w}_k\|^2 \leq P ,   \nonumber\\
 &~~~\text{Constraints on $\boldPhi$}. \nonumber
\end{aligned}$}
\end{equation}}}

\newsavebox{\minpowerunicastmultichannel}
\sbox{\minpowerunicastmultichannel}{%
\parbox{.2\textwidth}{%
\begin{equation} 
\resizebox{.3\textwidth}{!}{$\begin{aligned}
&\text{(P2-MA):}\nonumber\\
&\min_{\W, \boldPhi}~\sum\limits_{k=1}^{K}\|\bm{w}_k\|^2 \\
&\mathrm{s.t.}~\frac{\bm{w}_k^H\bm{H}_{k}(\boldPhi)\bm{w}_k |}{\sum\limits_{j\in\{1,\ldots,K\}\setminus\{k\}}|\bm{w}_j^H\bm{H}_{k}(\boldPhi)\bm{w}_j +  \sigma^2_k}\geq \gamma_k, \nonumber\\
&~~~~~~~~~~~\forall k\in\{1,\ldots,K\},  \nonumber\\
 &~~~\text{Constraints on $\boldPhi$}. \nonumber
\end{aligned}$}
\end{equation}}}

\newsavebox{\minpowerbroadcastmultichannel}
\sbox{\minpowerbroadcastmultichannel}{%
\parbox{.2\textwidth}{%
\begin{equation} 
\resizebox{.18\textwidth}{!}{$\begin{aligned}
&\text{(P3-MA):}\nonumber\\
&\min_{\bm{w}, \boldPhi}~ \|\bm{w}\|^2 \nonumber \\
&\mathrm{s.t.}~\frac{\bm{w}^H\bm{H}_{i}(\boldPhi)\bm{w}}{\sigma^2_i}\geq \gamma_i,\nonumber\\
&~~~~~~~~~\forall i \in \{1,\ldots,K\}, \nonumber\\
 &~~~\text{Constraints on $\boldPhi$}. \nonumber
\end{aligned}$}
\end{equation}}}

\newsavebox{\minpowermulticastmultichannel}
\sbox{\minpowermulticastmultichannel}{%
\parbox{.2\textwidth}{%
\begin{equation} 
\resizebox{.3\textwidth}{!}{$\begin{aligned}
&\text{(P1-MA):}\nonumber\\
&\min_{\W, \boldPhi}~\sum\limits_{k=1}^{g}\|\bm{w}_k\|^2  \nonumber \\
&\mathrm{s.t.}~\frac{\bm{w}_k^H\bm{H}_{i}(\boldPhi)\bm{w}_k}{\sum\limits_{j\in\{1,\ldots,g\}\setminus\{k\}}\bm{w}_j^H\bm{H}_{i}(\boldPhi)\bm{w}_j   +  \sigma^2_i}\geq \gamma_i,   \nonumber\\
&~~~~~~~~~~\forall k \in\{1,\ldots,g\},~\forall i \in \mathcal{G}_k, \nonumber\\
 &~~~\text{Constraints on $\boldPhi$}. \nonumber
\end{aligned}$}
\end{equation}}}

\newsavebox{\fairunicastmultichannel}
\sbox{\fairunicastmultichannel}{%
\parbox{.2\textwidth}{%
\begin{equation} 
\resizebox{.38\textwidth}{!}{$\begin{aligned}
&\text{(P5-MA):}\nonumber\\
&\max_{\W, \boldPhi} \min_{k\in\{1,\ldots,K\}}   \frac{\bm{w}_k^H\bm{H}_{k}(\boldPhi)\bm{w}_k}{\gamma_k\left[\sum\limits_{j\in\{1,\ldots,K\}\setminus\{k\}}\bm{w}_j^H\bm{H}_{k}(\boldPhi)\bm{w}_j +  \sigma^2_k\right]}  \nonumber\\
&\mathrm{s.t.}~\sum\limits_{k=1}^{K}\|\bm{w}_k\|^2 \leq P ,   \nonumber\\
 &~~~\text{Constraints on $\boldPhi$}. \nonumber
\end{aligned}$}
\end{equation}}}

\newsavebox{\fairbroadcastmultichannel}
\sbox{\fairbroadcastmultichannel}{%
\parbox{.2\textwidth}{%
\begin{equation} 
\resizebox{.2\textwidth}{!}{$\begin{aligned}
&\text{(P6-MA):}\nonumber\\
&\max_{\bm{w},\boldPhi} \min_{i \in \{1,\ldots,K\}}   \frac{\bm{w}^H\bm{H}_{i}(\boldPhi)\bm{w}}{\gamma_i\sigma^2_i}  \nonumber\\
&\mathrm{s.t.}~\|\bm{w}\|^2 \leq P ,    \nonumber\\
 &~~~\text{Constraints on $\boldPhi$}. \nonumber
\end{aligned}$}
\end{equation}}}

\newsavebox{\fairmulticastmultichannel}
\sbox{\fairmulticastmultichannel}{%
\parbox{.2\textwidth}{%
\begin{equation} 
\resizebox{.41\textwidth}{!}{$\begin{aligned}
&\text{(P4-MA):}\nonumber\\
&\max_{\W, \boldPhi} \min_{k\in\{1,\ldots,g\}}  \min_{i \in \mathcal{G}_k}  \frac{\bm{w}_k^H\bm{H}_{i}(\boldPhi)\bm{w}_k}{\gamma_i\left[\sum\limits_{j\in\{1,\ldots,g\}\setminus\{k\}}\bm{w}_j^H\bm{H}_{i}(\boldPhi)\bm{w}_j +  \sigma^2_i\right]}  \nonumber\\
&\mathrm{s.t.}~\sum\limits_{k=1}^{g}\|\bm{w}_k\|^2 \leq P ,  \nonumber\\
 &~~~\text{Constraints on $\boldPhi$}. \nonumber
\end{aligned}$}
\end{equation}}}
 
\newsavebox{\weightedunicastmultichannel}
\sbox{\weightedunicastmultichannel}{%
\parbox{.2\textwidth}{%
\begin{equation} 
\resizebox{.38\textwidth}{!}{$\begin{aligned}
&\text{(P8-MA):}\nonumber\\
&\max_{\W, \boldPhi} \sum_{k=1}^K c_k \log\left(1 +  \frac{\bm{w}_k^H\bm{H}_{k}(\boldPhi)\bm{w}_k}{\sum\limits_{j\in\{1,\ldots,K\}\setminus\{k\}}\bm{w}_j^H\bm{H}_{k}(\boldPhi)\bm{w}_j +  \sigma^2_k}\right) \nonumber\\
&\mathrm{s.t.}~\sum\limits_{k=1}^{K}\|\bm{w}_k\|^2 \leq P ,   \nonumber\\
 &~~~\text{Constraints on $\boldPhi$}. \nonumber
\end{aligned}$}
\end{equation}}}     

\newsavebox{\weightedmulticastmultichannel}
\sbox{\weightedmulticastmultichannel}{%
\parbox{.2\textwidth}{%
\begin{equation} 
\resizebox{.41\textwidth}{!}{$\begin{aligned}
&\text{(P7-MA):}\nonumber\\
&\max_{\W, \boldPhi} \sum_{k=1}^g c_k \log\left(1 +   \min_{i \in \mathcal{G}_k}  \frac{\bm{w}_k^H\bm{H}_{i}(\boldPhi)\bm{w}_k}{\sum\limits_{j\in\{1,\ldots,g\}\setminus\{k\}}\bm{w}_j^H\bm{H}_{i}(\boldPhi)\bm{w}_j +  \sigma^2_i} \right) \nonumber\\
&\mathrm{s.t.}~\sum\limits_{k=1}^{g}\|\bm{w}_k\|^2 \leq P ,   \nonumber\\
 &~~~\text{Constraints on $\boldPhi$}. \nonumber
\end{aligned}$}
\end{equation}}}

\newsavebox{\minpowerunicastmultiRIS}
\sbox{\minpowerunicastmultiRIS}{%
\parbox{.2\textwidth}{%
\begin{equation}
\resizebox{.35\textwidth}{!}{$\begin{aligned}
&\text{(P2-MR):}\nonumber\\
&\min_{\W, \bm{\Phi}_1,\ldots,\bm{\Phi}_L}~\sum\limits_{k=1}^{K}\|\bm{w}_k\|^2 \\
&\mathrm{s.t.}~\frac{|\bm{h}^H_{k}(\bm{\Phi}_1,\ldots,\bm{\Phi}_L)\bm{w}_k |^2}{\sum\limits_{j\in\{1,\ldots,K\}\setminus\{k\}}|\bm{h}^H_{k}(\bm{\Phi}_1,\ldots,\bm{\Phi}_L)\bm{w}_j |^2 +  \sigma^2_k}\geq \gamma_k, \nonumber\\
&~~~~~~~~~~~\forall k\in\{1,\ldots,K\},  \nonumber\\
 &~~~\text{Constraints on $\bm{\Phi}_1,\ldots,\bm{\Phi}_L$}.\nonumber 
\end{aligned}$}
\end{equation}}}

\newsavebox{\minpowerbroadcastmultiRIS}
\sbox{\minpowerbroadcastmultiRIS}{%
\parbox{.22\textwidth}{%
\begin{equation}
\resizebox{.22\textwidth}{!}{$\begin{aligned}
&\text{(P3-MR):}\nonumber\\
&\min_{\bm{w}, \bm{\Phi}_1,\ldots,\bm{\Phi}_L}~ \|\bm{w}\|^2 \nonumber \\
&\mathrm{s.t.}~\frac{|\bm{h}^H_{i}(\bm{\Phi}_1,\ldots,\bm{\Phi}_L)\bm{w}  |^2}{\sigma^2_i}\geq \gamma_i,\nonumber\\
&~~~~~~~~~\forall i \in \{1,\ldots,K\}, \nonumber\\
 &~~~\text{Constraints on $\bm{\Phi}_1,\ldots,\bm{\Phi}_L$}. \nonumber
\end{aligned}$}
\end{equation}}}

\newsavebox{\minpowermulticastmultiRIS}
\sbox{\minpowermulticastmultiRIS}{%
\parbox{.2\textwidth}{%
\begin{equation}
\resizebox{.37\textwidth}{!}{$\begin{aligned}
&\text{(P1-MR):}\nonumber\\
&\min_{\W, \bm{\Phi}_1,\ldots,\bm{\Phi}_L}~\sum\limits_{k=1}^{g}\|\bm{w}_k\|^2  \nonumber \\
&\mathrm{s.t.}~\frac{|\bm{h}^H_{i}(\bm{\Phi}_1,\ldots,\bm{\Phi}_L)\bm{w}_k |^2}{\sum\limits_{j\in\{1,\ldots,g\}\setminus\{k\}}|\bm{h}^H_{i}(\bm{\Phi}_1,\ldots,\bm{\Phi}_L)\bm{w}_j |^2 +  \sigma^2_i}\geq \gamma_i,   \nonumber\\
&~~~~~~~~~~\forall k \in\{1,\ldots,g\},~\forall i \in \mathcal{G}_k, \nonumber\\
 &~~~\text{Constraints on $\bm{\Phi}_1,\ldots,\bm{\Phi}_L$}. \nonumber
\end{aligned}$}
\end{equation}}}

\newsavebox{\fairunicastmultiRIS}
\sbox{\fairunicastmultiRIS}{%
\parbox{.2\textwidth}{%
\begin{equation}
\resizebox{.36\textwidth}{!}{$\begin{aligned}
&\text{(P5-MR):}\nonumber\\
&\max_{\W, \bm{\Phi}_1,\ldots,\bm{\Phi}_L} \min_{k\in\{1,\ldots,K\}}   \nonumber\\
&\frac{|\bm{h}^H_{k}(\bm{\Phi}_1,\ldots,\bm{\Phi}_L)\bm{w}_k |^2}{\gamma_k\left[\sum\limits_{j\in\{1,\ldots,K\}\setminus\{k\}}|\bm{h}^H_{k}(\bm{\Phi}_1,\ldots,\bm{\Phi}_L)\bm{w}_j |^2 +  \sigma^2_k\right]}  \nonumber\\
&\mathrm{s.t.}~\sum\limits_{k=1}^{K}\|\bm{w}_k\|^2 \leq P ,   \nonumber\\
 &~~~\text{Constraints on $\bm{\Phi}_1,\ldots,\bm{\Phi}_L$}. \nonumber
\end{aligned}$}
\end{equation}}}

\newsavebox{\fairbroadcastmultiRIS}
\sbox{\fairbroadcastmultiRIS}{%
\parbox{.22\textwidth}{%
\begin{equation}
\resizebox{.22\textwidth}{!}{$\begin{aligned}
&\text{(P6-MR):}\nonumber\\
&\max_{\bm{w},\bm{\Phi}_1,\ldots,\bm{\Phi}_L}\min_{i \in \{1,\ldots,K\}}   \nonumber\\
& \frac{|\bm{h}^H_{i}(\bm{\Phi}_1,\ldots,\bm{\Phi}_L)\bm{w}  |^2}{\gamma_i\sigma^2_i}  \nonumber\\
&\mathrm{s.t.}~\|\bm{w}\|^2 \leq P ,    \nonumber\\
 &~~~\text{Constraints on $\bm{\Phi}_1,\ldots,\bm{\Phi}_L$}. \nonumber
\end{aligned}$}
\end{equation}}}

\newsavebox{\fairmulticastmultiRIS}
\sbox{\fairmulticastmultiRIS}{%
\parbox{.2\textwidth}{%
\begin{equation}
\resizebox{.38\textwidth}{!}{$\begin{aligned}
&\text{(P4-MR):}\nonumber\\
&\max_{\W, \bm{\Phi}_1,\ldots,\bm{\Phi}_L} \min_{k\in\{1,\ldots,g\}}  \min_{i \in \mathcal{G}_k}  \nonumber\\
&\frac{|\bm{h}^H_{i}(\bm{\Phi}_1,\ldots,\bm{\Phi}_L)\bm{w}_k |^2}{\gamma_i\left[\sum\limits_{j\in\{1,\ldots,g\}\setminus\{k\}}|\bm{h}^H_{i}(\bm{\Phi}_1,\ldots,\bm{\Phi}_L)\bm{w}_j |^2 +  \sigma^2_i\right]}  \nonumber\\
&\mathrm{s.t.}~\sum\limits_{k=1}^{g}\|\bm{w}_k\|^2 \leq P ,  \nonumber\\
 &~~~\text{Constraints on $\bm{\Phi}_1,\ldots,\bm{\Phi}_L$}. \nonumber
\end{aligned}$}
\end{equation}}}

\newsavebox{\weightedunicastmultiRIS}
\sbox{\weightedunicastmultiRIS}{%
\parbox{.2\textwidth}{%
\begin{equation}
\resizebox{.385\textwidth}{!}{$\begin{aligned}
&\text{(P8-MR):}\nonumber\\
&\max_{\W, \bm{\Phi}_1,\ldots,\bm{\Phi}_L} \sum_{k=1}^K \nonumber\\
&c_k \log\left(1 +  \frac{|\bm{h}^H_{k}(\bm{\Phi}_1,\ldots,\bm{\Phi}_L)\bm{w}_k |^2}{\sum\limits_{j\in\{1,\ldots,K\}\setminus\{k\}}|\bm{h}^H_{k}(\bm{\Phi}_1,\ldots,\bm{\Phi}_L)\bm{w}_j |^2 +  \sigma^2_k}\right) \nonumber\\
&\mathrm{s.t.}~\sum\limits_{k=1}^{K}\|\bm{w}_k\|^2 \leq P ,   \nonumber\\
 &~~~\text{Constraints on $\bm{\Phi}_1,\ldots,\bm{\Phi}_L$} . \nonumber
\end{aligned}$}
\end{equation}}}

\newsavebox{\weightedmulticastmultiRIS}
\sbox{\weightedmulticastmultiRIS}{%
\parbox{.2\textwidth}{%
\begin{equation}
\resizebox{.415\textwidth}{!}{$\begin{aligned}
&\text{(P7-MR):}\nonumber\\
&\max_{\W, \bm{\Phi}_1,\ldots,\bm{\Phi}_L} \sum_{k=1}^g \nonumber\\
&c_k \log\left(1 +   \min_{i \in \mathcal{G}_k}  \frac{|\bm{h}^H_{i}(\bm{\Phi}_1,\ldots,\bm{\Phi}_L)\bm{w}_k |^2}{\sum\limits_{j\in\{1,\ldots,g\}\setminus\{k\}}|\bm{h}^H_{i}(\bm{\Phi}_1,\ldots,\bm{\Phi}_L)\bm{w}_j |^2 +  \sigma^2_i} \right) \nonumber\\
&\mathrm{s.t.}~\sum\limits_{k=1}^{g}\|\bm{w}_k\|^2 \leq P ,   \nonumber\\
 &~~~\text{Constraints on $\bm{\Phi}_1,\ldots,\bm{\Phi}_L$}. \nonumber
\end{aligned}$}
\end{equation}}}

\newsavebox{\minpowerunicastmultiAntennamultiRIS}
\sbox{\minpowerunicastmultiAntennamultiRIS}{%
\parbox{.2\textwidth}{%
\begin{equation}
\resizebox{.3\textwidth}{!}{$\begin{aligned}
&\text{(P2-MA-MR):}\nonumber\\
&\min_{\W, \boldPhi}~\sum\limits_{k=1}^{K}\|\bm{w}_k\|^2 \\
&\mathrm{s.t.}~\frac{\bm{w}_k^H\bm{H}_{k}(\boldPhi)\bm{w}_k }{\sum\limits_{j\in\{1,\ldots,K\}\setminus\{k\}}\bm{w}_j^H\bm{H}_{k}(\boldPhi)\bm{w}_j +  \sigma^2_k}\geq \gamma_k, \nonumber\\
&~~~~~~~~~~~\forall k\in\{1,\ldots,K\},  \nonumber\\
 &~~~\text{Constraints on $\bm{\Phi}_1,\ldots,\bm{\Phi}_L$}. \nonumber
\end{aligned}$}
\end{equation}}}

\newsavebox{\minpowerbroadcastmultiAntennamultiRIS}
\sbox{\minpowerbroadcastmultiAntennamultiRIS}{%
\parbox{.2\textwidth}{%
\begin{equation}
\resizebox{.21\textwidth}{!}{$\begin{aligned}
&\text{(P3-MA-MR):}\nonumber\\
&\min_{\bm{w}, \boldPhi}~ \|\bm{w}\|^2 \nonumber \\
&\mathrm{s.t.}~\frac{\bm{w}^H\bm{H}_{i}(\bm{\Phi}_1,\ldots,\bm{\Phi}_L)\bm{w}}{\sigma^2_i}\geq \gamma_i,\nonumber\\
&~~~~~~~~~\forall i \in \{1,\ldots,K\}, \nonumber\\
 &~~~\text{Constraints on $\bm{\Phi}_1,\ldots,\bm{\Phi}_L$}. \nonumber
\end{aligned}$}
\end{equation}}}

\newsavebox{\minpowermulticastmultiAntennamultiRIS}
\sbox{\minpowermulticastmultiAntennamultiRIS}{%
\parbox{.2\textwidth}{%
\begin{equation}
\resizebox{.35\textwidth}{!}{$\begin{aligned}
&\text{(P1-MA-MR):}\nonumber\\
&\min_{\W, \bm{\Phi}_1,\ldots,\bm{\Phi}_L}~\sum\limits_{k=1}^{g}\|\bm{w}_k\|^2  \nonumber \\
&\mathrm{s.t.}~\frac{\bm{w}_k^H\bm{H}_{i}(\bm{\Phi}_1,\ldots,\bm{\Phi}_L)\bm{w}_k}{\sum\limits_{j\in\{1,\ldots,g\}\setminus\{k\}}\bm{w}_j^H\bm{H}_{i}(\bm{\Phi}_1,\ldots,\bm{\Phi}_L)\bm{w}_j  +  \sigma^2_i}\geq \gamma_i,   \nonumber\\
&~~~~~~~~~~\forall k \in\{1,\ldots,g\},~\forall i \in \mathcal{G}_k, \nonumber\\
 &~~~\text{Constraints on $\bm{\Phi}_1,\ldots,\bm{\Phi}_L$}. \nonumber
\end{aligned}$}
\end{equation}}}

\newsavebox{\fairunicastmultiAntennamultiRIS}
\sbox{\fairunicastmultiAntennamultiRIS}{%
\parbox{.2\textwidth}{%
\begin{equation}
\resizebox{.38\textwidth}{!}{$\begin{aligned}
&\text{(P5-MA-MR):}\nonumber\\
&\max_{\W, \bm{\Phi}_1,\ldots,\bm{\Phi}_L} \min_{k\in\{1,\ldots,K\}}   \frac{\bm{w}_k^H\bm{H}_{k}(\bm{\Phi}_1,\ldots,\bm{\Phi}_L)\bm{w}_k}{\gamma_k\left[\sum\limits_{j\in\{1,\ldots,K\}\setminus\{k\}}\bm{w}_j^H\bm{H}_{k}(\bm{\Phi}_1,\ldots,\bm{\Phi}_L)\bm{w}_j +  \sigma^2_k\right]}  \nonumber\\
&\mathrm{s.t.}~\sum\limits_{k=1}^{K}\|\bm{w}_k\|^2 \leq P ,   \nonumber\\
 &~~~\text{Constraints on $\bm{\Phi}_1,\ldots,\bm{\Phi}_L$}. \nonumber
\end{aligned}$}
\end{equation}}}

\newsavebox{\fairbroadcastmultiAntennamultiRIS}
\sbox{\fairbroadcastmultiAntennamultiRIS}{%
\parbox{.2\textwidth}{%
\begin{equation}
\resizebox{.22\textwidth}{!}{$\begin{aligned}
&\text{(P6-MA-MR):}\nonumber\\
&\max_{\bm{w},\boldPhi} \min_{i \in \{1,\ldots,K\}}   \frac{\bm{w}^H\bm{H}_{i}(\bm{\Phi}_1,\ldots,\bm{\Phi}_L)\bm{w}}{\gamma_i\sigma^2_i}  \nonumber\\
&\mathrm{s.t.}~\|\bm{w}\|^2 \leq P ,    \nonumber\\
 &~~~\text{Constraints on $\bm{\Phi}_1,\ldots,\bm{\Phi}_L$}. \nonumber
\end{aligned}$}
\end{equation}}}

\newsavebox{\fairmulticastmultiAntennamultiRIS}
\sbox{\fairmulticastmultiAntennamultiRIS}{%
\parbox{.2\textwidth}{%
\begin{equation}
\resizebox{.41\textwidth}{!}{$\begin{aligned}
&\text{(P4-MA-MR):}\nonumber\\
&\max_{\W, \bm{\Phi}_1,\ldots,\bm{\Phi}_L} \min_{k\in\{1,\ldots,g\}}  \min_{i \in \mathcal{G}_k}  \frac{\bm{w}_k^H\bm{H}_{i}(\bm{\Phi}_1,\ldots,\bm{\Phi}_L)\bm{w}_k}{\gamma_i\left[\sum\limits_{j\in\{1,\ldots,g\}\setminus\{k\}}\bm{w}_j^H\bm{H}_{i}(\bm{\Phi}_1,\ldots,\bm{\Phi}_L)\bm{w}_j +  \sigma^2_i\right]}  \nonumber\\
&\mathrm{s.t.}~\sum\limits_{k=1}^{g}\|\bm{w}_k\|^2 \leq P ,  \nonumber\\
 &~~~\text{Constraints on $\bm{\Phi}_1,\ldots,\bm{\Phi}_L$}. \nonumber
\end{aligned}$}
\end{equation}}}

\newsavebox{\weightedunicastmultiAntennamultiRIS}
\sbox{\weightedunicastmultiAntennamultiRIS}{%
\parbox{.2\textwidth}{%
\begin{equation}
\resizebox{.38\textwidth}{!}{$\begin{aligned}
&\text{(P8-MA-MR):}\nonumber\\
&\max_{\W, \bm{\Phi}_1,\ldots,\bm{\Phi}_L} \sum_{k=1}^K c_k \log\left(1 +  \frac{\bm{w}_k^H\bm{H}_{k}(\bm{\Phi}_1,\ldots,\bm{\Phi}_L)\bm{w}_k}{\sum\limits_{j\in\{1,\ldots,K\}\setminus\{k\}}\bm{w}_j^H\bm{H}_{k}(\bm{\Phi}_1,\ldots,\bm{\Phi}_L)\bm{w}_j +  \sigma^2_k}\right) \nonumber\\
&\mathrm{s.t.}~\sum\limits_{k=1}^{K}\|\bm{w}_k\|^2 \leq P ,   \nonumber\\
 &~~~\text{Constraints on $\bm{\Phi}_1,\ldots,\bm{\Phi}_L$}. \nonumber
\end{aligned}$}
\end{equation}}}

\newsavebox{\weightedmulticastmultiAntennamultiRIS}
\sbox{\weightedmulticastmultiAntennamultiRIS}{%
\parbox{.2\textwidth}{%
\begin{equation}
\resizebox{.41\textwidth}{!}{$\begin{aligned}
&\text{(P7-MA-MR):}\nonumber\\
&\max_{\W, \bm{\Phi}_1,\ldots,\bm{\Phi}_L} \sum_{k=1}^g c_k \log\left(1 +   \min_{i \in \mathcal{G}_k}  \frac{\bm{w}_k^H\bm{H}_{i}(\bm{\Phi}_1,\ldots,\bm{\Phi}_L)\bm{w}_k}{\sum\limits_{j\in\{1,\ldots,g\}\setminus\{k\}}\bm{w}_j^H\bm{H}_{i}(\bm{\Phi}_1,\ldots,\bm{\Phi}_L)\bm{w}_j +  \sigma^2_i} \right) \nonumber\\
&\mathrm{s.t.}~\sum\limits_{k=1}^{g}\|\bm{w}_k\|^2 \leq P ,   \nonumber\\
 &~~~\text{Constraints on $\bm{\Phi}_1,\ldots,\bm{\Phi}_L$}. \nonumber
\end{aligned}$}
\end{equation}}}

\begin{document}

%



\title{\LARGE Optimizations with Intelligent Reflecting Surfaces (IRSs) \\in 6G Wireless Networks: Power Control, Quality of Service, Max-Min Fair Beamforming for Unicast, Broadcast, and Multicast  with Multi-antenna Mobile Users and Multiple IRSs} 
\titleheader{Under Submission}
%
%
%


\author{\IEEEauthorblockN{Jun Zhao}
\IEEEauthorblockA{Assistant Professor
\\School of Computer Science and Engineering
\\Nanyang Technological University, Singapore 
\\
\href{mailto:junzhao@ntu.edu.sg}{JunZhao@ntu.edu.sg}\\
\href{mailto:junzhao@alumni.cmu.edu}{JunZhao@alumni.cmu.edu}\\ \href{http://www.ntu.edu.sg/home/junzhao/}{http://www.ntu.edu.sg/home/junzhao/}
}}


%
%

\markboth{Journal of \LaTeX\ Class Files,~Vol.~14, No.~8, August~2015}%
{Shell \MakeLowercase{\textit{et al.}}: Bare Demo of IEEEtran.cls for IEEE Journals}
%



\maketitle

\begin{abstract}
Intelligent reflecting surfaces (IRSs) have received much attention recently and are envisioned to promote 6G communication networks.
In this paper, for wireless communications aided by IRS units, we formulate optimization problems for \textit{power control under quality of service (QoS)} and \textit{\mbox{max-min} fair QoS} under three kinds of traffic patterns from a base station (BS) to mobile users (MUs): unicast, broadcast, and multicast. The optimizations are achieved by jointly designing the transmit beamforming of the BS and the phase shift matrix of the IRS. For power control under QoS, existing IRS studies in the literature address only the unicast setting, whereas no IRS work has considered \mbox{max-min} fair QoS. Furthermore, we extend our above optimization studies to the novel settings of \textit{multi}-antenna mobile users or/and \textit{multiple} intelligent reflecting surfaces. For all the above optimizations, we provide detailed analyses to propose efficient algorithms.
 To summarize, our paper presents a comprehensive study of optimization problems involving power control, QoS,  and fairness in wireless networks enhanced by IRSs.
\end{abstract}

\begin{IEEEkeywords}
Intelligent reflecting surfaces, 6G communications, power control, quality of service, \mbox{max-min} fair design, wireless networks.
\end{IEEEkeywords}








\section{Introduction}




\textbf{Intelligent reflecting surfaces (IRSs) and IRS-aided communications}. An \textit{intelligent reflecting surface (IRS)}, or simply an \textit{intelligent surface}, can intelligently control the wireless environment to improve signal strength received at the destination. This is vastly different from prior techniques which improve wireless communications via optimizations at the sender or receiver. Specifically, an IRS consists of many IRS units, each of which can  reflect the incident signal at a reconfigurable angle. In such \textit{IRS-aided communications}, the wireless signal travels from the source to the IRS, is optimized at the IRS, and then travels from the IRS to the destination. Such communication method is particularly useful when the source and destination such as a base station (BS) and a mobile user  (MU) have a weak  wireless channel in between  due to obstacles or poor environmental conditions, or they do not have direct line of sights. 

Because of the ability to configure wireless environments, IRSs are envisioned by many experts in wireless communications to play an important role in 6G networks. In November 2018, the Japanese mobile operator NTT DoCoMo and a startup MetaWave demonstrated the use of IRS-like technology for assisting wireless communications in 28GHz band~\cite{DOCOMO}. IRSs have been compared with the massive MIMO technology used in 5G communications. IRSs reflect wireless signals and hence consume little power, whereas massive MIMO transmits signals and needs much more power~\cite{hu2018beyond}.





\textbf{Problems studied in this paper: Various optimizations in IRS-aided communications.} In this bold paper, we investigate how to jointly design the transmit beamforming of the BS and the phase shift matrix of the IRS, for the two optimization problems of \textbf{power control under quality of service (QoS)} and \textbf{\mbox{max-min} fair QoS}, under various traffic models from the BS to MUs including \textit{unicast}, \textit{broadcast}, and \textit{multicast}, in consideration of constraints of the phase shift matrix (e.g., with or without amplitude attenuation, \textit{continuous} or \textit{discrete} phase shifts), with extensions to \textit{multi}-antenna mobile users or/and \textit{multiple} IRSs.
We characterize the QoS for an MU by the received signal-to-interference-plus-noise ratio (SINR) at the MU.


\textbf{Contributions.} The contributions of this paper are summarized as follows:
\begin{enumerate}
\item We formulate optimization problems for \textit{power control under QoS} and \textit{\mbox{max-min} fair QoS} under three kinds of traffic patterns from the BS to the MUs: unicast, broadcast, and multicast. The former optimization problem is addressed only in the unicast setting by existing IRS studies~\cite{wu2018intelligent,wu2018intelligentfull,wu2019beamforming}, whereas no IRS work has considered the latter problem. 
\item Furthermore, we extend our  optimization problems to consider \textit{multi}-antenna mobile users or/and \textit{multiple} IRSs, where such settings are novel in their own rights.
\item For all the optimizations  discussed above, we present detailed analyses to propose efficient algorithms. 
\end{enumerate}

\textbf{Organization of this paper.} Section~\ref{sec:System}  presents the communication models. In Section~\ref{sec:Optimization}, we formulate the optimization problems. The analysis and algorithms for solving the problems are elaborated in Section~\ref{section-Solutions}. In Section~\ref{sec-Related-Work}, we survey related studies.
  Finally, Section~\ref{sec-Conclusion} concludes the paper.

\textbf{Notation.} Scalars are denoted by italic letters, while vectors and matrices are denoted by bold-face
lower-case and upper-case letters, respectively. $\mathbb{C}$ denotes the set of all complex numbers.
  For a matrix $\boldsymbol{M}$, its transpose and conjugate transpose are denoted by $\boldsymbol{M}^T$ and $\boldsymbol{M}^H$, while $\boldsymbol{M}_{i,j}$ (if not defined in other ways) means the element in the $i$th row and $j$th column of $\boldsymbol{M}$. For a vector $\boldsymbol{x}$, its transpose, conjugate transpose, and Euclidean norm are denoted by $\boldsymbol{x}^T$, $\boldsymbol{x}^H$, and $\|\boldsymbol{x}\|$, while $\boldsymbol{x}_{i}$ (if not defined in other ways) means the $i$th element of $\boldsymbol{x}$. 








\section{Communication Models} \label{sec:System}

We now present the IRS-aided wireless communication models.

In a typical system which we study, there are a base station (BS) with $M$ antennas, an intelligent reflecting surface (IRS) with $N$ IRS units, and $K$ single-antenna mobile users (MUs) numbered from $1$ to $K$. We will be clear when the system is extended to the cases of multi-antenna MUs or/and multiple IRSs.

We will discuss three kinds of traffic patterns from the BS to the MUs:
\begin{itemize}
\item \textbf{unicast}, where the BS sends an independent data stream to each MU,
\item \textbf{broadcast}, where the BS sends the same data stream to all $K$ MUs, and
\item \textbf{multicast}, where $K$ MUs are divided into $g$ groups $\mathcal{G}_1, \mathcal{G}_2, \ldots, \mathcal{G}_{g}$, and the BS sends an independent data stream to each group.
\end{itemize}

Since unicast and broadcast can be seen as special cases of multicast, we focus on    multicast below, where $k$ denotes the group index and $i$ denotes the MU index; i.e., $k\in\{1,\ldots,g\}$ and $i \in \mathcal{G}_k$. When multicast reduces to broadcast, there is only one group and $i$ still denotes the MU index. When multicast reduces to unicast, each MU is a group and $k$ denotes the MU index. 



We define the following notation for the wireless channels.
Let $\bm{H}_{\text{b},\text{r}} \in \mathbb{C}^{N \times M} $ be the  channel from the BS  to the IRS. For MU $i \in \mathcal{G}_k$ with $k\in\{1,\ldots,g\}$, we define $\boldsymbol{h}_{\text{r},i}^H \in \mathbb{C}^{1 \times N} $ as the  channel from the IRS to the $i$th MU, and define $\boldsymbol{h}_{\text{b},i}^H\in \mathbb{C}^{1 \times M} $ as the downlink channel from the BS to the $i$th MU. In the above notation, the subscript ``b'' represents the \underline{B}S, whereas the subscript ``r'' signifies \underline{R}IS. When we extend one IRS to multiple IRSs (say $L$), the subscript ``r'' in the channel notation will be replaced by the IRS index $\ell\in\{1,\ldots,L\}$ to denote the channel associated with the $\ell$th IRS; i.e., $\bm{H}_{\text{b},\text{r}}$ and $\boldsymbol{h}_{\text{r},i}^H$ will be changed to $\bm{H}_{\text{b},\ell}$ and $\boldsymbol{h}_{\ell,i}^H$. When we extend single-antenna MUs to multi-antenna MUs, we will add a subscript $q$ after the subscript $i$ in the channel notation to denote the channel associated with MU $i$'s $q$th antenna
 (note $q\in\{1,\ldots,Q_i\}$ if MU $i$ has $Q_i$ antennas). This means that in the case of multi-antenna MUs and one IRS, $\boldsymbol{h}_{\text{r},i}^H$ and $\boldsymbol{h}_{\text{b},i}^H$ will be replaced by $\boldsymbol{h}_{\text{r},i,q}^H$ and $\boldsymbol{h}_{\text{b},i,q}^H$, while in the case of multi-antenna MUs and $L$ IRSs, $\boldsymbol{h}_{\ell,i}^H$ and $\boldsymbol{h}_{\text{b},i}^H$ will be replaced by $\boldsymbol{h}_{\ell,i,q}^H$ and $\boldsymbol{h}_{\text{b},i,q}^H$. 

We now focus back on the case of single-antenna MUs and one IRS. 
 In an IRS with $N$ IRS units, for the $n$th IRS unit with $n \in\{1,\ldots,N\}$, we let  $\beta_n$ be its amplitude change factor and $\theta_n$ be its phase shift to the incident signal. Then we define the reflection coefficient matrix $\boldPhi$ as follows: 
\begin{align}
 & \boldPhi:= \text{diag}(\beta_1 e^{j \theta_1}, \ldots, \beta_N e^{j \theta_N}) , \label{eq-def-phase-shift-matrix-general}
\end{align}
which means an $N \times N$ diagonal matrix with the diagonal elements being $\beta_1 e^{j \theta_1}, \ldots, \beta_N e^{j \theta_N}$.

If the IRS units  change only phases of the incident signals but do not change their amplitudes, then $\beta_n=1$ for $n \in \{1,\ldots,N\}$, and each diagonal element of $\boldPhi$ lies in the complex unit circle, so that 
\begin{align}
 & \boldPhi:= \text{diag}(e^{j \theta_1}, \ldots, e^{j \theta_N}) . \label{eq-def-phase-shift-matrix}
\end{align}
Clearly, the reflection coefficient matrix in Eq.~(\ref{eq-def-phase-shift-matrix-general}) is a generalization of phase shift matrix Eq.~(\ref{eq-def-phase-shift-matrix}). Most studies~\cite{jung2018performance,hu2018beyond,hu2017potential,huang2019Reconfigurable,yu2019miso,jung2019performance} in the literature to date have assumed $\beta_n=1$ for $n \in \{1,\ldots,N\}$, with few work~\cite{guo2019weighted} considering $\beta_n \leq 1$ (i.e., amplitude attenuation is possible). For simplicity, we use reflection coefficient matrix and phase shift matrix interchangeably and they both can be in the form of Eq.~(\ref{eq-def-phase-shift-matrix-general}). 

About the values that the phase shifts $\theta_n|_{n \in \{1,\ldots,N\}}$ can take, the two simple variants are the \textit{continuous} and \textit{discrete} models below. In the continuous case, each of $\theta_n|_{n \in \{1,\ldots,N\}}$ can take any value in $[0,2\pi)$, as in~\cite{wu2018intelligent,wu2018intelligentfull}. In the discrete case, each of $\theta_n|_{n \in \{1,\ldots,N\}}$ can only take predefined discrete values; e.g., $\tau$ discrete values 
equally spaced on a circle for some positive integer $\tau$: $\big\{0, \frac{2\pi}{\tau}, \ldots, \frac{2\pi \cdot (\tau-1)}{\tau} \big\} $, as in~\cite{wu2019beamforming,guo2019weighted}.






We define $\boldsymbol{h}_i^H(\boldPhi)\in \mathbb{C}^{1 \times M}$  by
\begin{align}
 &\boldsymbol{h}_i^H(\boldPhi):=  \bm{h}^H_{\text{r},i}\boldPhi \bm{H}_{\text{b},\text{r}}+\bm{h}^H_{\text{b},i}, \label{eq-def-hk}
\end{align}
so that $\boldsymbol{h}_i^H(\boldPhi) $ means the overall downlink channel to MU $i$ by combining the direct channel with the indirect channels via all IRS units.

 
Let $\bm{w}_k \in \mathbb{C}^{M \times 1}$ be the BS transmit beamforming for group $\mathcal{G}_k$ with $k\in\{1,\ldots,g\}$. We also define $\boldsymbol{W} :=[\bm{w}_1, \ldots, \bm{w}_g]$. For BS's signal $s_k$ for group $\mathcal{G}_k$, when it arrives at MU $i$ of group $\mathcal{G}_k$, the received signal at  MU $i$ is given by $s_k \bm{h}^H_{i}(\boldPhi)\bm{w}_k$. The interference at MU $i$ consists of signals intended for other groups $j\in\{1,\ldots,g\}\setminus\{k\}$ and is given by $\sum\limits_{j\in\{1,\ldots,g\}\setminus\{k\}}s_j\bm{h}^H_{i}(\boldPhi)\bm{w}_j $. We consider that signals are normalized to unit power, so that the  signal-to-interference-plus-noise ratio (SINR) at MU $i$ is given by
\begin{align}
\text{SINR}_i   = \frac{|\bm{h}^H_{i}(\boldPhi)\bm{w}_k |^2}{\sum\limits_{j\in\{1,\ldots,g\}\setminus\{k\}}|\bm{h}^H_{i}(\boldPhi)\bm{w}_j |^2 +  \sigma^2_i}, \label{eq-SINR-i}
\end{align}
where $\sigma^2_i$ denotes the additive white Gaussian noise's 
 power spectral density at MU $i$.

We consider that the BS controls the IRS and obtains the channel information $\boldsymbol{h}_{\text{r},i}^H, \bm{H}_{\text{b},\text{r}} ,\boldsymbol{h}_{\text{b},i}^H $ at the channel estimation stage. For example, we can consider a time-division duplexing (TDD) protocol for the uplink and downlink, and exploit channel reciprocity to acquire the channel state information. After getting $\boldsymbol{h}_{\text{r},i}^H, \bm{H}_{\text{b},\text{r}} ,\boldsymbol{h}_{\text{b},i}^H $ and other parameters from the mobile users, the goal of  the BS is to design its transmit beamforming $\boldsymbol{W}$ and the IRS's phase shift matrix $\boldPhi$    for an optimization problem such as \textit{power control under QoS} and \textit{\mbox{max-min} fair QoS} discussed below. Afterwards, the BS will set the transmit beamforming as the obtained~$\boldsymbol{W}$, and remotely set the IRS phase shift as the obtained~$\boldPhi$.

In this paper, we focus on the following two optimization problems: \textbf{power control under QoS}, and \textbf{\mbox{max-min} fair QoS}. 
We briefly discuss them below and will present more details in Section~\ref{sec:Optimization}.


\textbf{Power control under QoS}.
The total power consumed by the BS to transmit the signals to all $g$ groups is given by
\begin{align}
\sum\limits_{k=1}^{g}\|\bm{w}_k\|^2 , \label{eq-BS-power}
\end{align}
where the operation $\|\cdot \|$ denotes the Euclidean norm.

Power control under QoS means 
minimizing the BS's total power consumption in~(\ref{eq-BS-power}) subject to the constraint that $\text{SINR}_i$ in Eq.~(\ref{eq-SINR-i}) is at  least some predefined requirement $\gamma_i$, for MU index $i \in \mathcal{G}_k$ with group index $k\in\{1,\ldots,g\}$.

\textbf{\mbox{Max-min} fair QoS}. In the optimization problem of \mbox{max-min} fair QoS, similar to the seminal work~\cite{karipidis2008quality} by Karipidis~\textit{et~al.}, we consider that the received SINR of each MU $i$ is scaled by a predetermined factor $1/\gamma_i$ for a positive real constant $\gamma_i$, to model possibly different grades
of services. 
 Then  the minimum scaled SINRs among all MUs is given by
\begin{align}
\min_{k\in\{1,\ldots,g\}}  \min_{i \in \mathcal{G}_k} \frac{\text{SINR}_i}{\gamma_i}   ,  \label{eq-SINR-min}
\end{align}
where $\text{SINR}_i$ is given by Eq.~(\ref{eq-SINR-i}).

In \mbox{max-min} fair QoS, the problem is to maximize the term in~(\ref{eq-SINR-min}) subject to that the BS's total power consumption $\sum\limits_{k=1}^{g}\|\bm{w}_k\|^2 $ in~(\ref{eq-BS-power}) is at most some value $P$. Maximizing (\ref{eq-SINR-min}) is more general than the problem of maximizing the minimum SINR among all MUs, since the former reduces to the latter in the special case of equal $\gamma_i$ for all $i$.

\section{Optimization Problems} \label{sec:Optimization}

In this section, we elaborate the following two optimization problems which have been briefly discussed in the previous section:
\begin{itemize}
\item \textbf{power control under QoS}, and 
\item \textbf{\mbox{max-min} fair QoS}.
\end{itemize} 

The optimizations are done by jointly designing the transmit beamforming of the BS and the phase shift matrix of the IRS. 


As already noted in the previous section, we discuss three kinds of traffic patterns from the BS to the MUs:
\begin{itemize}
\item \textbf{unicast}, where the BS sends an independent data stream to each MU,
\item \textbf{broadcast}, where the BS sends the same data stream to all $K$ MUs, and
\item \textbf{multicast}, where $K$ MUs are divided into $g$ groups $\mathcal{G}_1, \mathcal{G}_2, \ldots, \mathcal{G}_{g}$, and the BS sends an independent data stream to each group.
\end{itemize}
The combination of the two optimization problems and the three traffic patterns induce six settings.


For the above six settings, we further have the following variants:
\begin{itemize}
\item the reflection coefficient matrix $\boldPhi$ can be in the form of Eq.~(\ref{eq-def-phase-shift-matrix-general}) or~(\ref{eq-def-phase-shift-matrix}) (i.e., with or without amplitude attenuation), where the phase shifts $\theta_n|_{n \in \{1,\ldots,N\}}$ can further be continuous or discrete,
\item an MU can have single antenna or multiple antennas,
\item the system can have one IRS of $N$ IRS units, or $L$ IRSs comprising $N_1, \ldots, N_L$ IRS units. 
\end{itemize}

Below, we first discuss optimization problems for the multicast traffic with single-antenna MUs and one IRS, which will imply the corresponding problems for unicast and broadcast since unicast and broadcast can be seen as special cases of multicast. Later, we extend the problems to the cases of multi-antenna MUs or/and multiple IRSs.


\subsection{\textbf{Power control under QoS}} \label{subsec:Power-control-under-QoS}

For power control under QoS, we first present the multicast setting and then reduce it to the unicast and broadcast cases.

\textbf{Multicast.} We have defined the notation for the multicast  in Section~\ref{sec:System}.
For the multicast traffic, power control under QoS means 
minimizing the BS's total power consumption $\sum\limits_{k=1}^{g}\|\bm{w}_k\|^2 $ in~(\ref{eq-BS-power}) subject to the constraint that $\text{SINR}_i$ in Eq.~(\ref{eq-SINR-i}) is at least some predefined requirement $\gamma_i$, for MU index $i \in \mathcal{G}_k$ with group index $k\in\{1,\ldots,g\}$.
 Hence, power control under QoS for multicast traffic is given by the following optimization problem:
\begin{subequations} \label{eq-Prob-P1}
\begin{alignat}{2}
 \text{(P1):~}  
&\min_{\W, \boldPhi}~\sum\limits_{k=1}^{g}\|\bm{w}_k\|^2  \label{OptProb-power-multicast-obj}  \\
&~\mathrm{s.t.}~\frac{|\bm{h}^H_{i}(\boldPhi)\bm{w}_k |^2}{\sum\limits_{j\in\{1,\ldots,g\}\setminus\{k\}}|\bm{h}^H_{i}(\boldPhi)\bm{w}_j |^2 +  \sigma^2_i}\geq \gamma_i, \label{OptProb-power-multicast-SINR-constraint}   \\
&~~~~~~~~~~\forall k \in\{1,\ldots,g\},~\forall i \in \mathcal{G}_k, \nonumber\\
 &~~~~~~\text{Constraints on $\boldPhi$}. \label{OptProb-power-multicast-Phi-constraint} 
\end{alignat}
\end{subequations}

\textbf{Unicast.} When multicast reduces to unicast, each MU is a group, so the number $g$ of groups is $K$, and $k$ denotes the MU index. Then Problem~(P1) becomes
\begin{subequations} \label{eq-Prob-P2}
\begin{alignat}{2}
 \text{(P2):~}  
&\min_{\W, \boldPhi}~\sum\limits_{k=1}^{K}\|\bm{w}_k\|^2  \label{OptProb-power-unicast-obj}  \\
&~\mathrm{s.t.}~\frac{|\bm{h}^H_{k}(\boldPhi)\bm{w}_k |^2}{\sum\limits_{j\in\{1,\ldots,K\}\setminus\{k\}}|\bm{h}^H_{k}(\boldPhi)\bm{w}_j |^2 +  \sigma^2_k}\geq \gamma_k,   \label{OptProb-power-unicast-SINR-constraint} \\
&~~~~~~~~~~\forall k \in\{1,\ldots,K\}, \nonumber\\
 &~~~~~~\text{Constraints on $\boldPhi$}. \label{OptProb-power-unicast-Phi-constraint} 
\end{alignat}
\end{subequations}
Recent studies by Wu and Zhang~\cite{wu2018intelligent,wu2018intelligentfull,wu2019beamforming} have addressed Problem~(P2) with the constraints on $\boldPhi$ of~(\ref{OptProb-power-unicast-Phi-constraint}) given in the form of  Eq.~(\ref{eq-def-phase-shift-matrix}) (i.e., $\boldPhi = \text{diag}(e^{j \theta_1}, \ldots, e^{j \theta_N})$). In ~\cite{wu2018intelligent,wu2018intelligentfull}, each of $\theta_n|_{n \in \{1,\ldots,N\}}$ can take any value in $[0,2\pi)$. In contrast, in ~\cite{wu2019beamforming}, each of  $\theta_n|_{n \in \{1,\ldots,N\}}$ can only take the following $\tau$ discrete values 
equally spaced on a circle for some positive integer $\tau$: $\big\{0, \frac{2\pi}{\tau}, \ldots, \frac{2\pi \cdot (\tau-1)}{\tau} \big\} $.  


\textbf{Broadcast.} When multicast reduces to broadcast, there is only one group $\mathcal{G}_1$, so that $g=1$ and $\mathcal{G}_1=\{1,\ldots,K\}$. Then Problem~(P1) becomes
\begin{subequations} \label{eq-Prob-P3}
\begin{alignat}{2}
 \text{(P3):~}  
&\min_{\bm{w}, \boldPhi}~ \|\bm{w}\|^2 \label{OptProb-power-broadcast-obj}   \\
&~\mathrm{s.t.}~\frac{|\bm{h}^H_{i}(\boldPhi)\bm{w}  |^2}{\sigma^2_i}\geq \gamma_i, ~\forall i \in \{1,\ldots,K\}, \label{OptProb-power-broadcast-SINR-constraint}\\
 &~~~~~~\text{Constraints on $\boldPhi$}. \label{OptProb-power-broadcast-Phi-constraint} 
\end{alignat}
\end{subequations} 
We summarize Problems~(P1)--(P3) in Table~\ref{table:1} on Page~\pageref{table:1}.

\textbf{Extensions to multi-antenna MUs or/and multiple IRSs.} The above Problems~(P1)--(P3) consider single-antenna MUs and one IRS. We now extend the problems to the cases of multi-antenna MUs or/and multiple IRSs.
\begin{itemize}
\item \textbf{Multi-antenna MUs and one IRS.} When MU $i$ has $Q_i$ antennas for $i \in \mathcal{G}_k$ with $k\in\{1,\ldots,g\}$, with $q\in\{1,\ldots,Q_i\}$ indexing antennas of MU $i$, we define $\boldsymbol{h}_{\text{r},i,q}^H \in \mathbb{C}^{1 \times N} $ as the  channel from the IRS to the $q$th antenna of MU $i$, and define $\boldsymbol{h}_{\text{b},i,q}^H\in \mathbb{C}^{1 \times M} $ as the downlink channel from the BS to the $q$th antenna of MU $i$. Then we define $\bm{h}^H_{i,q}(\bm{\Phi})$ as follows to represent the overall downlink channel to MU $i$'s $q$th antenna by combining the direct channel with the indirect channels via all IRS units:
\begin{align}
\bm{h}^H_{i,q}(\bm{\Phi}):= \bm{h}^H_{\text{b},i,q} +  \bm{h}^H_{\text{r},i,q}\boldPhi \bm{H}_{\text{b},\text{r}}. \label{eq-define-MU-h-i-q}
\end{align}
Then at MU $i$, the power of the received signal associated with MU group $k$ is given by
\begin{align}
\bm{w}_k^H \bm{H}_{i}(\boldPhi) \bm{w}_k, \label{eq-define-MU-h-i-q-received-power}
\end{align}
where we define $\bm{H}_{i}(\boldPhi)$ as follows for notational simplicity: 
\begin{align}
\bm{H}_{i}(\boldPhi):= \sum_{q=1}^{Q_i}  \bm{h}_{i,q}(\bm{\Phi})  \bm{h}^H_{i,q}(\bm{\Phi}). \label{eq-define-MU-multi-antenna}
\end{align}
Similar to~(\ref{eq-define-MU-h-i-q-received-power}), at MU $i$, the power of the received interference associated with MU group $j \in\{1,\ldots,g\}\setminus\{k\} $ is given by $\bm{w}_j^H\bm{H}_{i}(\boldPhi)\bm{w}_j $.

Then
\begin{itemize}
\item[\textbullet] replacing $|\bm{h}^H_{i}(\boldPhi)\bm{w}_k |^2$ and $|\bm{h}^H_{i}(\boldPhi)\bm{w}_j |^2$ of Problem~(P1) by $\bm{w}_k^H\bm{H}_{i}(\boldPhi)\bm{w}_k$ and $\bm{w}_j^H\bm{H}_{i}(\boldPhi)\bm{w}_j $,
\item[\textbullet] replacing $|\bm{h}^H_{k}(\boldPhi)\bm{w}_k |^2$ and $|\bm{h}^H_{k}(\boldPhi)\bm{w}_j |^2$ of Problem~(P2) by $\bm{w}_k^H\bm{H}_{k}(\boldPhi)\bm{w}_k$ and $\bm{w}_j^H\bm{H}_{k}(\boldPhi)\bm{w}_j$, and
\item[\textbullet] replacing $|\bm{h}^H_{i}(\boldPhi)\bm{w}  |^2$ of Problem~(P3) by $\bm{w}^H\bm{H}_{i}(\boldPhi)\bm{w}$,
\end{itemize}
we obtain the corresponding optimization problems respectively with multi-antenna MUs and one IRS. They are denoted by Problems~(P1-MA)--(P3-MA) and presented in Table~\ref{table:2MA} on Page~\pageref{table:2MA}, where ``MA'' means \underline{m}ulti-\underline{a}ntenna.
\item \textbf{Single-antenna MUs and multiple IRSs.} When there are $L$ IRSs comprising $N_1, \ldots, N_L$ IRS units, we define for $\ell\in\{1,\ldots,L\}$ that $\bm{H}_{\text{b},\ell} \in \mathbb{C}^{N_{\ell} \times M} $ represents the  channel from the BS  to the $\ell$th IRS, and   $\boldsymbol{h}_{\ell,i}^H \in \mathbb{C}^{1 \times N_{\ell}} $ represents the  channel from the $\ell$th IRS to the $i$th MU, for MU $i \in \mathcal{G}_k$ with $k\in\{1,\ldots,g\}$. Then with the phase shift matrices of the $L$ IRSs denoted by $\bm{\Phi}_1,\ldots,\bm{\Phi}_L$, we define $\bm{h}^H_{i}(\bm{\Phi}_1,\ldots,\bm{\Phi}_L)$ as follows to represent the overall downlink channel to MU $i$ by combining the direct channel with the indirect channels via all IRSs:
\begin{align}
\bm{h}^H_{i}(\bm{\Phi}_1,\ldots,\bm{\Phi}_L):= \bm{h}^H_{\text{b},i}+\sum_{\ell=1}^L \bm{h}^H_{\ell,i}\bm{\Phi}_{\ell} \bm{H}_{\text{b},\ell} , \label{eq-define-multiple-IRSs}
\end{align}
where we can replace $i$ in $\bm{h}^H_{i}(\bm{\Phi}_1,\ldots,\bm{\Phi}_L)$ by $k$ to obtain the notation $\bm{h}^H_{k}(\bm{\Phi}_1,\ldots,\bm{\Phi}_L)$.
Then
\begin{itemize}
\item[\textbullet] replacing $|\bm{h}^H_{i}(\boldPhi)\bm{w}_k |^2$ and $|\bm{h}^H_{i}(\boldPhi)\bm{w}_j |^2$ of Problem~(P1) by $|\bm{h}^H_{i}(\bm{\Phi}_1,\ldots,\bm{\Phi}_L)\bm{w}_k |^2$ and $|\bm{h}^H_{i}(\bm{\Phi}_1,\ldots,\bm{\Phi}_L)\bm{w}_j |^2$,
\item[\textbullet] replacing $|\bm{h}^H_{k}(\boldPhi)\bm{w}_k |^2$ and $|\bm{h}^H_{k}(\boldPhi)\bm{w}_j |^2$ of Problem~(P2) by $|\bm{h}^H_{k}(\bm{\Phi}_1,\ldots,\bm{\Phi}_L)\bm{w}_k |^2$ and $|\bm{h}^H_{k}(\bm{\Phi}_1,\ldots,\bm{\Phi}_L)\bm{w}_j |^2$, 
\item[\textbullet] replacing $|\bm{h}^H_{i}(\boldPhi)\bm{w}  |^2$ of Problem~(P3) by $|\bm{h}^H_{i}(\bm{\Phi}_1,\ldots,\bm{\Phi}_L)\bm{w}  |^2$,
\end{itemize}
and also replacing the constraints on $\boldPhi$ in Eq.~(\ref{OptProb-power-multicast-Phi-constraint}) (\ref{OptProb-power-unicast-Phi-constraint}) or (\ref{OptProb-power-broadcast-Phi-constraint}) by the constraints on $\bm{\Phi}_1,\ldots,\bm{\Phi}_L$,
 we obtain the corresponding optimization problems respectively with single-antenna MUs and multiple IRSs. They are denoted by Problems~(P1-MR)--(P3-MR) and presented in Table~\ref{table:3MR} on Page~\pageref{table:3MR}, where ``MR'' means \underline{m}ultiple \underline{R}ISs.  
\item \textbf{Multi-antenna MUs and multiple IRSs.} Combining the above discussions, we now tackle the most general case of   multi-antenna MUs and multiple IRSs. Let $Q_i$ be MU $i$'s number of antennas,   for   $i \in \mathcal{G}_k$ with $k\in\{1,\ldots,g\}$.  Suppose there are $L$ IRSs comprising $N_1, \ldots, N_L$ IRS units. For $\ell\in\{1,\ldots,L\}$, we define    $\bm{H}_{\text{b},\ell} \in \mathbb{C}^{N_{\ell} \times M} $ as the  channel from the BS  to the $\ell$th IRS, and   $\boldsymbol{h}_{\ell,i,q}^H \in \mathbb{C}^{1 \times N_{\ell}} $ as the  channel from the $\ell$th IRS to MU $i$'s $q$th antenna, for   $q \in \{1,\ldots,Q_i\}$. Then with the phase shift matrices of the $L$ IRSs denoted by $\bm{\Phi}_1,\ldots,\bm{\Phi}_L$, we   define $\bm{h}^H_{i,q}(\bm{\Phi}_1,\ldots,\bm{\Phi}_L)$ as follows to represent the overall downlink channel to MU $i$'s $q$th antenna by combining the direct channel with the indirect channels via all IRSs:
\begin{align}
\bm{h}^H_{i,q}(\bm{\Phi}_1,\ldots,\bm{\Phi}_L):= \bm{h}^H_{\text{b},i,q} +  \sum_{\ell=1}^L\bm{h}^H_{\ell,i,q}\boldPhi_{\ell} \bm{H}_{\text{b},\ell}. \label{eq-define-MU-h-i-q-L}
\end{align}
Furthermore, we define
\begin{align}
 & \bm{H}_{i}(\bm{\Phi}_1,\ldots,\bm{\Phi}_L)  \nonumber  \\ &  := \sum_{q=1}^{Q_i}  \bm{h}_{i,q}(\bm{\Phi}_1,\ldots,\bm{\Phi}_L)  \bm{h}^H_{i,q}(\bm{\Phi}_1,\ldots,\bm{\Phi}_L). \label{eq-define-MU-h-i-q-L-H-matrix}
\end{align}
Then
\begin{itemize}
\item[\textbullet] replacing $|\bm{h}^H_{i}(\bm{\Phi})\bm{w}_k |^2$ and $|\bm{h}^H_{i}(\bm{\Phi})\bm{w}_j |^2$ of Problem~(P1) by $\bm{w}_k^H\bm{H}_{i}(\bm{\Phi}_1,\ldots,\bm{\Phi}_L)\bm{w}_k$ and $\bm{w}_j^H\bm{H}_{i}(\bm{\Phi}_1,\ldots,\bm{\Phi}_L)\bm{w}_j $,
\item[\textbullet] replacing $|\bm{h}^H_{k}(\bm{\Phi})\bm{w}_k |^2$ and $|\bm{h}^H_{k}(\bm{\Phi})\bm{w}_j |^2$ of Problem~(P2) by $\bm{w}_k^H\bm{H}_{k}(\bm{\Phi}_1,\ldots,\bm{\Phi}_L)\bm{w}_k$ and $\bm{w}_j^H\bm{H}_{k}(\bm{\Phi}_1,\ldots,\bm{\Phi}_L)\bm{w}_j$,  
\item[\textbullet] replacing $|\bm{h}^H_{i}(\bm{\Phi})\bm{w}  |^2$ of Problem~(P3) by $\bm{w}^H\bm{H}_{i}(\bm{\Phi}_1,\ldots,\bm{\Phi}_L)\bm{w}$,
\end{itemize}
and also replacing the constraints on $\boldPhi$ in Eq.~(\ref{OptProb-power-multicast-Phi-constraint}) (\ref{OptProb-power-unicast-Phi-constraint}) or (\ref{OptProb-power-broadcast-Phi-constraint}) by the constraints on $\bm{\Phi}_1,\ldots,\bm{\Phi}_L$,  we obtain the corresponding optimization problems respectively with multi-antenna MUs and multiple IRSs. They are denoted by Problems~(P1-MA-MR)--(P3-MA-MR) and presented in Table~\ref{table:4MA-MR} on Page~\pageref{table:4MA-MR}. 
\end{itemize}

\subsection{\textbf{\mbox{Max-min} fair QoS}} \label{subsec:Max-min-fair-QoS}

Similar to Section~\ref{subsec:Power-control-under-QoS},
for \mbox{max-min} fair QoS here, we first present the multicast setting and then reduce it to the unicast and broadcast cases.

\textbf{Multicast.} We have defined the notation for   multicast  in Section~\ref{sec:System}. Similar to the seminal work~\cite{karipidis2008quality} by Karipidis~\textit{et~al.}, we consider that the received SINR of each MU $i$ is scaled by a predetermined factor $1/\gamma_i$ for a positive real constant $\gamma_i$, to model possibly different grades
of services.
For the multicast traffic, \mbox{max-min} fair QoS means maximizing the minimum scaled SINRs among all MUs in~(\ref{eq-SINR-min}) subject to that the BS's total power consumption $\sum\limits_{k=1}^{g}\|\bm{w}_k\|^2 $ in~(\ref{eq-BS-power}) is at most some value $P$. 
 Then we obtain the following optimization problem: 
\begin{subequations} \label{eq-Prob-P4}
\begin{alignat}{2}
 \text{(P4):~}
&\max_{\W, \boldPhi} \min_{k\in\{1,\ldots,g\}}  \min_{i \in \mathcal{G}_k}  \frac{|\bm{h}^H_{i}(\boldPhi)\bm{w}_k |^2}{\gamma_i\left[\sum\limits_{j\in\{1,\ldots,g\}\setminus\{k\}}|\bm{h}^H_{i}(\boldPhi)\bm{w}_j |^2 +  \sigma^2_i\right]}  \label{OptProb-max-min-fair-multicast-obj}  \\
&~\mathrm{s.t.}~\sum\limits_{k=1}^{g}\|\bm{w}_k\|^2 \leq P , \label{OptProb-max-min-fair-multicast-power-constraint}   \\
 &~~~~~~\text{Constraints on $\boldPhi$}. \label{OptProb-max-min-fair-multicast-Phi-constraint}
\end{alignat}
\end{subequations}

\textbf{Unicast.} When multicast reduces to unicast, each MU is a group, so the number $g$ of groups is $K$, and $k$ denotes the MU index. Then Problem~(P4) becomes
\begin{subequations} \label{eq-Prob-P5}
\begin{alignat}{2}
 \text{(P5):~}
&\max_{\W, \boldPhi} \min_{k\in\{1,\ldots,K\}}   \frac{|\bm{h}^H_{k}(\boldPhi)\bm{w}_k |^2}{\gamma_k\left[\sum\limits_{j\in\{1,\ldots,K\}\setminus\{k\}}|\bm{h}^H_{k}(\boldPhi)\bm{w}_j |^2 +  \sigma^2_k\right]}   \label{OptProb-max-min-fair-unicast-obj}  \\
&~\mathrm{s.t.}~\sum\limits_{k=1}^{K}\|\bm{w}_k\|^2 \leq P ,  \label{OptProb-max-min-fair-unicast-power-constraint} \\ 
 &~~~~~~\text{Constraints on $\boldPhi$}. \label{OptProb-max-min-fair-unicast-Phi-constraint}
\end{alignat}
\end{subequations}

\textbf{Broadcast.} When multicast reduces to broadcast, there is only one group $\mathcal{G}_1$, so that $g=1$ and $\mathcal{G}_1=\{1,\ldots,K\}$. Then Problem~(P4) becomes
\begin{subequations} \label{eq-Prob-P6}
\begin{alignat}{2}
 \text{(P6):~}
&\max_{\bm{w},\boldPhi} ~\min_{i \in \{1,\ldots,K\}}   \frac{|\bm{h}^H_{i}(\boldPhi)\bm{w}  |^2}{\gamma_i\sigma^2_i}  \label{OptProb-max-min-fair-broadcast-obj}   \\
&~\mathrm{s.t.}~~\|\bm{w}\|^2 \leq P ,  \label{OptProb-max-min-fair-broadcast-power-constraint}\\
 &~~~~~~~\text{Constraints on $\boldPhi$}. \label{OptProb-max-min-fair-broadcast-Phi-constraint}
\end{alignat}
\end{subequations}

Similar to the multicast case above, for the unicast (resp.,~broadcast) setting, in Eq.~(\ref{OptProb-max-min-fair-unicast-obj}) (resp.,~(\ref{OptProb-max-min-fair-broadcast-obj}), the received SINR of MU $k$ (resp.,~MU $i$) is scaled by a predetermined factor $1/\gamma_k$ (resp.,~MU $1/\gamma_i$), to model possibly different grades
of services~\cite{karipidis2008quality}.

In the special case where all the $\gamma$ factors are the same and hence can be removed from the optimizations, Problems~(P4)--(P6) become maximizing the minimum SINR in the system.

We summarize Problems~(P4)--(P6) in Table~\ref{table:1} on Page~\pageref{table:1}.


\textbf{Extensions to multi-antenna MUs or/and multiple IRSs.} The above Problems~(P4)--(P6) consider single-antenna MUs and one IRS. We now extend the problems to the cases of multi-antenna MUs or/and multiple IRSs.
 Modifying Problems~(P4)--(P6) in a way similar to that of modifying Problems~(P1)--(P3) in Section~\ref{subsec:Power-control-under-QoS}, we have the following:
\begin{itemize}
\item Under multi-antenna MUs and one IRS, we modify Problems~(P4)--(P6) to Problems~(P4-MA)--(P6-MA), which are presented in Table~\ref{table:2MA} on Page~\pageref{table:2MA}. 
\item Under single-antenna MUs and multiple IRSs, we modify Problems~(P4)--(P6) to Problems~(P4-MR)--(P6-MR), which are presented in Table~\ref{table:3MR} on Page~\pageref{table:3MR}.
\item Under multi-antenna MUs and multiple IRSs , we modify Problems~(P4)--(P6) to Problems~(P4-MA-MR)--(P6-MA-MR), which are presented in Table~\ref{table:4MA-MR} on Page~\pageref{table:4MA-MR}.
\end{itemize}

\begin{table*}[h!] 
\setlength\tabcolsep{1pt}
\hspace{-30pt}
\begin{tabular}{|l||*{3}{l|}}\hline
\backslashbox{Problem\kern-3em}{\kern-.5em Traffic}
&\makebox[.41\textwidth]{Multicast}&\makebox[.38\textwidth]{Unicast}&\makebox[.2\textwidth]{Broadcast} \\\hline\hline
\begin{tabular}[c]{@{}l@{}}Power\\ control\\under\\QoS\end{tabular}  &\usebox{\minpowermulticast} &\usebox{\minpowerunicast} 
&\usebox{\minpowerbroadcast}  \\\hline
\begin{tabular}[c]{@{}l@{}}Max-min \\fair QoS  \end{tabular}  &\usebox{\fairmulticast}   &\usebox{\fairunicast} 
&\usebox{\fairbroadcast}  \\\hline
\end{tabular}
\caption{~(with \textbf{single-antenna mobile users and one intelligent reflecting surface}): Optimizing the transmit beamforming $\W$ (or $\bm{w}$) of the base station (BS) and the phase shift matrix $\boldPhi$ of the intelligent reflecting surface (IRS) comprising $N$ IRS units for \textit{power control under QoS} and \textit{\mbox{max-min} fair QoS}
 under various downlink traffic patterns (i.e., unicast, broadcast, and multicast) from a multi-antenna base station to $K$ single-antenna mobile users (MUs). In the unicast case, the BS sends an independent data stream to each MU. In the broadcast case, the BS sends the same data stream to all $K$ MUs. In the multicast case, $K$ MUs are divided into $g$ groups $\mathcal{G}_1, \mathcal{G}_2, \ldots, \mathcal{G}_{g}$, and the BS sends an independent data stream to each group.  The notation  $\bm{h}^H_{k}(\boldPhi)$ in the table means $\bm{h}^H_{\text{b},k}+ \bm{h}^H_{\text{r},k}\boldPhi \bm{H}_{\text{b},\text{r}}$, meaning the overall downlink channel to MU $k$ by combining the direct channel with the indirect channels via all IRS units. Similarly,   $\bm{h}^H_{i}(\boldPhi)$ in the table means $\bm{h}^H_{\text{b},i}+ \bm{h}^H_{\text{r},i}\boldPhi \bm{H}_{\text{b},\text{r}}$.
 The notation $P$ denotes the maximal power consumed by the BS. }
\label{table:1}
\end{table*}

\begin{table*}[h!] 
\setlength\tabcolsep{1pt}
\hspace{-30pt}
\begin{tabular}{|l||*{3}{l|}}\hline
\backslashbox{Problem\kern-3em}{\kern-.5em Traffic}
&\makebox[.41\textwidth]{Multicast}&\makebox[.38\textwidth]{Unicast}&\makebox[.2\textwidth]{Broadcast} \\\hline\hline
\begin{tabular}[c]{@{}l@{}}Power\\ control\\under\\QoS\end{tabular}  &\usebox{\minpowermulticastmultichannel}  &\usebox{\minpowerunicastmultichannel} 
&\usebox{\minpowerbroadcastmultichannel} \\\hline
\begin{tabular}[c]{@{}l@{}}Max-min  \\fair QoS  \end{tabular}    &\usebox{\fairmulticastmultichannel} &\usebox{\fairunicastmultichannel} 
&\usebox{\fairbroadcastmultichannel}  \\\hline
\end{tabular}
\caption{~(with \textbf{multi-antenna mobile users and one intelligent reflecting surface}): Optimizing the transmit beamforming $\W$ (or $\bm{w}$) of the base station (BS) and the phase shift matrix $\boldPhi$ of the intelligent reflecting surface (IRS) comprising $N$ IRS units for \textit{power control under QoS} and \textit{\mbox{max-min} fair QoS}
 under various downlink traffic patterns (i.e., unicast, broadcast, and multicast) from a multi-antenna base station to $K$ multi-antenna mobile users (MUs), where MU $i$ has $Q_i$ antennas for $i \in \mathcal{G}_k$ with $k\in\{1,\ldots,g\}$. In the unicast case, the BS sends an independent data stream to each MU. In the broadcast case, the BS sends the same data stream to all $K$ MUs. In the multicast case, $K$ MUs are divided into $g$ groups $\mathcal{G}_1, \mathcal{G}_2, \ldots, \mathcal{G}_{g}$, and the BS sends an independent data stream to each group. The notation  $\bm{H}_{k}(\bm{\Phi})$ in the table means $ \sum_{q=1}^{Q_k}  \bm{h}_{k,q}(\bm{\Phi})  \bm{h}^H_{k,q}(\bm{\Phi})$, where $\bm{h}^H_{k,q}(\bm{\Phi})$ denotes \mbox{$\bm{h}^H_{\text{b},k,q} +  \bm{h}^H_{\text{r},k,q}\boldPhi \bm{H}_{\text{b},\text{r}}$} and means the overall downlink channel to MU $k$'s $q$th antenna by combining the direct channel with the indirect channels via all IRS units.
  Similarly,   $\bm{H}_{i}(\bm{\Phi})$ in the table means $ \sum_{q=1}^{Q_i}  \bm{h}_{i,q}(\bm{\Phi})  \bm{h}^H_{i,q}(\bm{\Phi})$, where $\bm{h}^H_{i,q}(\bm{\Phi})$ denotes \mbox{$\bm{h}^H_{\text{b},i,q} +  \bm{h}^H_{\text{r},i,q}\boldPhi \bm{H}_{\text{b},\text{r}}$}.   
   The notation $P$ denotes the maximal power consumed by the BS.
   }
\label{table:2MA}
\end{table*}

\begin{table*}[h!]
\setlength\tabcolsep{1pt}
\hspace{-30pt}
\begin{tabular}{|l||*{3}{l|}}\hline
\backslashbox{Problem\kern-3em}{\kern-.5em Traffic}
&\makebox[.41\textwidth]{Multicast}&\makebox[.38\textwidth]{Unicast}&\makebox[.22\textwidth]{Broadcast} \\\hline\hline
\begin{tabular}[c]{@{}l@{}}Power\\ control\\under\\QoS\end{tabular}  &\usebox{\minpowermulticastmultiRIS} &\usebox{\minpowerunicastmultiRIS}
&\usebox{\minpowerbroadcastmultiRIS}  \\\hline
\begin{tabular}[c]{@{}l@{}}Max-min  \\fair QoS  \end{tabular}  &\usebox{\fairmulticastmultiRIS}   &\usebox{\fairunicastmultiRIS}
&\usebox{\fairbroadcastmultiRIS}  \\\hline
\end{tabular}
    \captionsetup{singlelinecheck=off} 
\caption{~(with \textbf{single-antenna mobile users and multiple intelligent reflecting surfaces}): Optimizing the transmit beamforming $\W$ (or $\bm{w}$) of the base station (BS) and the phase shift matrices $\bm{\Phi}_1,\ldots,\bm{\Phi}_L$ of $L$ intelligent reflecting surfaces (IRSs) comprising $N_1, \ldots, N_L$ IRS units for \textit{power control under QoS} and \textit{\mbox{max-min} fair QoS}
 under various downlink traffic patterns (i.e., unicast, broadcast, and multicast) from a multi-antenna base station to $K$ single-antenna mobile users (MUs). In the unicast case, the BS sends an independent data stream to each MU. In the broadcast case, the BS sends the same data stream to all $K$ MUs. In the multicast case, $K$ MUs are divided into $g$ groups $\mathcal{G}_1, \mathcal{G}_2, \ldots, \mathcal{G}_{g}$, and the BS sends an independent data stream to each group. The notation  $\bm{h}^H_{k}(\bm{\Phi}_1,\ldots,\bm{\Phi}_L)$ in the table means $\bm{h}^H_{\text{b},k}+\sum_{\ell=1}^L \bm{h}^H_{\ell,k}\bm{\Phi}_{\ell} \bm{H}_{\text{b},\ell}$, meaning the overall downlink channel to MU $k$ by combining the direct channel with the indirect channels via all IRSs. Similarly,   $\bm{h}^H_{i}(\bm{\Phi}_1,\ldots,\bm{\Phi}_L)$ in the table means $ \bm{h}^H_{\text{b},i}+\sum_{\ell=1}^L \bm{h}^H_{\ell,i}\bm{\Phi}_{\ell} \bm{H}_{\text{b},\ell} $. 
 The notation $P$ denotes the maximal power consumed by the BS. }
\label{table:3MR}
\end{table*}

\begin{table*}[h!]
\setlength\tabcolsep{1pt}
\hspace{-30pt}
\begin{tabular}{|l||*{3}{l|}}\hline
\backslashbox{Problem\kern-3em}{\kern-.5em Traffic}
&\makebox[.41\textwidth]{Multicast} &\makebox[.38\textwidth]{Unicast}&\makebox[.22\textwidth]{Broadcast}\\\hline\hline
\begin{tabular}[c]{@{}l@{}}Power\\ control\\under\\QoS\end{tabular}  &\usebox{\minpowermulticastmultiAntennamultiRIS} &\usebox{\minpowerunicastmultiAntennamultiRIS}
&\usebox{\minpowerbroadcastmultiAntennamultiRIS}  \\\hline
\begin{tabular}[c]{@{}l@{}}Max-min  \\fair QoS  \end{tabular}    &\usebox{\fairmulticastmultiAntennamultiRIS}  &\usebox{\fairunicastmultiAntennamultiRIS}
&\usebox{\fairbroadcastmultiAntennamultiRIS} \\\hline
\end{tabular}
    \captionsetup{singlelinecheck=off}
\caption{~(with \textbf{multi-antenna mobile users and multiple intelligent reflecting surfaces}): Optimizing the transmit beamforming $\W$ (or $\bm{w}$) of the base station (BS) and the phase shift matrix $\bm{\Phi}_1,\ldots,\bm{\Phi}_L$ of $L$ intelligent reflecting surfaces (IRSs) comprising $N_1, \ldots, N_L$ IRS units for \textit{power control under QoS} and \textit{\mbox{max-min} fair QoS}
 under various downlink traffic patterns (i.e., unicast, broadcast, and multicast) from a multi-antenna base station to $K$ multi-antenna mobile users (MUs), where MU $i$ has $Q_i$ antennas for $i \in \mathcal{G}_k$ with $k\in\{1,\ldots,g\}$. In the unicast case, the BS sends an independent data stream to each MU. In the broadcast case, the BS sends the same data stream to all $K$ MUs. In the multicast case, $K$ MUs are divided into $g$ groups $\mathcal{G}_1, \mathcal{G}_2, \ldots, \mathcal{G}_{g}$, and the BS sends an independent data stream to each group. The notation  $\bm{H}_{k}(\bm{\Phi}_1,\ldots,\bm{\Phi}_L)$ in the table means $ \sum_{q=1}^{Q_k}  \bm{h}_{k,q}(\bm{\Phi}_1,\ldots,\bm{\Phi}_L)  \bm{h}^H_{k,q}(\bm{\Phi}_1,\ldots,\bm{\Phi}_L)$, where $\bm{h}^H_{k,q}(\bm{\Phi}_1,\ldots,\bm{\Phi}_L)$ denotes \mbox{$\bm{h}^H_{\text{b},k,q} +  \sum_{\ell=1}^L\bm{h}^H_{\ell,k,q}\boldPhi_{\ell} \bm{H}_{\text{b},\ell}$} and means the overall downlink channel to MU $k$'s $q$th antenna by combining the direct channel with the indirect channels via all IRSs. 
  Similarly,   $\bm{H}_{i}(\bm{\Phi}_1,\ldots,\bm{\Phi}_L)$ in the table means $ \sum_{q=1}^{Q_i}  \bm{h}_{i,q}(\bm{\Phi}_1,\ldots,\bm{\Phi}_L)  \bm{h}^H_{i,q}(\bm{\Phi}_1,\ldots,\bm{\Phi}_L)$, where $\bm{h}^H_{i,q}(\bm{\Phi}_1,\ldots,\bm{\Phi}_L)$ denotes \mbox{$\bm{h}^H_{\text{b},i,q} +  \sum_{\ell=1}^L\bm{h}^H_{\ell,i,q}\boldPhi_{\ell} \bm{H}_{\text{b},\ell}$}. 
   The notation $P$ denotes the maximal power consumed by the BS. }
\label{table:4MA-MR}
\end{table*}

\section{Solving the Optimization Problems} \label{section-Solutions}

\subsection{Solutions to Problems~(P1)--(P3)} \label{subsection-P1-P3}

We now propose algorithms to solve Problems~(P1)--(P3). Since Problems~(P2) and~(P3) can be seen as special cases of Problem~(P1), we first focus on solving Problem~(P1), and then apply the solutions to Problems~(P2) and~(P3).






We use the idea of alternating optimization, which is widely used in multivariate optimization. Specifically, we will optimize $\W$ and $\boldPhi$ alternatively to solve Problem~(P1). Below, we discuss the optimization of $\W$ given $\boldPhi$ and the optimization of $\boldPhi$ given $\W$.

\textbf{Optimizing $\W$ given $\boldPhi$.} Given some $\boldPhi$ satisfying the constraints in (\ref{OptProb-power-multicast-Phi-constraint}),  the goal of optimizing $\W$ for Problem~(P1) is finding $\W$ to minimize the objective function in Eq.~(\ref{OptProb-power-multicast-obj}) subject to the constraint in Eq.~(\ref{OptProb-power-multicast-SINR-constraint}).

We use the idea of exchanging variables and convert Problem~(P1) to a form that is easier to analyze. Specifically, we define
\begin{align}
 & \bm{X}_k := \bm{w}_k \bm{w}_k^H,~\forall k \in\{1,\ldots,g\}. \label{eq-Xk-wk}
\end{align}
Then the objective function in~(\ref{eq-SINR-min}) of Problem~(P1) is given by
\begin{align}
 & \|\bm{w}_k\|^2 = \bm{w}_k^H \bm{w}_k =  \tr(\bm{w}_k \bm{w}_k^H)=  \tr(\bm{X}_k). \label{eq-Xk-wk-OptProb-power-multicast-obj}
\end{align}
To express the constraint in Inequality~(\ref{OptProb-power-multicast-SINR-constraint}) of Problem~(P1), we note
\begin{align}
|\bm{h}^H_{i}(\boldPhi)\bm{w}_k |^2  & = \bm{h}^H_{i}(\boldPhi)\bm{w}_k \bm{w}_k^H\bm{h}_{i}(\boldPhi) \nonumber  \\ & = \tr(\bm{w}_k \bm{w}_k^H\bm{h}_{i}(\boldPhi)\bm{h}^H_{i}(\boldPhi) ) \nonumber  \\ & = \tr(\bm{X}_k \bm{H}_{i}(\boldPhi) ) ,\label{eq-Xk-wk-OptProb-power-multicast-SINR-constraint}
\end{align}
where we define
\begin{align}
 & \bm{H}_{i}(\boldPhi) := \bm{h}_{i}(\boldPhi)\bm{h}^H_{i}(\boldPhi). \label{eq-Hk-Phi}
\end{align}
Replacing $k$ by $j$ in Eq.~(\ref{eq-Xk-wk-OptProb-power-multicast-SINR-constraint}), we also have $|\bm{h}^H_{i}(\boldPhi)\bm{w}_j |^2 = \tr(\bm{X}_k \bm{H}_{j}(\boldPhi) )$. Using this and Eq.~(\ref{eq-Xk-wk-OptProb-power-multicast-SINR-constraint}), we express the the constraint in Eq.~(\ref{OptProb-power-multicast-SINR-constraint}) of Problem~(P1) as
\begin{align}
\frac{\tr(\bm{X}_k \bm{H}_{i}(\boldPhi) ) }{\sum_{j\in\{1,\ldots,g\}\setminus\{k\}}\tr(\bm{X}_j \bm{H}_{i}(\boldPhi) )  +  \sigma^2_i}\geq \gamma_i.\label{eq-Xk-wk-OptProb-power-multicast-SINR-constraint2}
\end{align}

In addition, the definition of $\bm{X}_k$ in Eq.~(\ref{eq-Xk-wk}) implies that $\bm{X}_k$ is semi-definite and has rank one. Combining this with Eq.~(\ref{eq-Xk-wk-OptProb-power-multicast-obj}) and Inequality~(\ref{eq-Xk-wk-OptProb-power-multicast-SINR-constraint2}), the problem of optimizing $\W$ given $\boldPhi$ for Problem~(P1) is given by
\begin{subequations} \label{eq-Prob-P1a}
\begin{alignat}{2}
\text{(P1a)}: ~~\min_{\{\bm{X}_k\}|_{k=1}^g} ~~~& \sum_{k=1}^{g}\tr(\bm{X}_k) \\
\mathrm{s.t.}~~~~~~&   \tr(\bm{X}_k \bm{H}_{i}(\boldPhi)) \geq \nonumber \\
& \hspace{-15pt}\gamma_i \sigma^2_i+  \gamma_i \sum_{j\in\{1,\ldots,g\}\setminus\{k\}}\tr(\bm{X}_j \bm{H}_{i}(\boldPhi) ) ,  \label{eq:prob-P1-0-}\\
&~~~~~~~\forall k \in\{1,\ldots,g\},~\forall i \in \mathcal{G}_k,  \nonumber \\
& \bm{X}_k \succeq 0,~\forall k \in\{1,\ldots,g\}, \\
& \text{rank}(\bm{X}_k) =1,~\forall k \in\{1,\ldots,g\}. \label{eq:prob-P1-0-rank-constraint}
\end{alignat}
\end{subequations}

The only non-convex part in Problem~(P1a) is the rank constraint in Eq.~(\ref{eq:prob-P1-0-rank-constraint}). Hence, we adopt semidefinite relaxation (SDR) and drop Eq.~(\ref{eq:prob-P1-0-rank-constraint}) to obtain a semidefinite programming problem. After a candidate solution is obtained, appropriate post-processing such as Gaussian randomization~\cite{so2007approximating} (see also randA, randB, and randC coined by~\cite{sidiropoulos2006transmit}) is applied to convert the candidate solution into a solution which satisfies the rank constraint. The above method has been used to solve a problem similar to Problem~(P1a) by Karipidis~\textit{et~al.}~\cite{karipidis2008quality}, where the notation of Problem~$\mathcal{Q}$ is used. Hence, we can apply methods of~\cite{karipidis2008quality} to solve Problem~(P1a).






\textbf{Finding $\boldPhi$ given $\W$.} Given $\W$, Problem~(P1) becomes the following feasibility check problem of finding $\boldPhi$:
\begin{subequations} \label{eq-Prob-P1b}
\begin{alignat}{2}
\text{(P1b)}: \Find~~~& \boldPhi \\
\mathrm{s.t.}~~~~~&   \frac{|\bm{h}^H_{i}(\boldPhi)\bm{w}_k |^2}{\sum\limits_{j\in\{1,\ldots,g\}\setminus\{k\}}|\bm{h}^H_{i}(\boldPhi)\bm{w}_j |^2 +  \sigma^2_i}\geq \gamma_i, \label{eq:prob-P1a-SINR-constraint}\\
&~~~~~~~\forall k \in\{1,\ldots,g\},~\forall i \in \mathcal{G}_k, \nonumber  \\
 &\text{Constraints on $\boldPhi$}. \label{eq:prob-P1a-rank-constraint}
\end{alignat}
\end{subequations}

To solve Problem~(P1b), we start with analyzing the constraint in Inequality~(\ref{eq:prob-P1a-SINR-constraint}). To this end, we recall the definition of $\bm{h}^H_{i}(\boldPhi)$ in Eq.~(\ref{eq-def-hk}) to obtain $\bm{h}^H_{i}(\boldPhi)\bm{w}_k$ as $ \bm{h}^H_{\text{r},i}\boldPhi \bm{H}_{\text{b},\text{r}} \bm{w}_k+\bm{h}^H_{\text{b},i}\bm{w}_k$.  

If the constraint on $\boldPhi$ in Eq.~(\ref{eq:prob-P1a-rank-constraint}) is  Eq.~(\ref{eq-def-phase-shift-matrix-general}) (i.e., $\boldPhi:= \text{diag}(\beta_1 e^{j \theta_1}, \ldots, \beta_N e^{j \theta_N})$), we find it convenient to define  
\begin{align}
 & \bm{\phi} : = [\beta_1 e^{j \theta_1}, \ldots, \beta_N e^{j \theta_N}]^H. \label{eq-define-bm-phi}
\end{align} 
Then we change variables 
to have  
\begin{align}
\bm{h}^H_{i}(\boldPhi)\bm{w}_k  & =  ( \bm{h}^H_{\text{r},i}\boldPhi \bm{H}_{\text{b},\text{r}}+\bm{h}^H_{\text{b},i})\bm{w}_{k} \nonumber  \\ & = \bm{\phi}^H\bm{a}_{i}(\bm{w}_{k}) + {b}_{i}(\bm{w}_{k}) , \label{eq-hi-def-ai-wk-def-b-wk}   
\end{align}
where we define $\bm{a}_{i}(\bm{w}_{k}) \in \mathbb{C}^{N \times 1}$ by
\begin{align}
\bm{a}_{i}(\bm{w}_{k}):=\text{diag}(\bm{h}^H_{\text{r},i})\bm{H}_{\text{b},\text{r}}\bm{w}_{k} , \label{eq-def-ai-wk} 
\end{align}
and   complex numbers $b_{i}(\bm{w}_{k})$ by
\begin{align}
b_{i}(\bm{w}_{k}):=\bm{h}^H_{\text{b},i}\bm{w}_k. \label{eq-def-b-wk} 
\end{align}

From Eq.~(\ref{eq-hi-def-ai-wk-def-b-wk}), we further compute $|\bm{h}^H_{i}(\boldPhi)\bm{w}_k |^2$ which appears in Inequality~(\ref{eq:prob-P1a-SINR-constraint}):
\begin{align}
 & |\bm{h}^H_{i}(\boldPhi)\bm{w}_k |^2 \nonumber  \\ & = (\bm{\phi}^H\bm{a}_{i}(\bm{w}_{k}) + {b}_{i}(\bm{w}_{k}))(\bm{a}_{i}^H(\bm{w}_{k})\bm{\phi} + {b}_{i}^H(\bm{w}_{k})) \nonumber  \\ & =  \bm{\phi}^H\bm{a}_{i}(\bm{w}_{k})\bm{a}_{i}^H(\bm{w}_{k})\bm{\phi} +\bm{\phi}^H\bm{a}_{i}(\bm{w}_{k}){b}_{i}^H(\bm{w}_{k})\nonumber  \\ & \quad + {b}_{i}(\bm{w}_{k}) \bm{a}_{i}^H(\bm{w}_{k})\bm{\phi} + {b}_{i}(\bm{w}_{k}) {b}_{i}^H(\bm{w}_{k})\nonumber  \\ &=\begin{bmatrix} 
\bm{\phi}^H ,~
1
\end{bmatrix} \bm{A}_i(\bm{w}_{k}) \begin{bmatrix} 
\bm{\phi} \\
1
\end{bmatrix} + {b}_{i}(\bm{w}_{k}) {b}_{i}^H(\bm{w}_{k}), \label{eq-hi-def-ai-wk-def-b-wk-square}
\end{align}
where we define
\begin{align}
 & \bm{A}_i(\bm{w}_{k}) := \begin{bmatrix} 
\bm{a}_{i}(\bm{w}_{k})\bm{a}_{i}^H(\bm{w}_{k}), & \bm{a}_{i}(\bm{w}_{k}){b}_{i}^H(\bm{w}_{k}) \\
{b}_{i}(\bm{w}_{k})\bm{a}_{i}^H(\bm{w}_{k}), & 0 
\end{bmatrix}. \label{eq-def-Ai-matrix-wk} 
\end{align}
Replacing $k$ by $j$ in Eq.~(\ref{eq-hi-def-ai-wk-def-b-wk-square}), we also have 
\begin{align}
 & |\bm{h}^H_{i}(\boldPhi)\bm{w}_j |^2  \\ &=\begin{bmatrix} 
\bm{\phi}^H ,~
1
\end{bmatrix} \bm{A}_i(\bm{w}_{j}) \begin{bmatrix} 
\bm{\phi} \\
1
\end{bmatrix} + {b}_{i}(\bm{w}_{j}) {b}_{i}^H(\bm{w}_{j}). \label{eq-hi-def-ai-wj-def-b-wj-square}
\end{align}

Combining Eq.~(\ref{eq-hi-def-ai-wk-def-b-wk-square}) and Eq.~(\ref{eq-hi-def-ai-wj-def-b-wj-square}), we write the constraint in Inequality~(\ref{eq:prob-P1a-SINR-constraint}) of Problem~(P1b) as 
\begin{align}
 & \begin{bmatrix} 
\bm{\phi}^H ,~
1
\end{bmatrix} \bm{A}_i(\bm{w}_{k}) \begin{bmatrix} 
\bm{\phi} \\
1
\end{bmatrix} + {b}_{i}(\bm{w}_{k}) {b}_{i}^H(\bm{w}_{k}) \nonumber  \\ & \geq \hspace{-2pt} \gamma_i \hspace{-2pt}\left\{\hspace{-2pt}  \sigma^2_i\hspace{-2pt}+\hspace{-10pt} \sum\limits_{j\in\{1,\ldots,g\}\setminus\{k\}} \hspace{-2pt} \left\{\hspace{-2pt}\begin{bmatrix} 
\bm{\phi}^H ,\hspace{-2pt}~1
\end{bmatrix} \hspace{-2pt}\bm{A}_i(\bm{w}_{j})\hspace{-2pt} \begin{bmatrix} 
\bm{\phi} \\
1
\end{bmatrix} \hspace{-2pt}+\hspace{-2pt} {b}_{i}(\bm{w}_{j}) {b}_{i}^H(\bm{w}_{j})\hspace{-3pt} \right\}\hspace{-4pt}\right\}\hspace{-2pt}.\label{eq-hi-def-ai-wj-def-b-wj-square-v3}
\end{align}
We introduce an auxiliary  variable~$t$, which is a complex number satisfying $|t|=1$, and define 
\begin{align}
 & \bm{v}:= t  \begin{bmatrix} 
\bm{\phi}  \\
1
\end{bmatrix}=  \begin{bmatrix} 
\bm{\phi}t \\
t
\end{bmatrix}.
\end{align}
Then with $|t|=1$, Inequality~(\ref{eq-hi-def-ai-wj-def-b-wj-square-v3}) is equivalent to 
\begin{align}
 & \bm{v}^H \bm{A}_i(\bm{w}_{k}) \bm{v} + {b}_{i}(\bm{w}_{k}) {b}_{i}^H(\bm{w}_{k}) \nonumber  \\ & \geq  \gamma_i \left\{ \sigma^2_i+\sum\limits_{j\in\{1,\ldots,g\}\setminus\{k\}}  \left\{\bm{v}^H \bm{A}_i(\bm{w}_{j}) \bm{v} + {b}_{i}(\bm{w}_{j}) {b}_{i}^H(\bm{w}_{j})   \right\}  \right\}.\label{eq-hi-def-ai-wj-def-b-wj-square-v5}
\end{align}
We further define 
\begin{align}
\boldsymbol{V}: = \boldsymbol{v}  \boldsymbol{v}^H, \label{eq-boldsymbol-V-v}
\end{align}
If the constraint on $\boldPhi$ in Eq.~(\ref{eq:prob-P1a-rank-constraint}) is in the form of  Eq.~(\ref{eq-def-phase-shift-matrix-general}) (i.e., $\boldPhi:= \text{diag}(\beta_1 e^{j \theta_1}, \ldots, \beta_N e^{j \theta_N})$) so that $\bm{\phi} : = [\beta_1 e^{j \theta_1}, \ldots, \beta_N e^{j \theta_N}]^H$ from Eq.~(\ref{eq-define-bm-phi}), then the diagonal elements of $\boldsymbol{V}$ are given by $\boldsymbol{V}_{n,n} = (\beta_n e^{j \theta_n})^H \cdot \beta_n e^{j \theta_n}  = {\beta_n}^2  $ for $n\in \{1,\ldots, N\}$ and $\boldsymbol{V}_{N+1,N+1}=t  \cdot  t^H = 1$. Moreover, $\boldsymbol{V}$ is semi-definite and has rank one.

Using Eq.~(\ref{eq-boldsymbol-V-v}), we write Inequality~(\ref{eq-hi-def-ai-wj-def-b-wj-square-v5}) as
\begin{align}
 & \tr(\bm{A}_i(\bm{w}_{k})\boldsymbol{V} )  + {b}_{i}(\bm{w}_{k}) {b}_{i}^H(\bm{w}_{k}) \nonumber  \\ & \geq  \gamma_i \left\{\hspace{-2pt} \sigma^2_i +  \sum\limits_{j\in\{1,\ldots,g\}\setminus\{k\}} \hspace{-8pt}\left\{ \tr(\bm{A}_i(\bm{w}_{j})\boldsymbol{V} )  + {b}_{i}(\bm{w}_{j}) {b}_{i}^H(\bm{w}_{j})\right\} \hspace{-2pt}\right\}\hspace{-3pt}.\label{eq-hi-def-ai-wj-def-b-wj-square-v7}
\end{align}

From the above discussion,  we convert Problem~(P1b) into
\begin{subequations} \label{eq-Prob-P1c}
\begin{alignat}{2}
\text{(P1c)}: \Find~~~& \boldsymbol{V} \\
\mathrm{s.t.}~~~~~&   \tr(\bm{A}_i(\bm{w}_{k})\boldsymbol{V} )  + {b}_{i}(\bm{w}_{k}) {b}_{i}^H(\bm{w}_{k}) \nonumber  \\ & \geq  \gamma_i \bigg\{  \sigma^2_i + \nonumber  \\ & \hspace{-25pt}\sum\limits_{j\in\{1,\ldots,g\}\setminus\{k\}} \left[ \tr(\bm{A}_i(\bm{w}_{j})\boldsymbol{V} )  + {b}_{i}(\bm{w}_{j}) {b}_{i}^H(\bm{w}_{j})\right]\hspace{-2pt}\bigg\},  \label{eq:prob-Pc-SINR-constraint}\\
&~~~~~~~\forall k \in\{1,\ldots,g\},~\forall i \in \mathcal{G}_k, \nonumber \\
 &\boldsymbol{V}_{n,n} = {\beta_n}^2,~\forall n\in \{1,\ldots, N\}, \label{eq-Prob-P1c-constraints-Vnn} \\
 &\text{$\boldsymbol{V}_{N+1,N+1} = 1,$} \label{eq-Prob-P1c-constraints-Vnplus1}  \\
& \boldsymbol{V} \succeq 0, \\
& \text{rank}(\boldsymbol{V}) =1. \label{eq:prob-P1c-rank-constraint}
\end{alignat}
\end{subequations}

The above constraint in Eq.~(\ref{eq-Prob-P1c-constraints-Vnn}) is for the case where all $\beta_n$ are predefined constants. A special case of particular interest is the case of all $\beta_n$ being $1$ so that Eq.~(\ref{eq-Prob-P1c-constraints-Vnn}) and Eq.~(\ref{eq-Prob-P1c-constraints-Vnplus1}) can together be written as $\boldsymbol{V}_{n,n}  = 1  $ for $n\in \{1,\ldots, n+1\}$. 
If each $\beta_n$ can take any value in $[0,1]$, then Eq.~(\ref{eq-Prob-P1c-constraints-Vnn}) can be replaced by $\boldsymbol{V}_{n,n} \in [0,1] $ for $n\in \{1,\ldots, N\}$. Similarly, we may consider the most general case where  some $\beta_n$ are predefined constants while other $\beta_n$ can vary.

The only non-convex part in Problem~(P1c) is the rank constraint in Eq.~(\ref{eq:prob-P1c-rank-constraint}). Hence, we adopt semidefinite relaxation (SDR) and drop Eq.~(\ref{eq:prob-P1c-rank-constraint}) to obtain the following semidefinite programming Problem~(P1d): 
 \begin{subequations} \label{eq-Prob-P1d}
\begin{alignat}{2}
\text{(P1d)}: \Find~~~& \boldsymbol{V} \\
\mathrm{s.t.}~~~~~&   \tr(\bm{A}_i(\bm{w}_{k})\boldsymbol{V} )  + {b}_{i}(\bm{w}_{k}) {b}_{i}^H(\bm{w}_{k}) \nonumber  \\ & \geq  \gamma_i \bigg\{  \sigma^2_i + \nonumber  \\ & \hspace{-25pt}\sum\limits_{j\in\{1,\ldots,g\}\setminus\{k\}} \left[ \tr(\bm{A}_i(\bm{w}_{j})\boldsymbol{V} )  + {b}_{i}(\bm{w}_{j}) {b}_{i}^H(\bm{w}_{j})\right]\hspace{-2pt}\bigg\}, \label{eq:prob-P1c-SINR-constraint}\\
&~~~~~~~\forall k \in\{1,\ldots,g\},~\forall i \in \mathcal{G}_k,  \nonumber \\
 &\boldsymbol{V}_{n,n} = {\beta_n}^2,~\forall n\in \{1,\ldots, N\}, \label{eq-Prob-P1d-constraints-Vnn} \\
 &\text{$\boldsymbol{V}_{N+1,N+1} = 1,$} \label{eq-Prob-P1d-constraints-Vnplus1}  \\
& \boldsymbol{V} \succeq 0. 
\end{alignat}
\end{subequations}

Problem~(P1d) belongs to semidefinite programming and can be solved efficiently~\cite{luo2010semidefinite}. Wu and Zhang~\cite{wu2018intelligentfull} consider Problem~(P1d) in the unicast setting (i.e., the special case of $g=K$ with each MU being a group). Moreover, in a spirit similar to~\cite{wu2018intelligentfull}, Problem~(P1d) can be replaced by Problem~(P1d$'$) below which may find better $\boldPhi$ and hence $\boldsymbol{V}$ to accelerate the alternating optimization process:
\begin{subequations} \label{eq-Prob-P1d-prime}
\begin{alignat}{2}
\text{(P1d$'$)}:\max_{\boldsymbol{V},\boldsymbol{\alpha}}& \sum_{k=1}^g \sum_{i \in \mathcal{G}_k} \alpha_i \\
\mathrm{s.t.}~~~~~&   \tr(\bm{A}_i(\bm{w}_{k})\boldsymbol{V} )  + {b}_{i}(\bm{w}_{k}) {b}_{i}^H(\bm{w}_{k}) \nonumber  \\ & \geq \alpha_i + \gamma_i \bigg\{  \sigma^2_i + \nonumber  \\ & \hspace{-25pt}\sum\limits_{j\in\{1,\ldots,g\}\setminus\{k\}} \left[ \tr(\bm{A}_i(\bm{w}_{j})\boldsymbol{V} )  + {b}_{i}(\bm{w}_{j}) {b}_{i}^H(\bm{w}_{j})\right]\hspace{-2pt}\bigg\}, \label{eq:prob-P1c-prime-SINR-constraint}\\
&~~~~~~~\forall k \in\{1,\ldots,g\},~\forall i \in \mathcal{G}_k, \nonumber  \\
 &\hspace{-30pt}\text{$\boldsymbol{V}_{n,n}  = {\beta_n}^2  $ for $n\in \{1,\ldots, N\}$ and $\boldsymbol{V}_{N+1,N+1} = 1,$} \\
& \boldsymbol{V} \succeq 0, \\
& \alpha_i \geq 0, ~\forall k \in\{1,\ldots,g\},~\forall i \in \mathcal{G}_k. 
\end{alignat}
\end{subequations}
The quantity $\alpha_i$ can be understood as MU $i$'s ``SINR residual'' in the phase shift optimization~\cite{wu2018intelligentfull}.

After  a candidate solution $\boldsymbol{V}$ is obtained by solving Problem~(P1d) or Problem~(P1d$'$), appropriate post-processing such as Gaussian randomization~\cite{so2007approximating} can be applied to convert the candidate solution into a solution which satisfies the rank constraint in Eq.~(\ref{eq:prob-P1c-rank-constraint}).

Combining the above discussion of alternatively optimizing $\W$ and $\boldPhi$, we present our method to solve Problem~(P1) as Algorithm~\ref{Alg-Problem-P1}. 



\begin{algorithm}
\caption{via alternating optimization to find $\W$ and $\boldsymbol{\Phi}$ for Problem~(P1), which generalizes Problems~(P2) and~(P3).} \label{Alg-Problem-P1}
\begin{algorithmic}[1]
\STATE Initialize  $\boldsymbol{\Phi}$ as some initial (e.g., randomly generated) $ \boldsymbol{\Phi}^{(0)}:= \text{diag}(\beta_1 e^{j \theta_1^{(0)}}, \ldots, \beta_N e^{j \theta_N^{(0)}})$ which satisfies the constraints on $\boldPhi$;
\STATE Set the iteration number $r \leftarrow 1$;
\WHILE{1}
\item[] \COMMENT{The ``while'' loop will end if Line~\ref{Alg-Problem-P1-break1} or \ref{Alg-Problem-P1-break2} is executed.}
\item[] \COMMENT{Comment: Optimizing $\W$ given $\boldPhi$:}
\STATE Given $\boldPhi$ as $\boldsymbol{\Phi}^{(r-1)}$, use methods of Karipidis~\textit{et~al.}~\cite{karipidis2008quality} or other papers to solve Problem~(P1a) and post-process the obtained $\{\bm{X}_k\}|_{k=1}^g$ to set $\boldsymbol{W}$ as some $\boldsymbol{W}^{(r)}:=[\bm{w}_1^{(r)},\ldots,\bm{w}_g^{(r)}]$; \label{Alg-opt-P1-0}
\item[] \vspace{-5pt}
\STATE \mbox{Compute the object function value $f^{(r)} \leftarrow \sum_{k=1}^{g}\|\bm{w}_k^{(r)}\|^2$};
\IF{$r\geq 2$}
\IF{$1-\frac{f^{(r)}}{f^{(r-1)}}$ denoting the relative difference  between the object function values in consecutive iterations $r-1$ and $r$ is small}
\STATE \textbf{break};  \label{Alg-Problem-P1-break1}
\ENDIF
\ENDIF
\item[] \vspace{-5pt}
\item[] \COMMENT{Finding $\boldPhi$ given $\W$:}
\STATE Given $\W$ as $\boldsymbol{W}^{(r)}$, solve Problem~(P1d) in Eq.~(\ref{eq-Prob-P1d}) or Problem~(P1d$'$) in Eq.~(\ref{eq-Prob-P1d-prime}), and denote the obtained~$\boldsymbol{V}$ as~$\boldsymbol{V}^{(r)}_{\text{SDR}}$;
\item[] \COMMENT{Comment: Gaussian randomization:}
\STATE Perform the eigenvalue decomposition on $\boldsymbol{V}^{(r)}_{\text{SDR}}$ to obtain a unitary matrix $\boldsymbol{U}$ and a  diagonal  matrix $\boldsymbol{\Lambda}$ such that $ \boldsymbol{V}^{(r)}_{\text{SDR}} = \boldsymbol{U} \boldsymbol{\Lambda}  \boldsymbol{U}^H $;\label{Alg-opt-Gaussian-randomization-decomposition}
\FOR{$z$ from $1$ to some sufficiently large $Z$}
\STATE Generate a random vector $\boldsymbol{r}^{(r)}_{z}$ from a circularly-symmetric complex Gaussian distribution $\mathcal{CN}(\boldsymbol{0},\boldsymbol{I}_{N+1})$ with zero mean and covariance matrix $\boldsymbol{I}_{N+1}$ (the identity matrix with size $N+1$); \STATE Compute $ \boldsymbol{v}^{(r)}_{z}  \leftarrow \boldsymbol{U} \boldsymbol{\Lambda}^{\frac{1}{2}} \boldsymbol{r}^{(r)}_{z}$;
\STATE With $v_{N+1}$ being the $(N+1)$th element of $\boldsymbol{v}_{z}^{(r)}$, {take the first $N$ elements of $\frac{\boldsymbol{v}_{z}^{(r)}}{v_{N+1}}$ to form a vector $\boldsymbol{w}^{(r)}_{z}$};
\STATE 
Scale each component of $\boldsymbol{w}^{(r)}_{z}$ independently to obtain $\bm{\phi}^{(r)}_{z}$ such that the $n$th element of $\bm{\phi}^{(r)}_{z}$ has magnitude ${\beta_n}$; i.e., with $\boldsymbol{w}^{(r)}_{z}$ represented by $[w_1, \ldots, w_{N}]^T$, compute  $\bm{\phi}^{(r)}_{z} \leftarrow \left[{\beta_1}\frac{w_1}{|w_1|} , \ldots, {\beta_N}\frac{w_{N}}{|w_{N}|}\right]^T $;
\ENDFOR
\IF{for $z\in\{1,2,\ldots,Z\}$, there is no $\bm{\phi}^{(r)}_{z}$ to ensure Eq.~(\ref{eq-hi-def-ai-wj-def-b-wj-square-v3}) after setting $\bm{\phi}$ as $\bm{\phi}^{(r)}_{z}$}
\STATE \textbf{break};  \label{Alg-Problem-P1-break2}
\ELSE
\STATE Select one $\bm{\phi}^{(r)}_{z}$ according to some ordering among those ensuring Eq.~(\ref{eq-hi-def-ai-wj-def-b-wj-square-v3}) and with $\bm{\phi}^{(r)}_{z^*}$ denoting the selected one;
\STATE Map $\bm{\phi}^{(r)}_{z^*}$ to some ${{\widetilde{\bm{\phi}}}^{(r)}_{z^*}}$ to satisfy the constraint on $\bm{\phi}$ (e.g., discrete element values);
\STATE Set $\boldsymbol{\Phi} \leftarrow \boldsymbol{\Phi}^{(r)}_{z^*}$ for  $\boldsymbol{\Phi}^{(r)}_{z^*} := \text{diag}\left(({{\widetilde{\bm{\phi}}}^{(r)}_{z^*}})^H\right)$, and denote such $\boldsymbol{\Phi}$ as $\boldsymbol{\Phi}^{(r)}$ for notation convenience;
\ENDIF
\STATE Update the iteration number $r  \leftarrow r + 1$;  \label{Alg-opt-update-r}
\ENDWHILE
\end{algorithmic}
\end{algorithm}

\subsection{Solutions to Problems~(P4)--(P6)} \label{subsection-P4-P6}


We now propose algorithms to solve Problems~(P4)--(P6). Since Problems~(P5) and~(P6) can be seen as special cases of Problem~(P4), we first focus on solving Problem~(P4), and then apply the solutions to Problems~(P5) and~(P6).

We introduce an auxiliary variable $t$ and convert Problem~(P4) of Eq.~(\ref{eq-Prob-P4}) into the following equivalent Problem~(P4a):
\begin{subequations} \label{eq-Prob-P4a}
\begin{alignat}{2}
\text{(P4a):}~\max_{\W, \boldPhi, t}~
& t   \label{OptProb-P4a-fair-SINR-multicast-obj}   \\
\mathrm{s.t.}~& \frac{|\bm{h}^H_{i}(\boldPhi)\bm{w}_k |^2}{\gamma_i\left[\sum\limits_{j\in\{1,\ldots,g\}\setminus\{k\}}|\bm{h}^H_{i}(\boldPhi)\bm{w}_j |^2 +  \sigma^2_i\right]} \geq t, \label{OptProb-P4a-fair-SINR-multicast-obj-SINR-constraint} \\
&~~~~~~~\forall k \in\{1,\ldots,g\},~\forall i \in \mathcal{G}_k, \nonumber \\
 & \sum\limits_{k=1}^{g}\|\bm{w}_k\|^2 \leq P , \label{OptProb-P4a-fair-SINR-multicast-obj-power-constraint}   \\
 &\text{Constraints on $\boldPhi$},\label{OptProb-P4a-fair-SINR-multicast-obj-Phi-constraint}  \\
 & t \geq 0. \label{OptProb-P4a-fair-SINR-multicast-obj-t-constraint}
 \end{alignat}
\end{subequations}

We use the idea of alternating optimization, which is widely used in multivariate optimization. Specifically, we perform the following optimizations alternatively to solve Problem~(P4): optimizing $\W$ and $t$ given $\boldPhi$, and finding $\boldPhi$ given $\W$ and $t$.  The details are presented below.

\textbf{Optimizing $\W$ and $t$ given $\boldPhi$.} Given some $\boldPhi$ satisfying the constraints in (\ref{OptProb-P4a-fair-SINR-multicast-obj-Phi-constraint}), Problem~(P4a) means finding $\W$ and $t$ to maximize $t$ subject to the constraints in~(\ref{OptProb-P4a-fair-SINR-multicast-obj-SINR-constraint})~(\ref{OptProb-P4a-fair-SINR-multicast-obj-power-constraint})~(\ref{OptProb-P4a-fair-SINR-multicast-obj-t-constraint}).

We define $\bm{X}_k$ and $\bm{H}_{i}(\boldPhi)$ according to Eq.~(\ref{eq-Xk-wk}) and~(\ref{eq-Hk-Phi}); i.e.,
\begin{align}
&\bm{X}_k := \bm{w}_k \bm{w}_k^H   \label{eq-Xk-wk-v2}, \\
&\bm{H}_{i}(\boldPhi) := \bm{h}_{i}(\boldPhi)\bm{h}^H_{i}(\boldPhi). \label{eq-Hk-Phi-v2}
\end{align}
Replacing $k$ by $j$ in Eq.~(\ref{eq-Hk-Phi-v2}), we also have expressions for $\bm{H}_{j}(\boldPhi)$. Then similar to the process of writing Inequality~(\ref{OptProb-power-multicast-SINR-constraint}) of Problem~(P1) as Inequality~(\ref{eq:prob-P1-0-}) of Problem~(P1a), we write Inequality~(\ref{OptProb-P4a-fair-SINR-multicast-obj-SINR-constraint}) of Problem~(P4a) as Inequality~(\ref{eq:prob-P4b-}) below. Then given  $\boldPhi$, Problem~(P4a) becomes the following Problem~(P4b):
\begin{subequations} \label{eq-Prob-P4b}
\begin{alignat}{2}
\text{(P4b):}~\max_{\{\bm{X}_k\}|_{k=1}^g, t}~
& t   \label{OptProb-P4b-fair-SINR-multicast-obj}   \\
\mathrm{s.t.}~~~~~~&   \tr(\bm{X}_k \bm{H}_{i}(\boldPhi)) \geq \nonumber \\
& \hspace{-15pt}t\gamma_i \sigma^2_i+  t\gamma_i \sum_{j\in\{1,\ldots,g\}\setminus\{k\}}\tr(\bm{X}_j \bm{H}_{i}(\boldPhi) ) ,  \label{eq:prob-P4b-}\\
&~~~~~~~\forall k \in\{1,\ldots,g\},~\forall i \in \mathcal{G}_k, \nonumber  \\
& \sum\limits_{k=1}^{g}\tr(\bm{X}_k)\leq P ,   \\
& \bm{X}_k \succeq 0,~\forall k \in\{1,\ldots,g\},  \\
& \text{rank}(\bm{X}_k) =1,~\forall k \in\{1,\ldots,g\}, \label{eq:prob-P4b-rank-constraint} \\
 & t \geq 0. \label{OptProb-P4b-fair-SINR-multicast-obj-t-constraint}
\end{alignat}
\end{subequations}
 

A problem similar to Problem~(P4b) has been used to solved by Karipidis~\textit{et~al.}~\cite{karipidis2008quality}, where the notation of Problem~$\mathcal{F}_r$ is used to denote the problem after dropping Eq.~(\ref{eq:prob-P4b-rank-constraint}). Hence, we can apply methods of~\cite{karipidis2008quality} to solve Problem~(P4b).
 


\textbf{Finding $\boldPhi$ given $\W$ and $t$.} Given $\W$ and $t$, Problem~(P4) becomes the following feasibility check problem of finding $\boldPhi$:  
\begin{subequations} \label{eq-Prob-P4d}
\begin{alignat}{2}
\text{(P4d)}: \Find~~~& \boldPhi \\
\mathrm{s.t.}~~~~~&   \frac{|\bm{h}^H_{i}(\boldPhi)\bm{w}_k |^2}{\sum\limits_{j\in\{1,\ldots,g\}\setminus\{k\}}|\bm{h}^H_{i}(\boldPhi)\bm{w}_j |^2 +  \sigma^2_i}\geq t \gamma_i, \label{eq:prob-P4d-SINR-constraint}\\
&~~~~~~~\forall k \in\{1,\ldots,g\},~\forall i \in \mathcal{G}_k, \nonumber  \\
 &\text{Constraints on $\boldPhi$}. \label{eq:prob-P4d-rank-constraint}
\end{alignat}
\end{subequations}

The only difference between Problem~(P4d) of Eq.~(\ref{eq-Prob-P4d}) and Problem~(P1b) of Eq.~(\ref{eq-Prob-P1b}) is that the right hand side in Inequality~(\ref{eq:prob-P4d-SINR-constraint}) of Problem~(P4d) has $ t \gamma_i$, whereas the right hand side in Inequality~(\ref{eq:prob-P1a-SINR-constraint}) of Problem~(P1b) has $ \gamma_i$. Hence, we can apply the discussed approach of solving Problem~(P1b) to solve Problem~(P4d). Specifically, if the constraint on $\boldPhi$ in Eq.~(\ref{eq:prob-P1a-rank-constraint}) is  Eq.~(\ref{eq-def-phase-shift-matrix-general}) (i.e., $\boldPhi:= \text{diag}(\beta_1 e^{j \theta_1}, \ldots, \beta_N e^{j \theta_N})$), we define $\bm{\phi}$, $b_{i}(\bm{w}_{k})$, and $\bm{A}_i(\bm{w}_{k})$ according to Eq.~(\ref{eq-define-bm-phi})~(\ref{eq-def-b-wk}) and~(\ref{eq-def-Ai-matrix-wk}). Then in a way similar to the derivation leading to Inequality~(\ref{eq-hi-def-ai-wj-def-b-wj-square-v3}),  we write the constraint in Eq.~(\ref{eq:prob-P4d-SINR-constraint}) of Problem~(P4d) as 
\begin{align}
 & \begin{bmatrix} 
\bm{\phi}^H ,~
1
\end{bmatrix} \bm{A}_i(\bm{w}_{k}) \begin{bmatrix} 
\bm{\phi} \\
1
\end{bmatrix} + {b}_{i}(\bm{w}_{k}) {b}_{i}^H(\bm{w}_{k}) \nonumber  \\ & \geq \hspace{-2pt}t \gamma_i \hspace{-2pt}\left\{\hspace{-2pt}  \sigma^2_i\hspace{-2pt}+\hspace{-10pt} \sum\limits_{j\in\{1,\ldots,g\}\setminus\{k\}} \hspace{-4pt} \left\{\hspace{-2pt}\begin{bmatrix} 
\bm{\phi}^H \hspace{-2pt},\hspace{-2pt}~1
\end{bmatrix} \hspace{-2pt}\bm{A}_i(\bm{w}_{j})\hspace{-2pt} \begin{bmatrix} 
\bm{\phi} \\
1
\end{bmatrix} \hspace{-2pt}+\hspace{-2pt} {b}_{i}(\bm{w}_{j}) {b}_{i}^H(\bm{w}_{j})\hspace{-3pt} \right\}\hspace{-4pt}\right\}\hspace{-2pt}.\label{eq-hi-def-ai-wj-def-b-wj-square-v3-P4}
\end{align}

Then defining $\boldsymbol{V}$ according to (\ref{eq-boldsymbol-V-v}), similar to the process of formulating Problem~(P1c) of Eq.~(\ref{eq-Prob-P1c}), we can convert Problem~(P4d) into the following Problem~(P4e):
\begin{subequations} \label{eq-Prob-P4e}
\begin{alignat}{2}
\text{(P4e)}: \Find~~~& \boldsymbol{V} \\
\mathrm{s.t.}~~~~~&   \tr(\bm{A}_i(\bm{w}_{k})\boldsymbol{V} )  + {b}_{i}(\bm{w}_{k}) {b}_{i}^H(\bm{w}_{k}) \nonumber  \\ & \geq t \gamma_i \bigg\{  \sigma^2_i + \nonumber  \\ & \hspace{-25pt}\sum\limits_{j\in\{1,\ldots,g\}\setminus\{k\}} \left[ \tr(\bm{A}_i(\bm{w}_{j})\boldsymbol{V} )  + {b}_{i}(\bm{w}_{j}) {b}_{i}^H(\bm{w}_{j})\right]\hspace{-2pt}\bigg\}, \label{eq:prob-P4e-SINR-constraint}\\
&~~~~~~~\forall k \in\{1,\ldots,g\},~\forall i \in \mathcal{G}_k, \nonumber  \\
 &\hspace{-30pt}\text{$\boldsymbol{V}_{n,n}  = {\beta_n}^2  $ for $n\in \{1,\ldots, N\}$ and $\boldsymbol{V}_{N+1,N+1} = 1,$} \\
& \boldsymbol{V} \succeq 0, \\
& \text{rank}(\boldsymbol{V}) =1.\label{eq:prob-P4e-rank-constraint}
\end{alignat}
\end{subequations}

The only non-convex part in Problem~(P4e) is the rank constraint in Eq.~(\ref{eq:prob-P4e-rank-constraint}). Hence, we adopt semidefinite relaxation (SDR) and drop Eq.~(\ref{eq:prob-P4e-rank-constraint}) to obtain a semidefinite programming problem. The discussion is similar to that for Problem~(P1c) of Eq.~(\ref{eq-Prob-P1c}). Similar to the practice of replacing (P1d) by (P1d$'$), we can also replace (P4e) by (P4e$'$) below which may find better $\boldPhi$ and hence $\boldsymbol{V}$ to accelerate the alternating optimization process:  
\begin{subequations} \label{eq-Prob-P4e-prime}
\begin{alignat}{2}
\text{(P4e$'$)}:\max_{\boldsymbol{V},\boldsymbol{\alpha}}& \sum_{k=1}^g \sum_{i \in \mathcal{G}_k} \alpha_i \\
\mathrm{s.t.}~~~~~&   \tr(\bm{A}_i(\bm{w}_{k})\boldsymbol{V} )  + {b}_{i}(\bm{w}_{k}) {b}_{i}^H(\bm{w}_{k}) \nonumber  \\ & \geq \alpha_i + t \gamma_i \bigg\{  \sigma^2_i + \nonumber  \\ & \hspace{-25pt}\sum\limits_{j\in\{1,\ldots,g\}\setminus\{k\}} \left[ \tr(\bm{A}_i(\bm{w}_{j})\boldsymbol{V} )  + {b}_{i}(\bm{w}_{j}) {b}_{i}^H(\bm{w}_{j})\right]\hspace{-2pt}\bigg\}, \label{eq:prob-P1c-prime-SINR-constraint}\\
&~~~~~~~\forall k \in\{1,\ldots,g\},~\forall i \in \mathcal{G}_k, \nonumber  \\
 &\hspace{-30pt}\text{$\boldsymbol{V}_{n,n}  = {\beta_n}^2  $ for $n\in \{1,\ldots, N\}$ and $\boldsymbol{V}_{N+1,N+1} = 1,$} \\
& \boldsymbol{V} \succeq 0, \\
& \alpha_i \geq 0, ~\forall k \in\{1,\ldots,g\},~\forall i \in \mathcal{G}_k.
\end{alignat}
\end{subequations}

After  a candidate solution $\boldsymbol{V}$ is obtained by solving Problem~(P4e) or Problem~(P4e$'$), appropriate post-processing such as Gaussian randomization~\cite{so2007approximating} can be applied to convert the candidate solution into a solution which satisfies the rank constraint in Eq.~(\ref{eq:prob-P4e-rank-constraint}).

Combining the above discussion of alternatively optimizing $\W$ and $\boldPhi$, we present our method to solve Problem~(P4) as Algorithm~\ref{Alg-Problem-P4}.

\begin{algorithm}
\caption{via alternating optimization to find $\W$ and $\boldsymbol{\Phi}$ for Problem~(P4), which generalizes Problems~(P5) and~(P6).} \label{Alg-Problem-P4}
\begin{algorithmic}[1]
\STATE Initialize  $\boldsymbol{\Phi}$ as some initial (e.g., randomly generated) $ \boldsymbol{\Phi}^{(0)}:= \text{diag}(\beta_1 e^{j \theta_1^{(0)}}, \ldots, \beta_N e^{j \theta_N^{(0)}})$ which satisfies the constraints on $\boldPhi$;
\STATE Set the iteration number $r \leftarrow 1$;
\WHILE{1}
\item[] \COMMENT{The ``while'' loop will end if Line~\ref{Alg-Problem-P4-break1} or \ref{Alg-Problem-P4-break2} is executed.}
\item[] \COMMENT{Comment: Optimizing $\W$ and $t$ given $\boldPhi$:}
\STATE Given $\bm{\Phi}$ as $\bm{\Phi}^{(r-1)}$, use methods of Karipidis~\textit{et~al.}~\cite{karipidis2008quality} or other papers to solve Problem~(P4b) in Eq.~(\ref{eq-Prob-P4b}), and post-process the solution to set $\boldsymbol{W}$ as some $\boldsymbol{W}^{(r)}:=[\bm{w}_1^{(r)},\ldots,\bm{w}_g^{(r)}]$ and set $t$ as some $t^{(r)}$; \label{Alg-opt-P4-0}
\item[] \vspace{-5pt}
\IF{$r\geq 2$}
\IF{$ \frac{t^{(r)}}{t^{(r-1)}}-1$ denoting the relative difference  between the object function values in consecutive iterations $r-1$ and $r$ is small}
\STATE \textbf{break};  \label{Alg-Problem-P4-break1}
\ENDIF
\ENDIF
\item[] \vspace{-5pt}
\item[] \COMMENT{Finding $\boldPhi$ given $\W$ and $t$:}
\STATE Given $\W$ as $\boldsymbol{W}^{(r)}$ and $t$ as $t^{(r)}$, solve Problem~(P4e) in Eq.~(\ref{eq-Prob-P4e}) or Problem~(P4e$'$) in Eq.~(\ref{eq-Prob-P4e-prime}), and denote the obtained~$\boldsymbol{V}$ as~$\boldsymbol{V}^{(r)}_{\text{SDR}}$;
\item[] \COMMENT{Comment: Gaussian randomization:}
\STATE Perform the eigenvalue decomposition on $\boldsymbol{V}^{(r)}_{\text{SDR}}$ to obtain a unitary matrix $\boldsymbol{U}$ and a  diagonal  matrix $\boldsymbol{\Lambda}$ such that $ \boldsymbol{V}^{(r)}_{\text{SDR}} = \boldsymbol{U} \boldsymbol{\Lambda}  \boldsymbol{U}^H $;\label{Alg-opt-Gaussian-randomization-decomposition}
\FOR{$z$ from $1$ to some sufficiently large $Z$}
\STATE Generate a random vector $\boldsymbol{r}^{(r)}_{z}$ from a circularly-symmetric complex Gaussian distribution $\mathcal{CN}(\boldsymbol{0},\boldsymbol{I}_{N+1})$ with zero mean and covariance matrix $\boldsymbol{I}_{N+1}$ (the identity matrix with size $N+1$); \STATE Compute $ \boldsymbol{v}^{(r)}_{z}  \leftarrow \boldsymbol{U} \boldsymbol{\Lambda}^{\frac{1}{2}} \boldsymbol{r}^{(r)}_{z}$;
\STATE With $v_{N+1}$ being the $(N+1)$th element of $\boldsymbol{v}_{z}^{(r)}$, {take the first $N$ elements of $\frac{\boldsymbol{v}_{z}^{(r)}}{v_{N+1}}$ to form a vector $\boldsymbol{w}^{(r)}_{z}$};
\STATE 
Scale each component of $\boldsymbol{w}^{(r)}_{z}$ independently to obtain $\bm{\phi}^{(r)}_{z}$ such that the $n$th element of $\bm{\phi}^{(r)}_{z}$ has magnitude ${\beta_n}$; i.e., with $\boldsymbol{w}^{(r)}_{z}$ represented by $[w_1, \ldots, w_{N}]^T$, compute  $\bm{\phi}^{(r)}_{z} \leftarrow \left[{\beta_1}\frac{w_1}{|w_1|} , \ldots, {\beta_N}\frac{w_{N}}{|w_{N}|}\right]^T $;
\ENDFOR
\IF{for $z\in\{1,2,\ldots,Z\}$, there is no $\bm{\phi}^{(r)}_{z}$ to ensure Eq.~(\ref{eq-hi-def-ai-wj-def-b-wj-square-v3-P4}) after setting $\bm{\phi}$ as $\bm{\phi}^{(r)}_{z}$}
\STATE \textbf{break};  \label{Alg-Problem-P4-break2}
\ELSE
\STATE Select one $\bm{\phi}^{(r)}_{z}$ according to some ordering among those ensuring Eq.~(\ref{eq-hi-def-ai-wj-def-b-wj-square-v3-P4}) and with $\bm{\phi}^{(r)}_{z^*}$ denoting the selected one;
\STATE Map $\bm{\phi}^{(r)}_{z^*}$ to some ${{\widetilde{\bm{\phi}}}^{(r)}_{z^*}}$ to satisfy the constraint on $\bm{\phi}$ (e.g., discrete element values);
\STATE Set $\boldsymbol{\Phi} \leftarrow \boldsymbol{\Phi}^{(r)}_{z^*}$ for  $\boldsymbol{\Phi}^{(r)}_{z^*} := \text{diag}\left(({{\widetilde{\bm{\phi}}}^{(r)}_{z^*}})^H\right)$, and denote such $\boldsymbol{\Phi}$ as $\boldsymbol{\Phi}^{(r)}$ for notation convenience;
\ENDIF
\STATE Update the iteration number $r  \leftarrow r + 1$;  \label{Alg-opt-update-r}
\ENDWHILE
\end{algorithmic}
\end{algorithm}

\subsection{Solutions to Problems~(P1-MA)--(P3-MA),\\ \mbox{(P1-MR)--(P3-MR)}, and~(P1-MA-MR)--(P3-MA-MR)} \label{subsection-P1-MA-MR}


As Problems~(P1-MA)--(P3-MA),~(P1-MR)--(P3-MR), and~(P1-MA-MR)--(P3-MA-MR) are generalizations of Problems~(P1)--(P3), we will solve the former problems in ways similar to the solutions for the latter problems, which we have discussed in Section~\ref{subsection-P1-P3}. More specifically, since Problem~(P1-MA-MR) is in the most general form, we start with elaborating its solution below.

We restate Problem~(P1-MA-MR) given in Table~\ref{table:4MA-MR} of   Page~\pageref{table:4MA-MR}:
\begin{subequations} \label{eq-Prob-P1-MA-MR-restated}
\begin{alignat}{2}
&\text{(P1-MA-MR)}: \nonumber \\
&\min_{\W, \bm{\Phi}_1,\ldots,\bm{\Phi}_L}~\sum\limits_{k=1}^{g}\|\bm{w}_k\|^2   \label{OptProb-power-multicast-obj-P1-MA-MR}  \\
&~\mathrm{s.t.}~\frac{\bm{w}_k^H\bm{H}_{i}(\bm{\Phi}_1,\ldots,\bm{\Phi}_L)\bm{w}_k}{\sum\limits_{j\in\{1,\ldots,g\}\setminus\{k\}}\bm{w}_j^H\bm{H}_{i}(\bm{\Phi}_1,\ldots,\bm{\Phi}_L)\bm{w}_j  +  \sigma^2_i}\geq \gamma_i, \label{OptProb-power-multicast-SINR-constraint-P1-MA-MR}   \\
&~~~~~~~~~~\forall k \in\{1,\ldots,g\},~\forall i \in \mathcal{G}_k, \nonumber\\
 &~~~~~~\text{Constraints on $\bm{\Phi}_1,\ldots,\bm{\Phi}_L$}. \label{OptProb-power-multicast-Phi-constraint-P1-MA-MR}
\end{alignat}
\end{subequations}

We use the idea of alternating optimization. Specifically, we will optimize $\bm{\Phi}_1,\ldots,\bm{\Phi}_L$ and $\W$ alternatively to solve Problem~(P1-MA-MR). Below, we discuss the optimization of $\W$ given $\bm{\Phi}_1,\ldots,\bm{\Phi}_L$ and the optimization of $\bm{\Phi}_1,\ldots,\bm{\Phi}_L$ given $\W$.

\textbf{Optimizing $\W$ given $\bm{\Phi}_1,\ldots,\bm{\Phi}_L$.} Given some $\bm{\Phi}_1,\ldots,\bm{\Phi}_L$ satisfying the constraints in Eq.~(\ref{OptProb-power-multicast-Phi-constraint-P1-MA-MR}), Problem~(P1-MA-MR) means finding $\W$ given $\bm{\Phi}_1,\ldots,\bm{\Phi}_L$ to minimize the objective function $\sum\limits_{k=1}^{g}\|\bm{w}_k\|^2$ in~(\ref{OptProb-power-multicast-obj-P1-MA-MR}) subject to the constraints in~(\ref{OptProb-power-multicast-SINR-constraint-P1-MA-MR}).

As in Eq.~(\ref{eq-Xk-wk}), we define $\bm{X}_k$ as follows:
\begin{align}
 & \bm{X}_k := \bm{w}_k \bm{w}_k^H,~\forall k \in\{1,\ldots,g\}. \label{eq-Xk-wk-repeat}
\end{align}
Similar to the process of writing Inequality~(\ref{OptProb-power-multicast-SINR-constraint}) of Problem~(P1) as Inequality~(\ref{eq:prob-P1-0-}) of Problem~(P1a), we write Inequality~(\ref{OptProb-power-multicast-SINR-constraint-P1-MA-MR}) of Problem~(P1-MA-MR) as Inequality~(\ref{OptProb-power-multicast-SINR-constraint-P1-MA-MR-a}) below. Then optimizing $\W$ given $\bm{\Phi}_1,\ldots,\bm{\Phi}_L$ for Problem~(P1-MA-MR) becomes solving $\{\bm{X}_k\}|_{k=1}^g$ for the following Problem~(P1-MA-MR-a): 
\begin{subequations} \label{eq-Prob-P1-MA-MR-a}
\begin{alignat}{2}
&\hspace{-40pt}\text{(P1-MA-MR-a)}: \nonumber \\
 \min_{\{\bm{X}_k\}|_{k=1}^g} ~~~& \sum_{k=1}^{g}\tr(\bm{X}_k) \\
\mathrm{s.t.}~~~~~~&   \tr(\bm{X}_k \bm{H}_{i}(\bm{\Phi}_1,\ldots,\bm{\Phi}_L)) \geq \nonumber \\
& \hspace{-15pt}\gamma_i \sigma^2_i+  \gamma_i \sum_{j\in\{1,\ldots,g\}\setminus\{k\}}\tr(\bm{X}_j \bm{H}_{i}(\bm{\Phi}_1,\ldots,\bm{\Phi}_L) ) ,  \label{OptProb-power-multicast-SINR-constraint-P1-MA-MR-a}\\
&~~~~~~~\forall k \in\{1,\ldots,g\},~\forall i \in \mathcal{G}_k,  \nonumber \\
& \bm{X}_k \succeq 0,~\forall k \in\{1,\ldots,g\}, \\
& \text{rank}(\bm{X}_k) =1,~\forall k \in\{1,\ldots,g\}. \label{eq:prob-P1-MA-MR-a-rank-constraint}
\end{alignat}
\end{subequations}

The only non-convex part in Problem~(P1-MA-MR-a) is the rank constraint in Eq.~(\ref{eq:prob-P1-MA-MR-a-rank-constraint}). Hence, we adopt semidefinite relaxation (SDR) and drop Eq.~(\ref{eq:prob-P1-MA-MR-a-rank-constraint}) to obtain a semidefinite programming problem. After a candidate solution is obtained, appropriate post-processing such as Gaussian randomization~\cite{so2007approximating} (see also randA, randB, and randC coined by~\cite{sidiropoulos2006transmit}) is applied to convert the candidate solution into a solution which satisfies the rank constraint. The above method has been used to solve a problem similar to Problem~(P1-MA-MR-a) by Karipidis~\textit{et~al.}~\cite{karipidis2008quality}, where the notation of Problem~$\mathcal{Q}$ is used. Hence, we can apply methods of~\cite{karipidis2008quality} to solve Problem~(P1-MA-MR-a).

\textbf{Finding $\bm{\Phi}_1,\ldots,\bm{\Phi}_L$ given $\W$.} Given $\W$, Problem~(P1-MA-MR) becomes the following feasibility check problem of finding $\bm{\Phi}_1,\ldots,\bm{\Phi}_L$:
\begin{subequations} \label{eq-Prob-P1-MA-MR-b}
\begin{alignat}{2}
&\hspace{-40pt}\text{(P1-MA-MR-b)}: \nonumber \\
\Find~~~& \bm{\Phi}_1,\ldots,\bm{\Phi}_L \\
\mathrm{s.t.}~~~~~&   \frac{\bm{w}_k^H\bm{H}_{i}(\bm{\Phi}_1,\ldots,\bm{\Phi}_L)\bm{w}_k}{\sum\limits_{j\in\{1,\ldots,g\}\setminus\{k\}}\bm{w}_j^H\bm{H}_{i}(\bm{\Phi}_1,\ldots,\bm{\Phi}_L)\bm{w}_j  +  \sigma^2_i}\geq  \gamma_i, \label{eq:prob-P1-MA-MR-b-SINR-constraint}\\
&~~~~~~~\forall k \in\{1,\ldots,g\},~\forall i \in \mathcal{G}_k, \nonumber  \\
 &\text{Constraints on $\bm{\Phi}_1,\ldots,\bm{\Phi}_L$}. \label{eq:prob-P1-MA-MR-b-rank-constraint}
\end{alignat}
\end{subequations}

Recall from the caption of Table~\ref{table:4MA-MR} on   Page~\pageref{table:4MA-MR} that $\bm{H}_{i}(\bm{\Phi}_1,\ldots,\bm{\Phi}_L)$ appearing in Inequality~(\ref{eq:prob-P1-MA-MR-b-SINR-constraint}) is defined as
\begin{align}
\bm{H}_{i}(\bm{\Phi}_1,\ldots,\bm{\Phi}_L) : = \sum_{q=1}^{Q_i}  \bm{h}_{i,q}(\bm{\Phi}_1,\ldots,\bm{\Phi}_L)  \bm{h}^H_{i,q}(\bm{\Phi}_1,\ldots,\bm{\Phi}_L) , \label{eq:Hi-Phi1-Phi2-PhiL}
\end{align}
for
\begin{align}
\bm{h}^H_{i,q}(\bm{\Phi}_1,\ldots,\bm{\Phi}_L) : = \bm{h}^H_{\text{b},i,q} +  \sum_{\ell=1}^L\bm{h}^H_{\ell,i,q}\boldPhi_{\ell} \bm{H}_{\text{b},\ell}. \label{eq:h-H-k-q}
\end{align} 
Then $\bm{w}_k^H\bm{H}_{i}(\bm{\Phi}_1,\ldots,\bm{\Phi}_L)\bm{w}_k$ appearing in Inequality~(\ref{eq:prob-P1-MA-MR-b-SINR-constraint}) of Problem~(P1-MA-MR-b) is given by
\begin{align}
 &\bm{w}_k^H\bm{H}_{i}(\bm{\Phi}_1,\ldots,\bm{\Phi}_L)\bm{w}_k \nonumber \\
 & = \sum_{q=1}^{Q_i}  \bm{w}_k^H\bm{h}_{i,q}(\bm{\Phi}_1,\ldots,\bm{\Phi}_L)  \bm{h}^H_{i,q}(\bm{\Phi}_1,\ldots,\bm{\Phi}_L)\bm{w}_k. \label{eq:wk-Hi-Phi1-Phi2-PhiL}
\end{align}
Below we analyze $\bm{h}^H_{i,q}(\bm{\Phi}_1,\ldots,\bm{\Phi}_L)\bm{w}_k$ which appears in Eq.~(\ref{eq:wk-Hi-Phi1-Phi2-PhiL}).   

If the constraint on each $\boldPhi_{\ell}$ is in the form of Eq.~(\ref{eq-def-phase-shift-matrix-general}); i.e., 
\begin{align}
\boldPhi_{\ell}:= \text{diag}(\beta_{\ell,1}  e^{j \theta_{\ell,1}}, \ldots, \beta_{\ell,N_{\ell}} e^{j \theta_{\ell,N_{\ell}}}),~\forall \ell \in\{1,\ldots,L\}, \label{eq-constraint-Phi-ell}
\end{align}
then we define
\begin{align}
 & \bm{\phi}_{\ell} : = [\beta_{\ell,1} e^{j \theta_{\ell,1}}, \ldots, \beta_{\ell,N_{\ell}} e^{j \theta_{\ell,N_{\ell}}}]^H,~\forall \ell \in\{1,\ldots,L\}, \label{eq-define-bm-phi-ell-0-L}
\end{align} 
and change variables in Eq.~(\ref{eq:h-H-k-q}) to obtain for $\ell \in\{1,\ldots,L\}$, $k \in\{1,\ldots,g\}$, $i \in \mathcal{G}_k$ and $q  \in\{1,\ldots,Q_i\}$ that 
\begin{align}
 & \bm{h}^H_{i,q}(\bm{\Phi}_1,\ldots,\bm{\Phi}_L) \bm{w}_k  = b_{i,q}(\bm{w}_{k}) +  \sum_{\ell=1}^L \bm{\phi}_{\ell}^H \bm{a}_{\ell,i,q}(\bm{w}_{k}) , \label{eq-hi-def-ai-wk-def-b-wk-prob-P1-MA-MR} 
\end{align}
where we define $\bm{a}_{\ell,i,q}(\bm{w}_{k}) \in \mathbb{C}^{N_{\ell} \times 1}$ by
\begin{align}
\bm{a}_{\ell,i,q}(\bm{w}_{k}):=\text{diag}(\bm{h}^H_{\ell,i,q})\bm{H}_{\text{b},\ell}\bm{w}_{k} .\label{eq-def-ai-wk-P1-MA-MR} 
\end{align} 
and complex numbers $b_{i,q}(\bm{w}_{k})$ by 
\begin{align}
b_{i,q}(\bm{w}_{k}):=\bm{h}^H_{\text{b},i,q}\bm{w}_k. \label{eq-def-b-wk-P1-MA-MR} 
\end{align}

Applying Eq.~(\ref{eq-hi-def-ai-wk-def-b-wk-prob-P1-MA-MR}) to Eq.~(\ref{eq:wk-Hi-Phi1-Phi2-PhiL}), we have
\begin{widetext} 
\begin{align}
 & \bm{w}_k^H\bm{H}_{i}(\bm{\Phi}_1,\ldots,\bm{\Phi}_L)\bm{w}_k \nonumber 
 \\ & = \sum_{q=1}^{Q_i}  \left(b_{i,q}^H(\bm{w}_{k})   + \sum_{\ell=1}^L \bm{a}_{\ell,i,q}^H(\bm{w}_{k})\bm{\phi}_{\ell} \right)  \left(b_{i,q}(\bm{w}_{k})  +  \sum_{\ell=1}^L \bm{\phi}_{\ell}^H\bm{a}_{\ell,i,q}(\bm{w}_{k})  \right)\nonumber  \\ &= \sum_{q=1}^{Q_i} \left\{  \begin{bmatrix} 
\bm{\phi}_1^H ,~\ldots,~\bm{\phi}_L^H ,~
1
\end{bmatrix} \bm{A}_{i,q}(\bm{w}_{k}) \begin{bmatrix} 
\bm{\phi}_1 \\
\ldots \\
\bm{\phi}_L \\
1
\end{bmatrix} +   {b}_{i,q}(\bm{w}_{k}) {b}_{i,q}^H(\bm{w}_{k})\right\} , \label{eq-hi-def-ai-wj-def-b-wj-L}
\end{align}
where we define a matrix $\bm{A}_{i,q}(\bm{w}_{k}) \in \mathbb{C}^{(1 +\sum_{\ell=1}^L N_{\ell}) \times (1 +\sum_{\ell=1}^L N_{\ell})}$ as follows:
\begin{align}
 & \bm{A}_{i,q}(\bm{w}_{k}) := \begin{bmatrix} 
\bm{a}_{1,i,q}(\bm{w}_{k})\bm{a}_{1,i,q}^H(\bm{w}_{k}),& \bm{a}_{1,i,q}(\bm{w}_{k})\bm{a}_{2,i,q}^H(\bm{w}_{k}), &\ldots,& \bm{a}_{1,i,q}(\bm{w}_{k})\bm{a}_{L,i,q}^H(\bm{w}_{k}), & \bm{a}_{1,i,q}(\bm{w}_{k}){b}_{i,q}^H(\bm{w}_{k}) \\ \bm{a}_{2,i,q}(\bm{w}_{k})\bm{a}_{1,i,q}^H(\bm{w}_{k}),& \bm{a}_{2,i,q}(\bm{w}_{k})\bm{a}_{2,i,q}^H(\bm{w}_{k}), &\ldots,& \bm{a}_{2,i,q}(\bm{w}_{k})\bm{a}_{L,i,q}^H(\bm{w}_{k}), & \bm{a}_{2,i,q}(\bm{w}_{k}){b}_{i,q}^H(\bm{w}_{k}) \\ \ldots,& \ldots,& \ldots,& \ldots,& \ldots \\ \bm{a}_{\ell,i,q}(\bm{w}_{k})\bm{a}_{1,i,q}^H(\bm{w}_{k}), &\bm{a}_{\ell,i,q}(\bm{w}_{k})\bm{a}_{2,i,q}^H(\bm{w}_{k}),& \ldots,& \bm{a}_{\ell,i,q}(\bm{w}_{k})\bm{a}_{L,i,q}^H(\bm{w}_{k}), & \bm{a}_{\ell,i,q}(\bm{w}_{k}){b}_{i,q}^H(\bm{w}_{k}) \\ \ldots,& \ldots,& \ldots,& \ldots,& \ldots \\ \bm{a}_{L,i,q}(\bm{w}_{k})\bm{a}_{1,i,q}^H(\bm{w}_{k}),& \bm{a}_{L,i,q}(\bm{w}_{k})\bm{a}_{2,i,q}^H(\bm{w}_{k}),&\ldots,& \bm{a}_{L,i,q}(\bm{w}_{k})\bm{a}_{L,i,q}^H(\bm{w}_{k}), & \bm{a}_{L,i,q}(\bm{w}_{k}){b}_{i,q}^H(\bm{w}_{k}) \\
{b}_{i,q}(\bm{w}_{k})\bm{a}_{1,i,q}^H(\bm{w}_{k}),& {b}_{i,q}(\bm{w}_{k})\bm{a}_{2,i,q}^H(\bm{w}_{k}),& \ldots,& {b}_{i,q}(\bm{w}_{k})\bm{a}_{L,i,q}^H(\bm{w}_{k}),&   0 
\end{bmatrix}. \label{eq-def-Ai-matrix-wk-q} 
\end{align}
\end{widetext}
Replacing $k$ by $j$ in Eq.~(\ref{eq-def-Ai-matrix-wk-q}), we also define $\bm{A}_{i,q}(\bm{w}_{j}) $ and use it to compute $\bm{w}_j^H\bm{H}_{i}(\bm{\Phi}_1,\ldots,\bm{\Phi}_L)\bm{w}_j$. From this and Eq.~(\ref{eq-hi-def-ai-wj-def-b-wj-L}), after defining 
\begin{align}
\bm{\phi}: = \begin{bmatrix} 
\bm{\phi}_1 \\
\ldots \\
\bm{\phi}_L  
\end{bmatrix}, \label{eq-boldsymbol-all-phi}
\end{align} 
the constraint of Inequality~(\ref{eq:prob-P1-MA-MR-b-SINR-constraint}) in Problem~(P1-MA-b) becomes
\begin{align}
 & \sum_{q=1}^{Q_i} \left\{  \begin{bmatrix} 
\bm{\phi}^H,
1
\end{bmatrix} \bm{A}_{i,q}(\bm{w}_{k}) \begin{bmatrix} 
\bm{\phi} \\
1
\end{bmatrix} +   {b}_{i,q}(\bm{w}_{k}) {b}_{i,q}^H(\bm{w}_{k})\right\} \nonumber  \\ & \geq \hspace{-2pt} \gamma_i \hspace{-2pt}\left\{\hspace{-2pt}  \sigma^2_i\hspace{-2pt}+\hspace{-10pt} \sum\limits_{j\in\{1,\ldots,g\}\setminus\{k\}} \hspace{-1pt} \sum_{q=1}^{Q_i} \hspace{-2pt}\left\{\hspace{-6pt}  \begin{array}{l} \begin{bmatrix} 
\bm{\phi}^H,
1
\end{bmatrix} \bm{A}_{i,q}(\bm{w}_{j}) \begin{bmatrix} 
\bm{\phi} \\
1
\end{bmatrix} \\ +   {b}_{i,q}(\bm{w}_{j}) {b}_{i,q}^H(\bm{w}_{j})\end{array}\hspace{-6pt}\right\}\hspace{-3pt}\right\}\hspace{-2pt}.\label{eq-hi-def-ai-wj-def-b-wj-square-v3-L}
\end{align}

The analysis for Inequality~(\ref{eq-hi-def-ai-wj-def-b-wj-square-v3-L}) is similar to that for Inequality~(\ref{eq-hi-def-ai-wj-def-b-wj-square-v3}) in Section~\ref{subsection-P1-P3}. Specifically, 
We introduce an auxiliary  variable~$t$, which is a complex number satisfying $|t|=1$, and define 
\begin{align}
 & \bm{v}:= t \begin{bmatrix} 
\bm{\phi}  \\
1
\end{bmatrix}=  \begin{bmatrix} 
t \bm{\phi}  \\
t
\end{bmatrix}=  \begin{bmatrix} 
t\bm{\phi}_1 \\
\ldots \\
t\bm{\phi}_L \\
t
\end{bmatrix}. \label{eq-boldsymbol-v-K}
\end{align}
Then with $|t|=1$, Inequality~(\ref{eq-hi-def-ai-wj-def-b-wj-square-v3-L}) is equivalent to 
\begin{align}
 & \sum_{q=1}^{Q_i} \left\{ \bm{v}^H \bm{A}_{i,q}(\bm{w}_{k}) \bm{v} +   {b}_{i,q}(\bm{w}_{k}) {b}_{i,q}^H(\bm{w}_{k})\right\} \nonumber  \\ & \geq \hspace{-2pt} \gamma_i \hspace{-2pt}\left\{\hspace{-2pt}  \sigma^2_i\hspace{-2pt}+\hspace{-2pt} \sum\limits_{j\in\{1,\ldots,g\}\setminus\{k\}} \, \sum_{q=1}^{Q_i} \hspace{-1pt}\left\{\hspace{-3pt}  \begin{array}{l} \bm{v}^H \bm{A}_{i,q}(\bm{w}_{j}) \bm{v} \\ +   {b}_{i,q}(\bm{w}_{j}) {b}_{i,q}^H(\bm{w}_{j})\end{array}\hspace{-3pt}\right\}\hspace{-3pt}\right\}\hspace{-2pt}.\label{eq-hi-def-ai-wj-def-b-wj-square-v5-K}
\end{align}

We further define 
\begin{align}
\boldsymbol{V}: = \boldsymbol{v}  \boldsymbol{v}^H, \label{eq-boldsymbol-V-v-K}
\end{align}
Clearly, $\boldsymbol{V}$ is semi-definite and has rank one. For $\boldPhi_{\ell}$ in the form of Eq.~(\ref{eq-define-bm-phi-ell-0-L}), the diagonal elements of $\boldsymbol{V}$ are as follows:
\begin{align}
 &\forall x\in \{1,\ldots, L\}, ~ \forall y\in \{1,\ldots, N_{x}\}: \nonumber \\
 &\boldsymbol{V}_{(y+\sum_{\ell=0}^{x-1}N_{\ell}),\,(y+\sum_{\ell=0}^{x-1}N_{\ell})} \nonumber \\
 &=(\beta_{x,y} e^{j \theta_{x,y}})^H  \cdot \beta_{x,y} e^{j \theta_{x,y}} = {\beta_{x,y}}^2, \label{eq:V-x-y-element}
\end{align}
and
\begin{align}
 &\boldsymbol{V}_{(1 +\sum_{\ell=1}^L N_{\ell}),\,(1 +\sum_{\ell=1}^L N_{\ell})}=t  \cdot  t^H = 1. \label{eq:V-last-element}
\end{align}


Using Eq.~(\ref{eq-boldsymbol-V-v-K}), we write Eq.~(\ref{eq-hi-def-ai-wj-def-b-wj-square-v5-K}) as
\begin{align}
 & \sum_{q=1}^{Q_i} \left\{ \tr(\bm{A}_{i,q}(\bm{w}_{k})\boldsymbol{V} ) +   {b}_{i,q}(\bm{w}_{k}) {b}_{i,q}^H(\bm{w}_{k})\right\} \nonumber  \\ & \geq \hspace{-2pt} \gamma_i \hspace{-2pt}\left\{\hspace{-2pt}  \sigma^2_i\hspace{-2pt}+\hspace{-2pt} \sum\limits_{j\in\{1,\ldots,g\}\setminus\{k\}} \, \sum_{q=1}^{Q_i} \hspace{-1pt}\left\{\hspace{-3pt}  \begin{array}{l} \tr(\bm{A}_{i,q}(\bm{w}_{j})\boldsymbol{V} ) \\ +   {b}_{i,q}(\bm{w}_{j}) {b}_{i,q}^H(\bm{w}_{j})\end{array}\hspace{-3pt}\right\}\hspace{-3pt}\right\}\hspace{-2pt}.\label{eq-hi-def-ai-wj-def-b-wj-square-v5-K-L}
\end{align}

From the above discussion,  we convert Problem~(P1-MA-MR-b) of Eq.~(\ref{eq-Prob-P1-MA-MR-b}) into
\begin{subequations} \label{eq-Prob-P1-MA-MR-c}
\begin{alignat}{2}
&\hspace{-40pt}\text{(P1-MA-MR-c)}: \nonumber \\
\Find~~~& \boldsymbol{V} \\
\mathrm{s.t.}~~~~~&   \sum_{q=1}^{Q_i} \left\{ \tr(\bm{A}_{i,q}(\bm{w}_{k})\boldsymbol{V} ) +   {b}_{i,q}(\bm{w}_{k}) {b}_{i,q}^H(\bm{w}_{k})\right\} \nonumber  \\ & \geq  \gamma_i \bigg\{  \sigma^2_i + \nonumber  \\ & \hspace{-25pt}\sum\limits_{j\in\{1,\ldots,g\}\setminus\{k\}} \, \sum_{q=1}^{Q_i} \hspace{-1pt}\left\{ \tr(\bm{A}_{i,q}(\bm{w}_{j})\boldsymbol{V} )  +   {b}_{i,q}(\bm{w}_{j}) {b}_{i,q}^H(\bm{w}_{j})\right\}\hspace{-2pt}\bigg\},  \label{eq:prob-P1-MA-MR-c-SINR-constraint}\\
&~~~~~~~\forall k \in\{1,\ldots,g\},~\forall i \in \mathcal{G}_k, \nonumber \\
 &\boldsymbol{V}_{(y+\sum_{\ell=0}^{x-1}N_{\ell}),\,(y+\sum_{\ell=0}^{x-1}N_{\ell})}  = {\beta_{x,y}}^2, \label{eq-Prob-P1-MA-MR-c-constraints-Vnn} \\
&~~~~~~~\forall x\in \{1,\ldots, L\},~\forall y\in \{1,\ldots, N_{x}\}, \nonumber \\
 &\boldsymbol{V}_{(1 +\sum_{\ell=1}^L N_{\ell}),\,(1 +\sum_{\ell=1}^L N_{\ell})} = 1, \label{eq-Prob-P1-MA-MR-c-constraints-Vnplus1}  \\
& \boldsymbol{V} \succeq 0, \\
& \text{rank}(\boldsymbol{V}) =1. \label{eq:prob-P1-MA-MR-c-rank-constraint}
\end{alignat}
\end{subequations}

The above constraint in Eq.~(\ref{eq-Prob-P1-MA-MR-c-constraints-Vnn}) is for the case where all $\beta_{x,y}|_{\begin{subarray}{l}x\in \{1,\ldots, L\},\\ y\in \{1,\ldots, N_{x}\} \end{subarray}}$ are predefined constants. A special case of particular interest is the case of all $\beta_{x,y}|_{\begin{subarray}{l}x\in \{1,\ldots, L\},\\ y\in \{1,\ldots, N_{x}\} \end{subarray}}$ being $1$ so that Eq.~(\ref{eq-Prob-P1-MA-MR-c-constraints-Vnn}) and Eq.~(\ref{eq-Prob-P1-MA-MR-c-constraints-Vnplus1}) can together be written as $\boldsymbol{V}_{n,n}  = 1  $ for $n\in \{1,\ldots, 1 +\sum_{\ell=1}^L N_{\ell}\}$.
If each $\beta_{x,y}|_{\begin{subarray}{l}x\in \{1,\ldots, L\},\\ y\in \{1,\ldots, N_{x}\} \end{subarray}}$ can take any value in $[0,1]$, then Eq.~(\ref{eq-Prob-P1-MA-MR-c-constraints-Vnn}) can be replaced by $\boldsymbol{V}_{(y+\sum_{\ell=0}^{x-1}N_{\ell}),\,(y+\sum_{\ell=0}^{x-1}N_{\ell})}\in [0,1] $ for $x\in \{1,\ldots, L\}$ and $y\in \{1,\ldots, N_{x}\}$. Similarly, we may consider the most general case where  some $\beta_{x,y}$ are predefined constants while other $\beta_{x,y}$ can vary.

The only non-convex part in Problem~(P1-MA-MR-c) is the rank constraint in Eq.~(\ref{eq:prob-P1-MA-MR-c-rank-constraint}). Hence, we adopt semidefinite relaxation (SDR) and drop Eq.~(\ref{eq:prob-P1-MA-MR-c-rank-constraint}) to obtain the following semidefinite programming Problem~(P1-MA-MR-d): 
\begin{subequations} \label{eq-Prob-P1-MA-MR-d}
\begin{alignat}{2}
&\hspace{-30pt}\text{(P1-MA-MR-d)}: \nonumber \\
\Find~~~& \boldsymbol{V} \\
\mathrm{s.t.}~~~~~&   \sum_{q=1}^{Q_i} \left\{ \tr(\bm{A}_{i,q}(\bm{w}_{k})\boldsymbol{V} ) +   {b}_{i,q}(\bm{w}_{k}) {b}_{i,q}^H(\bm{w}_{k})\right\} \nonumber  \\ & \geq  \gamma_i \bigg\{  \sigma^2_i + \nonumber  \\ & \hspace{-25pt}\sum\limits_{j\in\{1,\ldots,g\}\setminus\{k\}} \, \sum_{q=1}^{Q_i} \hspace{-1pt}\left\{ \tr(\bm{A}_{i,q}(\bm{w}_{j})\boldsymbol{V} )  +   {b}_{i,q}(\bm{w}_{j}) {b}_{i,q}^H(\bm{w}_{j})\right\}\hspace{-2pt}\bigg\},  \label{eq:prob-P1-MA-MR-d-SINR-constraint}\\
&~~~~~~~\forall k \in\{1,\ldots,g\},~\forall i \in \mathcal{G}_k, \nonumber \\
  &\boldsymbol{V}_{(y+\sum_{\ell=0}^{x-1}N_{\ell}),\,(y+\sum_{\ell=0}^{x-1}N_{\ell})}  = {\beta_{x,y}}^2, \label{eq-Prob-P1-MA-MR-d-constraints-Vnn} \\
&~~~~~~~\forall x\in \{1,\ldots, L\},~\forall y\in \{1,\ldots, N_{x}\}, \nonumber \\
 &\boldsymbol{V}_{(1 +\sum_{\ell=1}^L N_{\ell}),\,(1 +\sum_{\ell=1}^L N_{\ell})} = 1, \label{eq-Prob-P1-MA-MR-d-constraints-Vnplus1}  \\
& \boldsymbol{V} \succeq 0.
\end{alignat}
\end{subequations}

Problem~(P1-MA-MR-d) belongs to semidefinite programming and can be solved efficiently~\cite{luo2010semidefinite}. Moreover, Problem~(P1-MA-MR-d) can be replaced by Problem~(P1-MA-MR-d$'$) below which may find better $\boldPhi$ and hence $\boldsymbol{V}$ to accelerate the alternating optimization process:
\begin{subequations} \label{eq-Prob-P1-MA-MR-d-prime}
\begin{alignat}{2}
&\hspace{-40pt}\text{(P1-MA-MR-d$'$)}: \nonumber \\ \max_{\boldsymbol{V},\boldsymbol{\alpha}}& \sum_{k=1}^g \sum_{i \in \mathcal{G}_k} \alpha_i \\
\mathrm{s.t.}~~~~~&   \sum_{q=1}^{Q_i} \left\{ \tr(\bm{A}_{i,q}(\bm{w}_{k})\boldsymbol{V} ) +   {b}_{i,q}(\bm{w}_{k}) {b}_{i,q}^H(\bm{w}_{k})\right\} \nonumber  \\ & \geq  \alpha_i + \gamma_i \bigg\{  \sigma^2_i + \nonumber  \\ & \hspace{-25pt}\sum\limits_{j\in\{1,\ldots,g\}\setminus\{k\}} \, \sum_{q=1}^{Q_i} \hspace{-1pt}\left\{ \tr(\bm{A}_{i,q}(\bm{w}_{j})\boldsymbol{V} )  +   {b}_{i,q}(\bm{w}_{j}) {b}_{i,q}^H(\bm{w}_{j})\right\}\hspace{-2pt}\bigg\},  \label{eq:prob-P1-MA-MR-d-prime-SINR-constraint}\\
&~~~~~~~\forall k \in\{1,\ldots,g\},~\forall i \in \mathcal{G}_k, \nonumber  \\
 &\boldsymbol{V}_{(y+\sum_{\ell=0}^{x-1}N_{\ell}),\,(y+\sum_{\ell=0}^{x-1}N_{\ell})}  = {\beta_{x,y}}^2, \label{eq-Prob-P1-MA-MR-d-prime-constraints-Vnn} \\
&~~~~~~~\forall x\in \{1,\ldots, L\},~\forall y\in \{1,\ldots, N_{x}\}, \nonumber \\
 &\boldsymbol{V}_{(1 +\sum_{\ell=1}^L N_{\ell}),\,(1 +\sum_{\ell=1}^L N_{\ell})} = 1, \label{eq-Prob-P1-MA-MR-d-prime-constraints-Vnplus1}  \\
& \boldsymbol{V} \succeq 0, \\
& \alpha_i \geq 0, ~\forall k \in\{1,\ldots,g\},~\forall i \in \mathcal{G}_k.
\end{alignat}
\end{subequations}

After  a candidate solution $\boldsymbol{V}$ is obtained by solving Problem~(P1-MA-MR-d) or Problem~(P1-MA-MR-d$'$), appropriate post-processing such as Gaussian randomization~\cite{so2007approximating} can be applied to convert the candidate solution into a solution which satisfies the rank constraint in Eq.~(\ref{eq:prob-P1c-rank-constraint}).

Combining the above discussion of alternatively optimizing $\W$ and $\boldPhi$, we present our method to solve Problem~(P1-MA-MR) as Algorithm~\ref{Alg-Problem-P1-MA-MR}. 

 \begin{algorithm*}
\caption{via alternating optimization to find $\bm{\Phi}_1,\ldots,\bm{\Phi}_L$ and $\W$ for Problem~(P1-MA-MR), which generalizes Problems~(P2-MA-MR)~(P3-MA-MR), \mbox{(P1-MA)--(P3-MA),} \mbox{(P1-MR)--(P3-MR),} and \mbox{(P1)--(P3).}  } \label{Alg-Problem-P1-MA-MR}
\begin{algorithmic}[1]
\STATE Initialize  $\boldsymbol{\Phi}_{\ell}$ for $\ell \in\{1,\ldots,L\}$ as some initial (e.g., randomly generated) $ \boldsymbol{\Phi}_{\ell}^{(0)}:= \text{diag}(\beta_{\ell,1}  e^{j \theta_{\ell,1}^{(0)}}, \ldots, \beta_{\ell,N_{\ell}} e^{j \theta_{\ell,N_{\ell}}^{(0)}})$ which satisfies the constraints on $\boldsymbol{\Phi}_{\ell}$;
\STATE Set the iteration number $r \leftarrow 1$;
\WHILE{1}
\item[] \COMMENT{The ``while'' loop will end if Line~\ref{Alg-Problem-P1-MA-MR-break1} or \ref{Alg-Problem-P1-MA-MR-break2} is executed.}
\item[] \COMMENT{Comment: Optimizing $\W$ given $\bm{\Phi}_1,\ldots,\bm{\Phi}_L$:}
\STATE Given $\bm{\Phi}_1,\ldots,\bm{\Phi}_L$ as $\bm{\Phi}_1^{(r-1)},\ldots,\bm{\Phi}_L^{(r-1)}$, use methods of Karipidis~\textit{et~al.}~\cite{karipidis2008quality} or other papers to solve Problem~(P1-MA-MR-a) in Eq.~(\ref{eq-Prob-P1-MA-MR-a}), and post-process the obtained $\{\bm{X}_k\}|_{k=1}^g$ to set $\boldsymbol{W}$ as some $\boldsymbol{W}^{(r)}:=[\bm{w}_1^{(r)},\ldots,\bm{w}_g^{(r)}]$; \label{Alg-opt-P1-MA-MR-0}
\item[] \vspace{-5pt}
\STATE \mbox{Compute the object function value $f^{(r)} \leftarrow \sum_{k=1}^{g}\|\bm{w}_k^{(r)}\|^2$};
\IF{$r\geq 2$}
\IF{$1-\frac{f^{(r)}}{f^{(r-1)}}$ denoting the relative difference  between the object function values in consecutive iterations $r-1$ and $r$ is small}
\STATE \textbf{break};  \label{Alg-Problem-P1-MA-MR-break1}
\ENDIF
\ENDIF
\item[] \vspace{-5pt}
\item[] \COMMENT{Finding $\bm{\Phi}_1,\ldots,\bm{\Phi}_L$ given $\W$:}
\STATE Given $\W$ as $\boldsymbol{W}^{(r)}$, solve Problem~(P1-MA-MR-d) in Eq.~(\ref{eq-Prob-P1-MA-MR-d}) or Problem~(P1-MA-MR-d$'$) in Eq.~(\ref{eq-Prob-P1-MA-MR-d-prime}), and denote the obtained~$\boldsymbol{V}$ as~$\boldsymbol{V}^{(r)}_{\text{SDR}}$;
\item[] \COMMENT{Comment: Gaussian randomization:}
\STATE Perform the eigenvalue decomposition on $\boldsymbol{V}^{(r)}_{\text{SDR}}$ to obtain a unitary matrix $\boldsymbol{U}$ and a  diagonal  matrix $\boldsymbol{\Lambda}$ such that $ \boldsymbol{V}^{(r)}_{\text{SDR}} = \boldsymbol{U} \boldsymbol{\Lambda}  \boldsymbol{U}^H $;\label{Alg-opt-Gaussian-randomization-decomposition}
\FOR{$z$ from $1$ to some sufficiently large $Z$}
\STATE Generate a random vector $\boldsymbol{r}^{(r)}_{z}$ from a circularly-symmetric complex Gaussian distribution $\mathcal{CN}(\boldsymbol{0},\boldsymbol{I}_{1 +\sum_{\ell=1}^L N_{\ell}})$ with zero mean and covariance matrix $\boldsymbol{I}_{1 +\sum_{\ell=1}^L N_{\ell}}$ (the identity matrix with size $1 +\sum_{\ell=1}^L N_{\ell}$); \STATE Compute $ \boldsymbol{v}^{(r)}_{z}  \leftarrow \boldsymbol{U} \boldsymbol{\Lambda}^{\frac{1}{2}} \boldsymbol{r}^{(r)}_{z}$;
\STATE With $v_{1 +\sum_{\ell=1}^L N_{\ell}}$ being the $(1 +\sum_{\ell=1}^L N_{\ell})$th element of $\boldsymbol{v}_{z}^{(r)}$, {take the first $\sum_{\ell=1}^L N_{\ell}$ elements of $\frac{\boldsymbol{v}_{z}^{(r)}}{v_{1 +\sum_{\ell=1}^L N_{\ell}}}$ to form a vector $\boldsymbol{w}^{(r)}_{z}$};
\STATE 
Scale each component of $\boldsymbol{w}^{(r)}_{z}$ independently to obtain $\bm{\phi}^{(r)}_{z}$ such that the $(y+\sum_{\ell=0}^{x-1}N_{\ell})$th element of $\bm{\phi}^{(r)}_{z}$ has magnitude ${\beta_{x,y}}$, for $x\in \{1,\ldots, L\}$ and $y\in \{1,\ldots, N_{x}\}$;
\ENDFOR
\IF{for $z\in\{1,2,\ldots,Z\}$, there is no $\bm{\phi}^{(r)}_{z}$ to ensure Eq.~(\ref{eq-hi-def-ai-wj-def-b-wj-square-v3-L}) after setting $\bm{\phi}$ as $\bm{\phi}^{(r)}_{z}$}
\STATE \textbf{break};  \label{Alg-Problem-P1-MA-MR-break2}
\ELSE
\STATE Select one $\bm{\phi}^{(r)}_{z}$ according to some ordering among those ensuring Eq.~(\ref{eq-hi-def-ai-wj-def-b-wj-square-v3-L})   and denote the selected one by $\bm{\phi}^{(r)}_{z^*}$;
\STATE Map $\bm{\phi}^{(r)}_{z^*}$ to some ${\bm{\phi}^{(r)}}$ to satisfy the constraint on $\bm{\phi}$ (e.g., discrete element values); 
\STATE For $\ell\in\{1,2,\ldots,L\}$, set $\bm{\Phi}_{\ell}$ as  $\bm{\Phi}_{\ell}^{(r)}:=\text{diag}\left((\bm{\phi}_{\ell}^{(r)})^H\right)$, where $\bm{\phi}_{\ell}^{(r)}|_{\ell\in\{1,2,\ldots,L\}}$ are defined such that $\bm{\phi}^{(r)} = \begin{bmatrix}
\bm{\phi}_1^{(r)} \\
\ldots \\
\bm{\phi}_L^{(r)}
\end{bmatrix}$;
\ENDIF
\STATE Update the iteration number $r  \leftarrow r + 1$;  \label{Alg-opt-update-r}
\ENDWHILE
\end{algorithmic}
\end{algorithm*}

\subsection{Solutions to Problems~(P4-MA)--(P6-MA),\\ \mbox{(P4-MR)--(P6-MR)}, and~(P4-MA-MR)--(P6-MA-MR)} \label{subsection-P4-MA-MR}

As Problems~(P4-MA)--(P6-MA),~(P4-MR)--(P6-MR), and~(P4-MA-MR)--(P6-MA-MR) are generalizations of Problems~(P4)--(P6), we will solve the former problems in ways similar to the solutions for the latter problems, which we have discussed in Section~\ref{subsection-P4-P6}. More specifically, since Problem~(P4-MA-MR) is in the most general form, we start with elaborating its solution below.

We restate Problem~(P4-MA-MR) given in Table~\ref{table:4MA-MR} of   Page~\pageref{table:4MA-MR}:
\begin{subequations} \label{eq-Prob-P4-MA-MR-restated}
\begin{alignat}{2}
&\text{(P4-MA-MR)}: \nonumber \\
&\max_{\W, \bm{\Phi}_1,\ldots,\bm{\Phi}_L}~\min_{k\in\{1,\ldots,g\}}  \min_{i \in \mathcal{G}_k} \nonumber  \\ 
 &~~~~~~\frac{\bm{w}_k^H\bm{H}_{i}(\bm{\Phi}_1,\ldots,\bm{\Phi}_L)\bm{w}_k}{\gamma_i\left[\sum\limits_{j\in\{1,\ldots,g\}\setminus\{k\}}\bm{w}_j^H\bm{H}_{i}(\bm{\Phi}_1,\ldots,\bm{\Phi}_L)\bm{w}_j +  \sigma^2_i\right]}   \label{OptProb-power-multicast-obj-P4-MA-MR}  \\
&~\mathrm{s.t.}~\sum\limits_{k=1}^{g}\|\bm{w}_k\|^2 \leq P, \label{OptProb-power-multicast-SINR-constraint-P4-MA-MR}   \\ 
 &~~~~~~\text{Constraints on $\bm{\Phi}_1,\ldots,\bm{\Phi}_L$}. \label{OptProb-power-multicast-Phi-constraint-P4-MA-MR}
\end{alignat}
\end{subequations}

We introduce an auxiliary variable $t$ and convert Problem~(P4) into the following equivalent \mbox{Problem~(P4-MA-MR-a):} 
\begin{subequations} \label{eq-Prob-P4-MA-MR-a}
\begin{alignat}{2}
 &\hspace{-50pt}\text{(P4-MA-MR-a):}\nonumber \\
 \max_{\W, \bm{\Phi}_1,\ldots,\bm{\Phi}_L, t}~&t   \label{OptProb-P4-MA-MR-a-fair-SINR-multicast-obj}   \\
&\hspace{-50pt}\mathrm{s.t.}~ \frac{\bm{w}_k^H\bm{H}_{i}(\bm{\Phi}_1,\ldots,\bm{\Phi}_L)\bm{w}_k}{\gamma_i\left[\sum\limits_{j\in\{1,\ldots,g\}\setminus\{k\}}\bm{w}_j^H\bm{H}_{i}(\bm{\Phi}_1,\ldots,\bm{\Phi}_L)\bm{w}_j +  \sigma^2_i\right]}  \geq t, \label{OptProb-P4-MA-MR-a-fair-SINR-multicast-obj-SINR-constraint} \\
&~~~~~~~\forall k \in\{1,\ldots,g\},~\forall i \in \mathcal{G}_k, \nonumber \\
 & \sum\limits_{k=1}^{g}\|\bm{w}_k\|^2 \leq P , \label{OptProb-P4-MA-MR-a-fair-SINR-multicast-obj-power-constraint}   \\
 &\text{Constraints on $\bm{\Phi}_1,\ldots,\bm{\Phi}_L$},\label{OptProb-P4-MA-MR-a-fair-SINR-multicast-obj-Phi-constraint}  \\
 & t \geq 0. \label{OptProb-P4-MA-MR-a-fair-SINR-multicast-obj-t-constraint}
 \end{alignat}
\end{subequations}

We use the idea of alternating optimization, which is widely used in multivariate optimization. Specifically, we perform the following optimizations alternatively to solve Problem~(P4-MA-MR-a): optimizing $\W$ and $t$ given $\boldPhi$, and finding $\boldPhi$ given $\W$ and $t$.  The details are presented below.

\textbf{Optimizing $\W$ and $t$ given $\bm{\Phi}_1,\ldots,\bm{\Phi}_L$.} Given some $\bm{\Phi}_1,\ldots,\bm{\Phi}_L$ satisfying the constraints in Eq.~(\ref{OptProb-P4-MA-MR-a-fair-SINR-multicast-obj-Phi-constraint}), Problem~(P4-MA-MR) means finding $\W$ and $t$ given $\bm{\Phi}_1,\ldots,\bm{\Phi}_L$ to maximize $t$ subject to the constraints in~(\ref{OptProb-P4-MA-MR-a-fair-SINR-multicast-obj-SINR-constraint})~(\ref{OptProb-P4-MA-MR-a-fair-SINR-multicast-obj-power-constraint})~(\ref{OptProb-P4-MA-MR-a-fair-SINR-multicast-obj-t-constraint}).

We define $\bm{X}_k$ according to Eq.~(\ref{eq-Xk-wk}); i.e.,
\begin{align}
\bm{X}_k := \bm{w}_k \bm{w}_k^H,~\forall k \in\{1,\ldots,g\}. \label{eq-Xk-wk-v2-P4-MA-MR-a}  
\end{align}
 Then similar to the process of writing Inequality~(\ref{OptProb-power-multicast-SINR-constraint}) of Problem~(P1) as Inequality~(\ref{eq:prob-P1-0-}) of Problem~(P1a), we write Inequality~(\ref{OptProb-P4-MA-MR-a-fair-SINR-multicast-obj-SINR-constraint}) of Problem~(P4-MA-MR-a) as Inequality~(\ref{OptProb-P4-MA-MR-b-fair-SINR-multicast-obj-SINR-constraint}) below. Then optimizing $\W$ given $\bm{\Phi}_1,\ldots,\bm{\Phi}_L$ for Problem~(P4-MA-MR-a) becomes solving $\{\bm{X}_k\}|_{k=1}^g$ for the following Problem~(P4-MA-MR-b):
\begin{subequations} \label{eq-Prob-P4-MA-MR-b}
\begin{alignat}{2}
 &\hspace{-50pt}\text{(P4-MA-MR-b):}\nonumber \\
 \max_{\W, \bm{\Phi}_1,\ldots,\bm{\Phi}_L, t}~&t   \label{OptProb-P4-MA-MR-b-fair-SINR-multicast-obj}   \\
&\hspace{-50pt}\mathrm{s.t.}~ \tr(\bm{X}_k \bm{H}_{i}(\bm{\Phi}_1,\ldots,\bm{\Phi}_L)) \geq \nonumber \\
& \hspace{-25pt}t\gamma_i \sigma^2_i+  t\gamma_i \sum_{j\in\{1,\ldots,g\}\setminus\{k\}}\tr(\bm{X}_j \bm{H}_{i}(\bm{\Phi}_1,\ldots,\bm{\Phi}_L) ) , \label{OptProb-P4-MA-MR-b-fair-SINR-multicast-obj-SINR-constraint} \\
&~~~~~~~\forall k \in\{1,\ldots,g\},~\forall i \in \mathcal{G}_k, \nonumber \\
 &  \sum\limits_{k=1}^{g}\tr(\bm{X}_k)\leq P ,   \\
& \bm{X}_k \succeq 0,~\forall k \in\{1,\ldots,g\},  \\
& \text{rank}(\bm{X}_k) =1,~\forall k \in\{1,\ldots,g\}, \label{eq:prob-P4-MA-MR-b-rank-constraint}  \\
 & t \geq 0. \label{OptProb-P4-MA-MR-b-fair-SINR-multicast-obj-t-constraint}
 \end{alignat}
\end{subequations}


A problem similar to Problem~(P4-MA-MR-b) has been used to solved by  Karipidis~\textit{et~al.}~\cite{karipidis2008quality}, where the notation of Problem~$\mathcal{F}_r$ is used to denote the problem after dropping Eq.~(\ref{eq:prob-P4-MA-MR-b-rank-constraint}). Hence, we can apply methods of~\cite{karipidis2008quality} to solve Problem~(P4-MA-MR-b).

\textbf{Finding $\bm{\Phi}_1,\ldots,\bm{\Phi}_L$ given $\W$ and $t$.} Given $\W$ and $t$, Problem~(P4-MA-MR) becomes the following feasibility check problem of finding $\bm{\Phi}_1,\ldots,\bm{\Phi}_L$:
\begin{subequations} \label{eq-Prob-P4-MA-MR-c}
\begin{alignat}{2}
 &\hspace{-30pt}\text{(P4-MA-MR-c):}\nonumber \\
\Find~~~& \bm{\Phi}_1,\ldots,\bm{\Phi}_L \\
&\hspace{-30pt}\mathrm{s.t.}~ \frac{\bm{w}_k^H\bm{H}_{i}(\bm{\Phi}_1,\ldots,\bm{\Phi}_L)\bm{w}_k}{\sum\limits_{j\in\{1,\ldots,g\}\setminus\{k\}}\bm{w}_j^H\bm{H}_{i}(\bm{\Phi}_1,\ldots,\bm{\Phi}_L)\bm{w}_j +  \sigma^2_i}  \geq t \gamma_i,  \label{eq:prob-P4-MA-MR-c-SINR-constraint}\\
&~~~~~~~\forall k \in\{1,\ldots,g\},~\forall i \in \mathcal{G}_k, \nonumber  \\
 &\text{Constraints on $\bm{\Phi}_1,\ldots,\bm{\Phi}_L$}. \label{eq:prob-P4-MA-MR-c-rank-constraint}
\end{alignat}
\end{subequations}

The only difference between Problem~(P4-MA-MR-c) of Eq.~(\ref{eq-Prob-P4-MA-MR-c}) and Problem~(P1-MA-MR-b) of Eq.~(\ref{eq-Prob-P1-MA-MR-b}) is that the right hand side in Inequality~(\ref{eq:prob-P4-MA-MR-c-SINR-constraint}) of Problem~(P4-MA-MR-c) has $ t \gamma_i$, whereas the right hand side in Inequality~(\ref{eq:prob-P1-MA-MR-b-SINR-constraint}) of Problem~(P1-MA-MR-b) has $ \gamma_i$. Hence, we can apply the discussed approach of solving Problem~(P1-MA-MR-b) to solve Problem~(P4-MA-MR-c). Specifically, if the constraints on   $\boldPhi_{\ell}$ in Eq.~(\ref{eq:prob-P1a-rank-constraint}) are  in the form of Eq.~(\ref{eq-constraint-Phi-ell}) (i.e., $\boldPhi_{\ell}:= \text{diag}(\beta_{\ell,1}  e^{j \theta_{\ell,1}}, \ldots, \beta_{\ell,N_{\ell}} e^{j \theta_{\ell,N_{\ell}}})$ for $\ell \in\{1,\ldots,L\}$), we define $\bm{\phi}_{\ell}$, $\bm{a}_{\ell,i,q}(\bm{w}_{k})$,  $b_{i,q}(\bm{w}_{k})$, $\bm{A}_i(\bm{w}_{k})$, and $\bm{\phi}$ according to Eq.~(\ref{eq-define-bm-phi-ell-0-L})~(\ref{eq-def-ai-wk-P1-MA-MR})~(\ref{eq-def-b-wk-P1-MA-MR})~(\ref{eq-def-Ai-matrix-wk-q}) and~(\ref{eq-boldsymbol-all-phi}). Then in a way similar to the derivation leading to Inequality~(\ref{eq-hi-def-ai-wj-def-b-wj-square-v3-L}),  we write the constraint in Eq.~(\ref{eq:prob-P4-MA-MR-c-SINR-constraint}) of Problem~(P4-MA-MR-c) as
\begin{align}
 & \sum_{q=1}^{Q_i} \left\{  \begin{bmatrix} 
\bm{\phi}^H,
1
\end{bmatrix} \bm{A}_{i,q}(\bm{w}_{k}) \begin{bmatrix} 
\bm{\phi} \\
1
\end{bmatrix} +   {b}_{i,q}(\bm{w}_{k}) {b}_{i,q}^H(\bm{w}_{k})\right\} \nonumber  \\ & \geq \hspace{-2pt} t \gamma_i \hspace{-2pt}\left\{\hspace{-2pt}  \sigma^2_i\hspace{-2pt}+\hspace{-10pt} \sum\limits_{j\in\{1,\ldots,g\}\setminus\{k\}} \hspace{-1pt} \sum_{q=1}^{Q_i} \hspace{-2pt}\left\{\hspace{-6pt}  \begin{array}{l} \begin{bmatrix} 
\bm{\phi}^H,
1
\end{bmatrix} \bm{A}_{i,q}(\bm{w}_{j}) \begin{bmatrix} 
\bm{\phi} \\
1
\end{bmatrix} \\ +   {b}_{i,q}(\bm{w}_{j}) {b}_{i,q}^H(\bm{w}_{j})\end{array}\hspace{-6pt}\right\}\hspace{-3pt}\right\}\hspace{-2pt}.\label{eq-hi-def-ai-wj-def-b-wj-square-v3-L-P4-MA-MR-c}
\end{align}

Then defining $\boldsymbol{V}$ according to Eq.~(\ref{eq-boldsymbol-V-v-K}) with $\bm{v}$ defined in Eq.~(\ref{eq-boldsymbol-v-K}), similar to the process of formulating Problem~(P1-MA-MR-c) of Eq.~(\ref{eq-Prob-P1-MA-MR-c}), we can convert Problem~(P4-MA-MR-c) into the following Problem~(P4-MA-MR-d):
\begin{subequations} \label{eq-Prob-P4-MA-MR-d}
\begin{alignat}{2}
&\hspace{-40pt}\text{(P4-MA-MR-d)}: \nonumber \\
\Find~~~& \boldsymbol{V} \\
\mathrm{s.t.}~~~~~&   \sum_{q=1}^{Q_i} \left\{ \tr(\bm{A}_{i,q}(\bm{w}_{k})\boldsymbol{V} ) +   {b}_{i,q}(\bm{w}_{k}) {b}_{i,q}^H(\bm{w}_{k})\right\} \nonumber  \\ & \geq t \gamma_i \bigg\{  \sigma^2_i + \nonumber  \\ & \hspace{-25pt}\sum\limits_{j\in\{1,\ldots,g\}\setminus\{k\}} \, \sum_{q=1}^{Q_i} \hspace{-1pt}\left\{ \tr(\bm{A}_{i,q}(\bm{w}_{j})\boldsymbol{V} )  +   {b}_{i,q}(\bm{w}_{j}) {b}_{i,q}^H(\bm{w}_{j})\right\}\hspace{-2pt}\bigg\},  \label{eq:prob-P4-MA-MR-d-SINR-constraint}\\
&~~~~~~~\forall k \in\{1,\ldots,g\},~\forall i \in \mathcal{G}_k, \nonumber \\
 &\boldsymbol{V}_{(y+\sum_{\ell=0}^{x-1}N_{\ell}),\,(y+\sum_{\ell=0}^{x-1}N_{\ell})}  = {\beta_{x,y}}^2, \label{eq-Prob-P4-MA-MR-d-constraints-Vnn} \\
&~~~~~~~\forall x\in \{1,\ldots, L\},~\forall y\in \{1,\ldots, N_{x}\}, \nonumber \\
 &\boldsymbol{V}_{(1 +\sum_{\ell=1}^L N_{\ell}),\,(1 +\sum_{\ell=1}^L N_{\ell})} = 1, \label{eq-Prob-P4-MA-MR-d-constraints-Vnplus1}  \\
& \boldsymbol{V} \succeq 0, \\
& \text{rank}(\boldsymbol{V}) =1. \label{eq:prob-P4-MA-MR-d-rank-constraint}
\end{alignat}
\end{subequations}

The above constraint in Eq.~(\ref{eq-Prob-P4-MA-MR-d-constraints-Vnn}) is for the case where all $\beta_{x,y}|_{\begin{subarray}{l}x\in \{1,\ldots, L\},\\ y\in \{1,\ldots, N_{x}\} \end{subarray}}$ are predefined constants. A special case of particular interest is the case of all $\beta_{x,y}|_{\begin{subarray}{l}x\in \{1,\ldots, L\},\\ y\in \{1,\ldots, N_{x}\} \end{subarray}}$ being $1$ so that Eq.~(\ref{eq-Prob-P4-MA-MR-d-constraints-Vnn}) and Eq.~(\ref{eq-Prob-P4-MA-MR-d-constraints-Vnplus1}) can together be written as $\boldsymbol{V}_{n,n}  = 1  $ for $n\in \{1,\ldots, 1 +\sum_{\ell=1}^L N_{\ell}\}$.
If each $\beta_{x,y}|_{\begin{subarray}{l}x\in \{1,\ldots, L\},\\ y\in \{1,\ldots, N_{x}\} \end{subarray}}$ can take any value in $[0,1]$, then Eq.~(\ref{eq-Prob-P4-MA-MR-d-constraints-Vnn}) can be replaced by $\boldsymbol{V}_{(y+\sum_{\ell=0}^{x-1}N_{\ell}),\,(y+\sum_{\ell=0}^{x-1}N_{\ell})}\in [0,1] $ for $x\in \{1,\ldots, L\}$ and $y\in \{1,\ldots, N_{x}\}$. Similarly, we may consider the most general case where  some $\beta_{x,y}$ are predefined constants while other $\beta_{x,y}$ can vary.

The only non-convex part in Problem~(P4-MA-MR-d) is the rank constraint in Eq.~(\ref{eq:prob-P4-MA-MR-d-rank-constraint}). Hence, we adopt semidefinite relaxation (SDR) and drop Eq.~(\ref{eq:prob-P4-MA-MR-d-rank-constraint}) to obtain the following semidefinite programming Problem~(P4-MA-MR-e):
\begin{subequations} \label{eq-Prob-P4-MA-MR-e}
\begin{alignat}{2}
&\hspace{-30pt}\text{(P4-MA-MR-e)}: \nonumber \\
\Find~~~& \boldsymbol{V} \\
\mathrm{s.t.}~~~~~&   \sum_{q=1}^{Q_i} \left\{ \tr(\bm{A}_{i,q}(\bm{w}_{k})\boldsymbol{V} ) +   {b}_{i,q}(\bm{w}_{k}) {b}_{i,q}^H(\bm{w}_{k})\right\} \nonumber  \\ & \geq  t \gamma_i \bigg\{  \sigma^2_i + \nonumber  \\ & \hspace{-25pt}\sum\limits_{j\in\{1,\ldots,g\}\setminus\{k\}} \, \sum_{q=1}^{Q_i} \hspace{-1pt}\left\{ \tr(\bm{A}_{i,q}(\bm{w}_{j})\boldsymbol{V} )  +   {b}_{i,q}(\bm{w}_{j}) {b}_{i,q}^H(\bm{w}_{j})\right\}\hspace{-2pt}\bigg\},  \label{eq:prob-P4-MA-MR-e-SINR-constraint}\\
&~~~~~~~\forall k \in\{1,\ldots,g\},~\forall i \in \mathcal{G}_k, \nonumber \\
  &\boldsymbol{V}_{(y+\sum_{\ell=0}^{x-1}N_{\ell}),\,(y+\sum_{\ell=0}^{x-1}N_{\ell})}  = {\beta_{x,y}}^2, \label{eq-Prob-P4-MA-MR-e-constraints-Vnn} \\
&~~~~~~~\forall x\in \{1,\ldots, L\},~\forall y\in \{1,\ldots, N_{x}\}, \nonumber \\
 &\boldsymbol{V}_{(1 +\sum_{\ell=1}^L N_{\ell}),\,(1 +\sum_{\ell=1}^L N_{\ell})} = 1, \label{eq-Prob-P4-MA-MR-e-constraints-Vnplus1}  \\
& \boldsymbol{V} \succeq 0.
\end{alignat}
\end{subequations}

Problem~(P4-MA-MR-e) belongs to semidefinite programming and can be solved efficiently~\cite{luo2010semidefinite}. Moreover, Problem~(P4-MA-MR-e) can be replaced by Problem~(P4-MA-MR-e$'$) below which may find better $\boldPhi$ and hence $\boldsymbol{V}$ to accelerate the alternating optimization process:
\begin{subequations} \label{eq-Prob-P4-MA-MR-e-prime}
\begin{alignat}{2}
&\hspace{-30pt}\text{(P4-MA-MR-e$'$)}: \nonumber \\ \max_{\boldsymbol{V},\boldsymbol{\alpha}}& \sum_{k=1}^g \sum_{i \in \mathcal{G}_k} \alpha_i \\
\mathrm{s.t.}~~~~~&   \sum_{q=1}^{Q_i} \left\{ \tr(\bm{A}_{i,q}(\bm{w}_{k})\boldsymbol{V} ) +   {b}_{i,q}(\bm{w}_{k}) {b}_{i,q}^H(\bm{w}_{k})\right\} \nonumber  \\ & \geq  \alpha_i + t\gamma_i \bigg\{  \sigma^2_i + \nonumber  \\ & \hspace{-25pt}\sum\limits_{j\in\{1,\ldots,g\}\setminus\{k\}} \, \sum_{q=1}^{Q_i} \hspace{-1pt}\left\{ \tr(\bm{A}_{i,q}(\bm{w}_{j})\boldsymbol{V} )  +   {b}_{i,q}(\bm{w}_{j}) {b}_{i,q}^H(\bm{w}_{j})\right\}\hspace{-2pt}\bigg\},  \label{eq:prob-P4-MA-MR-e-prime-SINR-constraint}\\
&~~~~~~~\forall k \in\{1,\ldots,g\},~\forall i \in \mathcal{G}_k, \nonumber  \\
 &\boldsymbol{V}_{(y+\sum_{\ell=0}^{x-1}N_{\ell}),\,(y+\sum_{\ell=0}^{x-1}N_{\ell})}  = {\beta_{x,y}}^2, \label{eq-Prob-P4-MA-MR-e-prime-constraints-Vnn} \\
&~~~~~~~\forall x\in \{1,\ldots, L\},~\forall y\in \{1,\ldots, N_{x}\}, \nonumber \\
 &\boldsymbol{V}_{(1 +\sum_{\ell=1}^L N_{\ell}),\,(1 +\sum_{\ell=1}^L N_{\ell})} = 1, \label{eq-Prob-P4-MA-MR-e-prime-constraints-Vnplus1}  \\
& \boldsymbol{V} \succeq 0, \\
& \alpha_i \geq 0, ~\forall k \in\{1,\ldots,g\},~\forall i \in \mathcal{G}_k.
\end{alignat}
\end{subequations}

After  a candidate solution $\boldsymbol{V}$ is obtained by solving Problem~(P4-MA-MR-e) or Problem~(P4-MA-MR-e$'$), appropriate post-processing such as Gaussian randomization~\cite{so2007approximating} can be applied to convert the candidate solution into a solution which satisfies the rank constraint in Eq.~(\ref{eq:prob-P4-MA-MR-d-rank-constraint}).

Combining the above discussion of alternating optimization, we present our method to solve Problem~(P4-MA-MR) as Algorithm~\ref{Alg-Problem-P4-MA-MR}. 

\begin{algorithm*}
\caption{via alternating optimization to find $\W$ and $\bm{\Phi}_1,\ldots,\bm{\Phi}_L$ for Problem~(P4-MA-MR), which generalizes Problems~(P5-MA-MR)~(P6-MA-MR), \mbox{(P4-MA)--(P6-MA),} \mbox{(P4-MR)--(P6-MR),} and \mbox{(P4)--(P6).}} \label{Alg-Problem-P4-MA-MR}
\begin{algorithmic}[1]
\STATE Initialize  $\boldsymbol{\Phi}_{\ell}$ for $\ell \in\{1,\ldots,L\}$ as some initial (e.g., randomly generated) $ \boldsymbol{\Phi}_{\ell}^{(0)}:= \text{diag}(\beta_{\ell,1}  e^{j \theta_{\ell,1}^{(0)}}, \ldots, \beta_{\ell,N_{\ell}} e^{j \theta_{\ell,N_{\ell}}^{(0)}})$ which satisfies the constraints on $\boldsymbol{\Phi}_{\ell}$;
\STATE Set the iteration number $r \leftarrow 1$;
\WHILE{1}
\item[] \COMMENT{The ``while'' loop will end if Line~\ref{Alg-Problem-P4-MA-MR-break1} or \ref{Alg-Problem-P4-MA-MR-break2} is executed.}
\item[] \COMMENT{Comment: Optimizing $\W$ and $t$ given $\bm{\Phi}_1,\ldots,\bm{\Phi}_L$:}
\STATE Given $\bm{\Phi}_1,\ldots,\bm{\Phi}_L$ as $\bm{\Phi}_1^{(r-1)},\ldots,\bm{\Phi}_L^{(r-1)}$, use methods of Karipidis~\textit{et~al.}~\cite{karipidis2008quality} or other papers to solve \mbox{Problem~(P4-MA-MR-b)}  in Eq.~(\ref{eq-Prob-P4-MA-MR-b}), and post-process the obtained $\{\bm{X}_k\}|_{k=1}^g$ and $t$ to set $\boldsymbol{W}$ as some $\boldsymbol{W}^{(r)}:=[\bm{w}_1^{(r)},\ldots,\bm{w}_g^{(r)}]$ and set $t$ as some $t^{(r)}$; \label{Alg-opt-P4-MA-MR-0}
\item[] \vspace{-5pt}
\IF{$r\geq 2$}
\IF{$ \frac{t^{(r)}}{t^{(r-1)}}-1$ denoting the relative difference  between the object function values in consecutive iterations $r-1$ and $r$ is small}
\STATE \textbf{break};  \label{Alg-Problem-P4-MA-MR-break1}
\ENDIF
\ENDIF
\item[] \vspace{-5pt}
\item[] \COMMENT{Finding $\bm{\Phi}_1,\ldots,\bm{\Phi}_L$ given $\W$ and $t$:}
\STATE Given $\W$ as $\boldsymbol{W}^{(r)}$ and $t$ as $t^{(r)}$, solve Problem~(P4-MA-MR-e) in Eq.~(\ref{eq-Prob-P4-MA-MR-e}) or Problem~(P4-MA-MR-e$'$) in Eq.~(\ref{eq-Prob-P4-MA-MR-e-prime}), and denote the obtained~$\boldsymbol{V}$ as~$\boldsymbol{V}^{(r)}_{\text{SDR}}$;
\item[] \COMMENT{Comment: Gaussian randomization:}
\STATE Perform the eigenvalue decomposition on $\boldsymbol{V}^{(r)}_{\text{SDR}}$ to obtain a unitary matrix $\boldsymbol{U}$ and a  diagonal  matrix $\boldsymbol{\Lambda}$ such that $ \boldsymbol{V}^{(r)}_{\text{SDR}} = \boldsymbol{U} \boldsymbol{\Lambda}  \boldsymbol{U}^H $;\label{Alg-opt-Gaussian-randomization-decomposition}
\FOR{$z$ from $1$ to some sufficiently large $Z$}
\STATE Generate a random vector $\boldsymbol{r}^{(r)}_{z}$ from a circularly-symmetric complex Gaussian distribution $\mathcal{CN}(\boldsymbol{0},\boldsymbol{I}_{1 +\sum_{\ell=1}^L N_{\ell}})$ with zero mean and covariance matrix $\boldsymbol{I}_{1 +\sum_{\ell=1}^L N_{\ell}}$ (the identity matrix with size $1 +\sum_{\ell=1}^L N_{\ell}$); \STATE Compute $ \boldsymbol{v}^{(r)}_{z}  \leftarrow \boldsymbol{U} \boldsymbol{\Lambda}^{\frac{1}{2}} \boldsymbol{r}^{(r)}_{z}$;
\STATE With $v_{1 +\sum_{\ell=1}^L N_{\ell}}$ being the $(1 +\sum_{\ell=1}^L N_{\ell})$th element of $\boldsymbol{v}_{z}^{(r)}$, {take the first $\sum_{\ell=1}^L N_{\ell}$ elements of $\frac{\boldsymbol{v}_{z}^{(r)}}{v_{1 +\sum_{\ell=1}^L N_{\ell}}}$ to form a vector $\boldsymbol{w}^{(r)}_{z}$};
\STATE 
Scale each component of $\boldsymbol{w}^{(r)}_{z}$ independently to obtain $\bm{\phi}^{(r)}_{z}$ such that the $(y+\sum_{\ell=0}^{x-1}N_{\ell})$th element of $\bm{\phi}^{(r)}_{z}$ has magnitude ${\beta_{x,y}}$, for $x\in \{1,\ldots, L\}$ and $y\in \{1,\ldots, N_{x}\}$;
\ENDFOR
\IF{for $z\in\{1,2,\ldots,Z\}$, there is no $\bm{\phi}^{(r)}_{z}$ to ensure Eq.~(\ref{eq-hi-def-ai-wj-def-b-wj-square-v3-L-P4-MA-MR-c}) after setting $\bm{\phi}$ as $\bm{\phi}^{(r)}_{z}$}
\STATE \textbf{break};  \label{Alg-Problem-P4-MA-MR-break2}
\ELSE
\STATE Select one $\bm{\phi}^{(r)}_{z}$ according to some ordering among those ensuring Eq.~(\ref{eq-hi-def-ai-wj-def-b-wj-square-v3-L-P4-MA-MR-c}) and denote the selected one by $\bm{\phi}^{(r)}_{z^*}$;
\STATE Map $\bm{\phi}^{(r)}_{z^*}$ to some ${\bm{\phi}^{(r)}}$ to satisfy the constraint on $\bm{\phi}$ (e.g., discrete element values); 
\STATE For $\ell\in\{1,2,\ldots,L\}$, set $\bm{\Phi}_{\ell}$ as  $\bm{\Phi}_{\ell}^{(r)}:=\text{diag}\left((\bm{\phi}_{\ell}^{(r)})^H\right)$, where $\bm{\phi}_{\ell}^{(r)}|_{\ell\in\{1,2,\ldots,L\}}$ are defined such that $\bm{\phi}^{(r)} = \begin{bmatrix}
\bm{\phi}_1^{(r)} \\
\ldots \\
\bm{\phi}_L^{(r)}
\end{bmatrix}$;
\ENDIF
\STATE Update the iteration number $r  \leftarrow r + 1$;  \label{Alg-opt-update-r}
\ENDWHILE
\end{algorithmic}
\end{algorithm*}

%

 







\section{Related Work} \label{sec-Related-Work}

We discuss both related studies in wireless communications aided by IRSs and those without IRSs.

\textbf{IRS-aided wireless communications}. Since IRSs can be controlled to reflect incident wireless signals in a desired way, IRS-aided communications have recently received much attention in the literature~\cite{Junginproceedings,jung2018performance,guo2019weighted,huang2019Reconfigurable,huang2018largeenergy,fu2019intelligent,yu2019miso,nadeem2019intelligent,taha2019enabling,he2019cascaded,mishra2019channel}. The studies include analyses of data rates~\cite{Junginproceedings,jung2018performance,guo2019weighted}, optimizations of power or spectral efficiency~\cite{huang2019Reconfigurable,huang2018largeenergy,fu2019intelligent,yu2019miso}, and channel estimation~\cite{nadeem2019intelligent,taha2019enabling,he2019cascaded,mishra2019channel}. In these studies, IRSs are also referred to as \textit{large intelligent surface}~\cite{hu2018beyond,jung2018performance,huang2018largeenergy}, \textit{reconfigurable intelligent surface}~\cite{huang2019Reconfigurable,basar2019wireless}, \textit{software-defined surface}~\cite{basar2019large}, and \textit{passive intelligent mirrors}~\cite{huang2018achievable}~\cite{di2019reflection}. Interested readers can refer to~\cite{zhao2019survey,basar2019wireless,di2019smart2,qingqing2019towards} for surveys of IRS-aided  communications.

For IRS-aided communications between a base  station and mobile users, downlinks are investigated in~\cite{wu2018intelligent,wu2018intelligentfull,wu2019beamforming,guo2019weighted,nadeem2019large,huang2018largeenergy,huang2019Reconfigurable,fu2019intelligent,yu2019miso,liaskos2019interpretable}, while uplinks are studied in~\cite{Junginproceedings,jung2018performance}. In particular, for \textit{power control under QoS}, Problem~(P2) for the unicast setting has been addressed by Wu and Zhang~\cite{wu2018intelligent,wu2018intelligentfull,wu2019beamforming}, with the constraints on $\boldPhi$ of~(\ref{OptProb-power-unicast-Phi-constraint}) given in the form of  Eq.~(\ref{eq-def-phase-shift-matrix}) (i.e., $\boldPhi = \text{diag}(e^{j \theta_1}, \ldots, e^{j \theta_N})$). In~\cite{wu2018intelligent,wu2018intelligentfull}, each of $\theta_n|_{n \in \{1,\ldots,N\}}$ can take any value in $[0,2\pi)$. In contrast, in~\cite{wu2019beamforming}, each of  $\theta_n|_{n \in \{1,\ldots,N\}}$ can only take the following $\tau$ discrete values 
equally spaced on a circle for some positive integer $\tau$: $\big\{0, \frac{2\pi}{\tau}, \ldots, \frac{2\pi \cdot (\tau-1)}{\tau} \big\} $.


In addition to the above settings where IRSs aid communications between a base  station and mobile users,  direct communications between IRSs and mobile users   are analyzed in~\cite{jung2019performance,hu2017potential,hu2018beyond,hu2018capacity}.





%


\textbf{Previous studies on power control and QoS for wireless communications without IRSs.} Power control and QoS for wireless communications without IRSs have been investigated extensively in the literature. We now discuss some representative studies. First, for \textit{power control under QoS}, the unicast setting is considered by~\cite{bengtsson1999optimal,bengtsson2001optimal}, and the broadcast setting is studied by~\cite{sidiropoulos2006transmit,tran2013conic,luo2007approximation}, whereas the multicast setting is addressed by~\cite{karipidis2008quality}. Second, for \textit{\mbox{max-min} fair QoS}, the unicast setting is considered by~\cite{wiesel2005linear}, and  the broadcast setting is studied by~\cite{sidiropoulos2006transmit}, whereas the multicast setting is addressed by~\cite{karipidis2008quality,chang2008approximation,christopoulos2015multicast}.



\section{Conclusion} \label{sec-Conclusion}
In this paper, we formulate a comprehensive set of optimization problems for \textit{power control under QoS} and \textit{\mbox{max-min} fair QoS}. The optimizations are done by jointly designing the transmit beamforming of the BS and the phase shift matrix of the IRS. We address three kinds of traffic patterns from the BS to the MUs: unicast, broadcast, and multicast. We also consider the novel settings of \textit{multi}-antenna mobile users or/and \textit{multiple} intelligent reflecting surfaces. For all the optimizations  discussed above, we present detailed analyses to propose efficient algorithms.
 There are many future research directions to investigate. We list some as follows:   1) discussing the NP-hardness of our formulated optimization problems, 2) proving approximation bounds
 of our proposed algorithms, 3) conducting extensive experiments to validate our theoretical analyses and algorithms, 4) extending the models to take into account   channel estimation errors and mobilities of MUs  or/and LISs,  5) extending current studies of data transfer to wireless power transfer~\cite{mishra2019channel,wu2019weighted}, 6) applying our analyses to other optimization problems such as those for weighted sum-rate~\cite{guo2019weighted} or weighted power transfer~\cite{wu2019weighted}.
 




 

 \end{document}